\documentclass[aps,10pt,prd,nofootinbib,twocolumn,superscriptaddress,preprintnumbers,balancelastpage,longbibliography]{revtex4-1}

\usepackage{bm}
\usepackage{booktabs}
\usepackage{siunitx}
\usepackage{placeins}
\usepackage{amsmath,amssymb,mathtools,bm}
\usepackage{graphicx, color, hepunits}
\usepackage[dvipsnames]{xcolor}
\usepackage{cancel}
\usepackage[normalem]{ulem}
\usepackage{float}
\usepackage{multirow}
\usepackage{comment}
\usepackage{hyperref} %Automatically links \label and \ref commands; Always load last
\hypersetup{
    colorlinks=true,       % false: boxed links; true: colored links
    linkcolor=blue,        % color of internal links
    citecolor=blue,        % color of links to bibliography
    filecolor=magenta,     % color of file links
    urlcolor=blue          % color of external links
}
\usepackage[utf8]{inputenc}
\usepackage[english]{babel}
\usepackage{physics}
\usepackage{isomath}

\newcommand{\alphat}{\alpha}
\newcommand{\betat}{\beta}
\newcommand{\gammat}{\gamma}
\newcommand{\lambdat}{\lambda}
\newcommand{\kappat}{\kappa}
\newcommand{\mut}{\mu}

\newcommand{\tuple}{{{\Upsilon}}}
\newcommand{\Mt}{M}
\newcommand{\Ft}{F}
\newcommand{\Phit}{\Phi}
\newcommand{\Lambdat}{\Lambda}
\newcommand{\First}{\Xi}
\newcommand{\cc}{}

\makeatletter
\newcounter{savesection}
\newcounter{apdxsection}
\renewcommand\appendix{\par
  \setcounter{savesection}{\value{section}}%
  \setcounter{section}{\value{apdxsection}}%
  \setcounter{subsection}{0}%
  \gdef\thesection{\@Alph\c@section}}
\newcommand\unappendix{\par
  \setcounter{apdxsection}{\value{section}}%
  \setcounter{section}{\value{savesection}}%
  \setcounter{subsection}{0}%
  \gdef\thesection{\@arabic\c@section}}
\makeatother

\begin{document}

\title{Prospects for gravitational wave and ultra-light dark matter detection with binary resonances beyond the secular approximation}
\author{Joshua~W.~Foster}
\affiliation{Astrophysics Theory Department, Theory Division, Fermilab, Batavia, IL 60510, USA}
\affiliation{Kavli Institute for Cosmological Physics, University of Chicago, Chicago, IL 60637, USA}
\email{jwfoster@fnal.gov}
\author{Diego~Blas}
\affiliation{Institut de F\'{i}sica d’Altes Energies (IFAE), The Barcelona Institute of Science and Technology,
Campus UAB, 08193 Bellaterra (Barcelona), Spain}
\affiliation{Instituci\'{o} Catalana de Recerca i Estudis Avan\c{c}ats (ICREA), Passeig Llu\'{i}s Companys 23, 08010 Barcelona, Spain}
\author{Adrien~Bourgoin}
\affiliation{LTE, Observatoire de Paris, Universit\'e PSL, Sorbonne Universit\'e, Universit\'e de Lille, LNE, CNRS 61 Avenue de l'Observatoire, 75014 Paris, France}
\author{Aurelien~Hees}
\affiliation{LTE, Observatoire de Paris, Universit\'e PSL, Sorbonne Universit\'e, Universit\'e de Lille, LNE, CNRS 61 Avenue de l'Observatoire, 75014 Paris, France}
\author{M\'iriam~Herrero-Valea}
\affiliation{Institut de F\'{i}sica d’Altes Energies (IFAE), The Barcelona Institute of Science and Technology,
Campus UAB, 08193 Bellaterra (Barcelona), Spain}
\affiliation{Barcelona Supercomputing Center (BSC), Plaça d'Eusebi Güell 1-3, 08034 Barcelona, Spain}
\author{Alexander~C.~Jenkins}
\affiliation{Kavli Institute for Cosmology, University of Cambridge, Madingley Road, Cambridge CB3 0HA, UK}
\affiliation{DAMTP, University of Cambridge, Wilberforce Road, Cambridge CB3 0WA, UK}
\author{Xiao~Xue}
\affiliation{Institut de F\'{i}sica d’Altes Energies (IFAE), The Barcelona Institute of Science and Technology,
Campus UAB, 08193 Bellaterra (Barcelona), Spain}

\preprint{FERMILAB-PUB-25-0092-T}

\date{\today}
\begin{abstract}
    Precision observations of orbital systems have recently emerged as a promising new means of detecting gravitational waves and ultra-light dark matter, offering sensitivity in new regimes with significant discovery potential.
    Such searches rely critically on precise modeling of the dynamical effects of these signals on the observed system; however, previous analyses have mainly only relied on the secularly-averaged part of the response.
    We introduce here a fundamentally different approach that allows for a fully time-resolved description of the effects of oscillatory metric perturbations on orbital dynamics.
    We find that gravitational waves and ultra-light dark matter can induce large oscillations in the orbital parameters of realistic binaries, enhancing the sensitivity to such signals by orders of magnitude compared to previous estimates.
\end{abstract}
\maketitle
\tableofcontents

%%%%%%%%%%%%%%%%%%%%%%%%%%%%%%%%%%%%%%%%%%%%%%%%%%%%%%%%%%%%%%%%%%%%%%%%
\section{Introduction}
%%%%%%%%%%%%%%%%%%%%%%%%%%%%%%%%%%%%%%%%%%%%%%%%%%%%%%%%%%%%%%%%%%%%%%%%

The direct detections of gravitational waves at frequencies $\nu \approx 100$\,Hz by ground-based interferometers, and the current strong evidence of gravitational waves at $\nu\approx 2-10\,\mathrm{nHz}$ from pulsar timing arrays, have transformed our capacity to observe the Universe. Still, when compared with the much wider frequency band where astrophysical and cosmological signals may be lurking, it is clear that many discoveries are still to be uncovered in the coming years. The most relevant effort in this front will be LISA, an ESA-led proposal to cover the mHz band with space-based laser interferometers, while next-generation ground-based detectors and new radio telescopes may continue expanding their reach.

Despite these prospects, some bands of frequencies will not be efficiently covered by `standard' approaches. In particular, GWs from tens of nHz to the lower edge of the LISA band (0.1 mHz) represent a gap in future sensitivities where many interesting signals are expected\footnote{The band above 10 kHz, corresponding to `high-frequency gravitational waves' is also poorly covered. Ideas to explore it efficiently are currently under exploration \cite{Aggarwal:2025noe}.} \cite{Sesana:2019vho,Blas:2021mqw}. This `$\mu$Hz gap' is expected to be pervaded by signals from a broad range of astrophysical and cosmological phenomena \cite{Sesana:2019vho}, which has motivated various ideas to try to access these frequencies \cite{Sesana:2019vho,Fedderke:2021kuy,Wang:2022sxn,Blas:2016ddr,Blas:2021mqw,Jaraba:2023djs,Lu:2024yuo}. 

In this work, we revisit the proposal of \cite{Hui:2012yp,Blas:2021mqw,Blas:2021mpc} to detect GWs via their resonant absorption by binary systems. GWs in the $\mu$Hz band oscillate at frequencies similar to the orbital frequencies of objects in the Solar system (from days to years). As a result, their influence on the gravitational dynamics of orbiting bodies may result in resonant effects that increase their impact to observable levels.
In particular, precise tracking of the Moon's orbit or artificial satellites around the Earth via laser ranging, as well as the timing of binary pulsars, have recently been investigated as promising approaches to filling the $\mu$Hz gap in the near future~\cite{Hui:2012yp,Blas:2021mpc,Blas:2021mqw}.

Similar approaches can also be used to study a wider variety of external forces on binary systems arising in fundamental physics. 
For example, the local gravitational field of dark matter (DM) may induce perturbations that could generate measurable effects. 
As in the case of GWs, these perturbations are particularly relevant if the DM-induced gravitational potential fluctuates at frequencies $f\sim\mu\mathrm{Hz}$. This is the case for ultra-light DM (ULDM) candidates with masses $m_\mathrm{DM}\sim10^{-20}\,\mathrm{eV}/c^2$, which undergo coherent narrowband oscillations at $\mu\mathrm{Hz}$ frequencies in large spatial regions, as well as for candidates with (much) higher masses, which give rise to stochastic, broadband oscillations in this frequency band~\cite{Kim:2023pkx}.
As a result, as suggested in Refs.~\cite{Blas:2016ddr,Rozner:2019gba,Blas:2019hxz}, binary pulsars can also be a sensitive probe of ULDM.  

Previous investigations of these effects have averaged over the orbital timescale to study the secular response of the binary to GW/ULDM perturbations.
While convenient for analytical calculations, these approaches discard valuable observational information encoded in sub-orbital timescales (see however \cite{Rozner:2019gba,Desjacques:2020fdi}).
Here we present a new theoretical framework in which we are able to capture the \textit{fully time-resolved evolution} of a generic binary system under oscillatory perturbations such as these.
Our results substantially surpass previous methods based on secular approximations, yielding observational sensitivities that are orders of magnitude stronger.
Barring possible degeneracies and noise systematics (an analysis of which will follow in a future paper), we find that we may be on the verge of GW detection with \emph{existing} laser-ranging/pulsar-timing data.
This article is devoted to the formalism behind this claim, while Ref.~\cite{Blas:2024PRLForward} summarizes our forecast sensitivity with current and future datasets.

The remainder of the paper is organized as follows. In Sec.~\ref{sec:pert_bin}, we review the formalism of the osculating element treatment of orbital dynamics and develop a perturbative treatment of two-body orbits acted upon by external accelerations based on the method of variation of parameters. In Sec.~\ref{sec:AnalyticResults}, we use our perturbative treatment to demonstrate the scaling behavior of orbital perturbations due to either on- or off-resonance external accelerations. We then proceed to specialize to evaluate orbital perturbations due to either GWs or ULDM, beginning with a treatment of deterministic GWs in Sec.~\ref{sec:det_GW}, followed by a treatment of stochastic GWs and stochastic fluctuations of ULDM in Sec.~\ref{sec:stochastic_gw} and Sec.~\ref{sec:stochastic_uldm}, respectively. Using these results, in Sec.~\ref{sec:ObservationalSensitivities}, we evaluate observational sensitivities via lunar laser ranging and via pulsar timing to GWs and ULDM. We then validate these analytic calculations of the observational sensitivities with numerical and Monte Carlo case studies in Sec.~\ref{sec:CaseStudies}. In Sec.~\ref{sec:beyond_isolation}, we relax the assumption of otherwise isolated binary orbits for GW or ULDM detection by considering the effect of the three-body force of the Sun's gravity on the Earth-Moon system and the Earth-Moon tidal interaction, demonstrating the flexibility and suitability of the computational framework we develop here for the study of realistic systems. Finally, we provide some concluding remarks in Sec.~\ref{sec:Conclusion}.

Beyond the main results presented in this manuscript, we provide some additional results of interest as appendices. Apps.~\ref{app:GWs}-\ref{app:FisherInformation} contain some technical calculations for the main body of the paper. In App.~\ref{app:RecoveringSecular}, we demonstrate how the results and sensitivity formalism we develop here may be used to recover the results of \cite{Blas:2021mpc, Blas:2021mqw}. In App.~\ref{app:MonochromaticCoherence} we develop the coherent limit of the ULDM background in the Milky Way.
Finally, in App.~\ref{app:NetworkCorrelations}, we present the first steps to generalize the results presented in this work to the case of multiple binary orbits that experience correlated perturbations. Though a detailed study is beyond the scope of this work, a network of laser-ranged satellites orbiting, \textit{e.g.}, the Earth or the Moon, could enable a robust detection with enhanced sensitivity via the particular correlation structure imparted by either GWs or ULDM, in analogy to the Hellings-Down curve for PTAs. 

%%%%%%%%%%%%%%%%%%%%%%%%%%%%%%%%%%%%%%%%%%%%%%%%%%%%%%%%%%%%%%%%%%%%%%%%
\section{Perturbed binary dynamics}\label{sec:pert_bin}
%%%%%%%%%%%%%%%%%%%%%%%%%%%%%%%%%%%%%%%%%%%%%%%%%%%%%%%%%%%%%%%%%%%%%%%%

We begin with a general treatment of orbital dynamics, following the method of osculating elements, see \textit{e.g.}~\cite{Poisson_Will_2014}. The variables we use are the semilatus rectum $p$, the eccentricity $e$, the inclination $\iota$, the longitude of the ascending node $\Omega$, the longitude of pericenter $\omega$, and the true anomaly $f$.\footnote{For low eccentricity orbits, alternate choices of orbital element parametrizations may be desirable, but the general computational approach of this work remains valid.} We use these parameters to form the $6$-tuple $\tuple^{\alphat}\equiv \{p,e,\iota,\Omega,\omega,f\}$. For an isolated Newtonian binary, only the true anomaly $f$ is dynamic, representing the time-evolving phase of the orbit; the remaining five orbital elements are constant. In the presence of additional external forces, the time-evolving orbital elements describe the instantaneous Keplerian orbit of the system that would describe the subsequent orbital dynamics if the external force were to abruptly vanish.

We consider a general external or non-Newtonian force acting on the binary, which has the effect of producing an acceleration of the form 
\begin{equation}
    \bm{a} = \mathcal{R} \hat{\bm{r}} + \mathcal{S} \hat{\bm{\theta}} +  \mathcal{W} \hat{\bm{z}},
\end{equation}
where $\hat{\bm{r}}$, $ \hat{\bm{\theta}}$, and $\hat{\bm{z}}$ are the basis vectors defined in the binary system's cylindrical coordinates. In terms of the six orbital elements, these unit vectors are given by 
\begin{equation}
\begin{gathered}
    \hat{\bm{r}} = 
    \begin{bmatrix}
    \cos \Omega \cos (\omega + f) - \cos \iota \sin \Omega \sin (\omega + f) \\
    \sin \Omega \cos (\omega + f) + \cos \iota \cos \Omega \sin (\omega + f) \\
    \sin \iota \sin (\omega + f)    
    \end{bmatrix}, \\
      \hat{\bm{\theta}} = 
    \begin{bmatrix}-\cos \Omega \sin (\omega + f) - \cos \iota \sin \Omega \cos (\omega + f) \\ 
    -\sin \Omega \sin (\omega + f) + \cos \iota \cos \Omega \cos (\omega + f) \\
    \sin \iota \cos (\omega + f)
    \end{bmatrix},\\
    \hat{\bm{z}} = 
\begin{bmatrix}
\sin \iota \sin \Omega \\
-\sin \iota \cos \Omega \\
\cos \iota
\end{bmatrix}.
\end{gathered}
\label{eq:BasisVectors}
\end{equation}

\begin{figure}[!ht]
\includegraphics[width=0.4\textwidth]{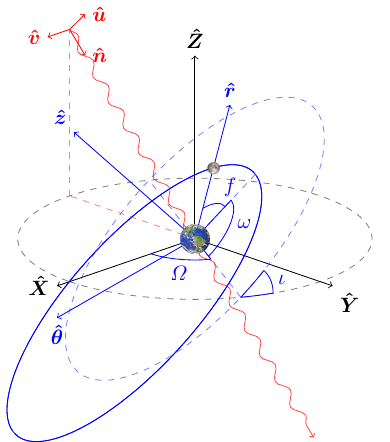}
\caption{A diagram of the coordinate systems and dynamical variables for orbital systems referenced in this work. The observer coordinates are defined by the Cartesian basis vectors $\hat{\bm{X}}$, $\hat{\bm{Y}}$ and $\hat{\bm{Z}}$. The binary frame is defined as centered on one of the two binary elements, with the dynamical variables $\iota$, $\Omega$, and $\omega$ defining the orientation of the orbit in the observer frame. Meanwhile, the semilatus rectum $p$ and the eccentricity $e$, not shown here, jointly term in the size and ellipticity of the orbit, while the true anomaly $f$ serves as an angular variable describing the phase of the orbit. Our basis vectors $\hat{\bm{r}}$, $\hat{\bm{\theta}}$, and $\hat{\bm{z}}$ are those associated with the cylindrical coordinates of the binary frame oriented such that $\hat{\bm{z}}$ is orthogonal to the orbital plane. A gravitational wave, not shown to scale, propagating in the $\hat{\bm{n}}$ direction is also shown; the collectively orthonormal unit vectors $\hat{u}$ and $\hat{v}$ will be used to define the polarization basis for this propagating tensor mode.}
\label{fig:CoordinatesFrames}
\vspace{-2ex}
\end{figure}

The equations of motion for the orbital elements are
\begin{equation}
\begin{aligned}
    \dot p &= 2 \sqrt{\frac{p^3}{G M}} \frac{1}{1+e\cos(f)} \mathcal{S}, \\
    \dot e &= \sqrt{\frac{p}{G M}} \\
    &\times\left[ \sin(f) \mathcal{R} + \frac{2 \cos(f) + e[1+\cos^2(f)]}{1 + e \cos(f)} \mathcal{S} \right], \\
    \dot \iota &= \sqrt{\frac{p}{G M}} \frac{\cos(\omega + f)}{1+e\cos(f)} \mathcal{W}, \\
    \dot \Omega &= \csc(\iota) \sqrt{\frac{p}{G M}} \frac{\sin(\omega + f)}{1+e\cos(f)} \mathcal{W}, \\
    \dot \omega &= \frac{1}{e} \sqrt{\frac{p}{G M}} \left[ -\cos(f)\mathcal{R} + \frac{2+ e \cos(f)}{1+e\cos(f)} \sin(f)\mathcal{S} \right. \\
    &\quad \left. - e \cot(\iota) \frac{\sin(\omega + f)}{1+e\cos(f)} \mathcal{W} \right], \\
    \dot f &= \sqrt{\frac{G M}{p^3}} (1+e\cos(f))^2\\
    &+ \frac{1}{e} \sqrt{\frac{p}{G M}} \left[ \cos(f)\mathcal{R} 
     - \frac{2+e\cos(f)}{1+ e\cos(f)} \sin(f)\mathcal{S} \right],
\end{aligned}
\label{eq:EoM}
\end{equation}
where $M$ is the total mass of the two objects in the binary system and $G$ is the gravitational Newton's constant. Note that these equations of motion make no assumption of the magnitude of $\bm{a}$ relative to the Newtonian gravitational interaction $ G M \bm{r} / r^3$. These coordinates and orbital elements are illustrated in Fig.~\ref{fig:CoordinatesFrames}.

To facilitate a more compact calculation, it is convenient to define the matrix $\Mt^{\alphat b}$ which is the response of $\tuple^{\alphat}$ to an acceleration, with $b$ indexing its $\hat{\bm{r}}$, $\hat{\bm{\theta}}$, and $\hat{\bm{z}}$ directions. To work in a Cartesian basis for our external forces, we define $\bm{e}_{ab}$ as the change of basis matrix from Cartesian to cylindrical coordinates. Explicitly, we have
\begin{equation}
\begin{split}
&\Mt(\tuple) = \sqrt{\frac{p}{GM}} \times \\
&\begin{bmatrix}
0 & \frac{2p}{1 + e \cos f} & 0 \\
\sin f & \frac{2 \cos f + e \left[1 + \cos^2(f)\right]}{1 + e \cos f} & 0 \\
0 & 0 & \frac{\cos (\omega + f)}{1 + e \cos f} \\
0 & 0 & \frac{\sin (\omega + f)}{(1 + e \cos f) \sin (\iota)} \\
-\frac{\cos f}{e} & \frac{\sin f \left(2 + e \cos f\right)}{e (1 + e \cos f)} & -\frac{\cot(\iota) \sin (\omega + f)}{1 + e \cos f} \\
\frac{\cos f}{e} & -\frac{\sin(f) \left(2 + e \cos f\right)}{e (1 + e \cos f)} & 0
\end{bmatrix},
\end{split}
\end{equation}
and, from Eq.~\eqref{eq:BasisVectors},
\begin{equation}
\bm{e}(\tuple)= \begin{bmatrix} \hat{\bm{r}} & \hat{\bm{\theta}} & \hat{\bm{z}}
\end{bmatrix}^T.
\end{equation}
These definitions allow us to write Eq.~\eqref{eq:EoM} in the simple form
\begin{equation}
\begin{aligned}
    \dot{\tuple}^{\alphat}(t) &= \Mt^{\alphat b}(\tuple) \bm{e}_{bc}(\tuple) \bm{a}^c(\tuple, t) \\
    &\quad + \sqrt{\frac{GM}{p^3}}(1+e\cos(f))^2 \delta^{\alphat 6},
\end{aligned}
\end{equation} 
where $\bm{a}^c$ is $c^\mathrm{th}$ Cartesian component of the acceleration and where $\delta$ is the Kronecker delta. We have allowed $\bm{a}$ to depend on both the state vector $\tuple^\alpha$ and the time~$t$. Hereafter, we use the convention that Greek indices denote elements of the 6-tuple $\tuple^\alpha$, while Latin indices $b,c$ indicate coordinate labels. 

In this work, our interest is the detection prospects for new accelerations associated with GWs and ULDM that are expected to be perturbatively weak, motivating a linearized treatment of Eqs.~\eqref{eq:EoM}. However, there may be other substantial forces beyond just the binary gravitational interaction and possible effects from GWs/ULDM acting on the system.  In later sections, we will consider the impact of the tidal force on the Earth-Moon system and that of a third external body. We will therefore separate our forces into the ``signal'' acceleration $\bm{a}_\mathrm{sig}$, which is treated at linear order, and all ``background'' accelerations $\bm{a}_\mathrm{bkg}$, which are fully treated at the zeroth-order of a perturbative approach we now outline. 

We begin with the equations of motion at zeroth order in $\bm{a}_\mathrm{sig}$, which evolve in the fully nonlinear system as
\begin{equation}
    \dot{\tuple}_0^{\alphat}(t) = \Ft_0^{\alphat}(\tuple_0, t), 
\end{equation}
where
\begin{equation}
    \begin{aligned}
         \Ft_0^{\alphat}(\tuple_0, t) &=  \Mt^{\alphat b}(\tuple_0) \bm{e}_{bc}(\tuple_0) \bm{a}_\mathrm{bkg}(\tuple_0, t)^c \\
    &\quad + \sqrt{\frac{GM}{p_0^3}} [1 + e_0 \cos(f_0)]^2 \delta^{\alphat 6},
    \label{Eq:ZerothOrder}
\end{aligned}
\end{equation}
represents the dynamics in the absence of $\bm{a}_\mathrm{sig}$. 
A numerical (sometimes even analytical) solution $\tuple_0(t)$ can be straightforwardly evaluated for simple forms of $\bm{a}_\mathrm{bkg}$. Once we have obtained the zeroth-order solution, we proceed to evaluate the solution at linear order in $\bm{a}_\mathrm{sig}$. We have
\begin{equation}
   \dot{\tuple}_1^{\alphat}  = \Ft_0^{\alphat \betat}(t) \tuple_1^{\betat}+ \Ft_1^{{\alphat}}(\tuple_0(t), t),
   \label{eq:general_linear}
\end{equation}
where
\begin{equation}
    \begin{gathered}
        \Ft_0^{\alphat \betat}(t) = \frac{\partial \Ft_0^{\alphat}}{\partial \tuple^{\betat}} \bigg|_{\tuple = \tuple_0(t)}, 
        \end{gathered}
\end{equation}
and
\begin{equation}
    \begin{gathered}
   \Ft_1^{\alphat}(t) = \Mt^{\alphat b}(\tuple_0(t)) \bm{e}_{bc}(\tuple_0(t)) \bm{a}_\mathrm{sig}(\tuple_0(t), t)^c.
   \label{Eq:FirstOrder}
\end{gathered}
\end{equation}
Qualitatively, $\Ft^{\alphat \betat}_0$ captures the change in the background forces due to any accumulated perturbation to the state of the system, while $\Ft^{\alphat}_1$ captures the leading order contribution of the perturbative signal acceleration. This perturbative approach remains valid while the accumulated perturbations are small compared to the corresponding unperturbed orbital elements and, in the case of the true anomaly, small with respect to $2 \pi$. This perturbative linearized approach can not reproduce other effects that may happen for larger modifications of the orbits that may be induced by external perturbations. Notably, it was recently remarked in \cite{Sharpe:2025voe} that resonant locking may be possible and yield new effects for strong enough backgrounds. In the following, we assume that the binaries of interest will be subjected only to perturbations where our premises are valid; see e.g. App.~\ref{app:PerturbativeDemonstration} for more details about the validity of our approach.

For deterministic signal accelerations, a direct solution of Eqs.~\eqref{Eq:ZerothOrder} and \eqref{Eq:FirstOrder} suffices to make predictions. However, for accelerations generated by stochastic gravitational wave backgrounds (SGWB) and the stochastic contribution of ULDM scenarios, only the defining statistics of $\bm{a}_\mathrm{sig}$ are known.
To handle these scenarios, it is useful to develop integral expressions for the solution of $\tuple_1$ given $\bm{a}_\mathrm{sig}$. % that facilitate taking expectation values of $\tuple_1$.

To do so, we begin by considering only the homogeneous part of Eq.~\eqref{eq:general_linear}  given by
\begin{equation}
\dot{\tuple}_1^{\alphat}  = \Ft_0^{\alphat\betat}(t) \tuple_1^{\betat}.
\label{Eq:FirstOrderHomogeneous}
\end{equation}
Next, we introduce the fundamental matrix $\Phit$, which is the linear mapping from the initial condition of $\tuple_1$ at time $t =0$ to its value $\tuple_1(t)$ as it evolves under Eq.~\eqref{Eq:FirstOrderHomogeneous},
\begin{equation}
    \tuple_1(t) = \Phit(t) \tuple_1(t = 0).
    \end{equation}
The fundamental matrix is governed by the differential equation and associated boundary condition
\begin{equation}
\dot{\Phit}_{\alphat \betat} = \Ft_0^{\alphat \gammat}(t) \Phit_{\gammat \betat}, \qquad \Phit(t = 0) = \bm{I}.
\label{Eq:FundamentalMatrix}
\end{equation}
After evaluating the fundamental matrix, we can use it to solve the inhomogeneous system in Eq.~\eqref{Eq:FirstOrder} in a manner that generalizes the standard method of integrating factors to the case of systems of differential equations \cite{BoyceDiprima}. The ultimate result is
\begin{equation}
\tuple_1^{\alphat}(t) = \Phit_{\alphat\betat}(t) \int_0^{t} \mathrm{d}\tau \, \Phit^{-1}_{\betat \gammat}(\tau) \Ft_1^{\gammat}(\tau).
\label{eq:solgen}
\end{equation}
This is particularly convenient because we can take moments of the solution, which will take the form of integrals of moments of the instantaneous signal accelerations. We emphasize that the entirety of this calculation has been, up to the assumption of perturbativity, fully agnostic regarding the form or origin of the signal acceleration. When we consider the specific cases of GWs and ULDM, these scenarios merely prescribe $\Ft_1$.

%%%%%%%%%%%%%%%%%%%%%%%%%%%%%%%%%%%%%%%%%%%%%%%%%%%%%%%%%%%%%%%%%%%%%%%%%%%%%%%%
\section{Analytic results for perturbed Keplerian binaries}
\label{sec:AnalyticResults}
%%%%%%%%%%%%%%%%%%%%%%%%%%%%%%%%%%%%%%%%%%%%%%%%%%%%%%%%%%%%%%%%%%%%%%%%%%%%%%%%

In this section, we study in detail the behavior of otherwise isolated Keplerian binaries subject to an external perturbation, following the approach of Sec.~\ref{sec:pert_bin}.  For this,  we will first evaluate the fundamental matrix for first-order perturbations to the Keplerian binary; as we will demonstrate, the periodicity of the unperturbed will admit a Floquet decomposition of the fundamental matrix. Next, we will use our Floquet decomposition to determine the time-dependence of perturbations which grow as they are driven by resonant and non-resonant first-order accelerations.  Later, we will use these results to develop an analytic understanding of how binary systems respond to both stochastic and deterministic sources, and the scaling behavior of our observational sensitivities.

%%%%%%%%%%%%%%%%%%%%%%%%%%%%%%%%%%%%%%%%%%%%%%%%%%%%%%%%%%%%%%%%%%%%%%%%%%%%%%%%
\subsection{Evaluating the fundamental matrix}
%%%%%%%%%%%%%%%%%%%%%%%%%%%%%%%%%%%%%%%%%%%%%%%%%%%%%%%%%%%%%%%%%%%%%%%%%%%%%%%%

For an isolated Keplerian binary, the $p$, $e$, $\iota$, $\Omega$, and $\omega$ parameters are constant at zeroth-order, with only the true anomaly that specifies the phase of the orbit being dynamical. Remaining agnostic regarding the details of the external perturbation, we can proceed to evaluate the fundamental matrix of the system, which evolves according to Eq.~\eqref{Eq:FundamentalMatrix}, with 
\begin{widetext}
\begin{equation}
\Ft_0^{\alphat \betat} = \sqrt{\frac{G M}{p_0^3}} \left(1 + e_0 \cos{f_0}\right)
\begin{pmatrix}
0 & 0 & 0 & 0 & 0 & 0 \\
0 & 0 & 0 & 0 & 0 & 0 \\
0 & 0 & 0 & 0 & 0 & 0 \\
0 & 0 & 0 & 0 & 0 & 0 \\
0 & 0 & 0 & 0 & 0 & 0 \\
-\frac{3}{2 p_0} \left(1 + e_0 \cos{f_0}\right) & 2 \cos{f_0}  &0&0&0& -2 e_0   \sin{f_0}
\end{pmatrix}.
\label{eq:SimpleF}
\end{equation}
\end{widetext}
The solution of $\Phit$ is of the form 
\begin{equation}
\Phit = 
\begin{pmatrix}
1 & 0 & 0 & 0 & 0 & 0\\ 
0 & 1 & 0 & 0 & 0 & 0\\ 
0 & 0 & 1 & 0 & 0 & 0\\ 
0 & 0 & 0 & 1 & 0 & 0\\ 
0 & 0 & 0 & 0 & 1 & 0\\ 
\Phit_{fp} & \Phit_{fe} & 0 & 0 & 0 & 1+ \Phit_{ff}
\end{pmatrix},
\end{equation}
with the nontrivial components of $\Phit$ evolving as 
\begin{equation}
\begin{gathered}
\dot{\Phit}_{f\alpha} = \Ft_0^{f\alpha} + \Ft_0^{ff}\Phit_{f\alpha},\\
\Phit_{f\alpha}(t = 0) = 0.
\label{eq:FundamentalMatrixODE}
\end{gathered}
\end{equation}
We can obtain the general solution for each of these components for the case $\bm{a}_\mathrm{bkg}(\tuple_0, t)=0$, though first, it is useful to introduce the zeroth-order eccentric anomaly defined by 
\begin{equation}
\begin{split}
    \tan \left(\frac{E_0(t)}{2}\right) = \sqrt{\frac{1-e_0}{1+e_0}} \tan \left(\frac{f_0(t)}{2}\right),
\end{split}
\end{equation}
and satisfying Kepler's equation 
\begin{equation}
t=\frac{P_0}{2\pi}(E_0(t)-e_0\sin(E_0(t))),
\end{equation}
where $P_0$ is the unperturbed orbital period.

The general solutions of Eq.~\eqref{eq:FundamentalMatrixODE} are then given by 
\begin{equation}
\begin{split}
\Phit_{fp}(t)&=-\frac{3\pi  \left[1+e_0\cos f_0(t)\right]^2 }{ p_0 (1-e_0^2)^{3/2}}  \frac{t}{P_0}, \\
\Phit_{fe}(t) &= - \frac{6 \pi e_0   [1+e_0 \cos f_0(t)]^2}{\left(1-e_0^2\right)^{5/2} } \frac{t}{P_0}\\
&+\frac{2-e_0^2}{\sqrt{1-e_0^2}} \frac{\sin E_0(t)-\sin E_0(0)}{[1-e_0 \cos E_0(t)]^2}\\
&-\frac{e_0}{2\sqrt{1-e_0^2}}\frac{\sin [2E_0(t)]-\sin [2 E_0(0)]}{[1-e_0 \cos E_0(t)]^2},\\
\Phit_{ff}(t) &= -1+\left(\frac{1+e_0 \cos f_0(t)}{1+e_0 \cos f_0(0)} \right)^2.
\end{split}
\end{equation}

These expressions can be framed in the context of Floquet's theorem \cite{barone1977floquet,Viebahn2020IntroductionTF}, which tells us that since $\Ft_0$ is periodic with period $P_0$, then $\Phit$ has a decomposition
\begin{equation}
\Phit(t) = \tilde{\Phit}(t) \exp[\Lambdat t],
\end{equation}
such that $\tilde{\Phit}(t)$ is a periodic matrix with period $P_0$ and $\Lambdat $ is a constant-valued matrix. Let us now inspect the fundamental solution matrix at $t = P_0$. Periodicity of the unperturbed orbit then tells us $f_0(P_0) = f_0(0) + 2 \pi$ and $E_0(P_0) = E_0(0) + 2 \pi$. Because our fundamental solution matrix is the identity at $t = 0$, we then have
\begin{equation}
\Phit(P_0) = \tilde{\Phit}(P_0) \exp[P_0\Lambdat ] = \exp[P_0\Lambdat ].
\end{equation}
Using our explicit evaluation of the fundamental solution matrix, we then determine $\Lambdat=\Lambdat[f_0(0)]$ where
\begin{equation}
\Lambdat[f]  =  -\frac{3 \pi}{P_0}\frac{[1 + e_0 \cos f]^2}{(1-e_0^2)^{3/2}}
 \begin{pmatrix}
0 & 0 & 0 & 0 & 0 & 0 \\
0 & 0 & 0 & 0 & 0 & 0 \\
0 & 0 & 0 & 0 & 0 & 0 \\
0 & 0 & 0 & 0 & 0 & 0 \\
0 & 0 & 0 & 0 & 0 & 0 \\
0 & 0 & 0 & 0 & 0 & 0 \\
\frac{1}{p_0} & \frac{2 e_0}{1-e_0^2}
& 0
& 0 & 0 & 0 
\end{pmatrix}.
\label{eq:Lambdat}
\end{equation}
Since $\Lambdat ^2 = 0$, we have
\begin{equation}
\exp[\Lambdat t] = \bm{I}+t\Lambdat . 
\end{equation}
It then follows that
\begin{equation}
\begin{gathered}
\tilde{\Phit}(t) =  \Phit(t)(\bm{I} - t \Lambdat ), \\
\Phit^{-1}(t) = \exp[-\Lambdat  t] \tilde{\Phit}^{-1}(t) = [\bm{I}-t\Lambdat ] \tilde{\Phit}^{-1}(t) 
\end{gathered}
\end{equation}
The critical observation here is that both the leading behavior of the fundamental solution matrix and its inverse scale with $t$.

%%%%%%%%%%%%%%%%%%%%%%%%%%%%%%%%%%%%%%%%%%%%%%%%%%%%%%%%%%%%%%%%%%%%%%%%
\subsection{The growth of perturbations}
%%%%%%%%%%%%%%%%%%%%%%%%%%%%%%%%%%%%%%%%%%%%%%%%%%%%%%%%%%%%%%%%%%%%%%%%

With our new understanding of the fundamental matrix, we are prepared to consider how an observable $\Delta_1$ which arises for the perturbing force evolves in time. We can now use Eq.~\eqref{eq:solgen} to determine\footnote{For compactness, we have denoted $\Mt^{\alphat d}(\tau) \equiv \Mt^{\alphat d}(\tuple_0(\tau))$, $\bm{e}_{dj}(\tau) \equiv \bm{e}_{dj}(\tuple_0(\tau))$, and $\bm{a}_\mathrm{sig}(\tau) = \bm{a}_\mathrm{sig}(\tuple_0(\tau), \tau)$.} $\tuple_1^{\alphat}(t)$. 
Note that $\tuple_0(t)$ and therefore all functions that depend exclusively on it will be periodic. 

Now, let us express $t = \bar t + n P_0$ where $n$ is a nonnegative integer that measures the number of complete periods undergone in time $t$ and $\bar t < P_0$. We can then rewrite the integral expression~\eqref{eq:solgen} for the perturbation as
\begin{widetext}
\begin{equation}
\begin{split}
\tuple_1^{\alphat}(t) =\tilde{\Phit}_{\alphat\betat} (\bar t)\bigg[&
\sum_{m=0}^{n-1}  \int_{\bar t}^{\bar t +P_0}\mathrm{d}\tau  \left[\left[ \bm{I} + \left((\bar t - \tau)  + (n-m) P_0\right) \Lambdat \right] \tilde{\Phit}^{-1}(\tau)\right]_{\betat  \gammat  }  \Mt^{\gammat  d}(\tau)\bm{e}_{dj}(\tau) \bm{a}_\mathrm{sig}(\tau + m P_0)^j \\
&+ \int_0^{\bar t} \mathrm{d}\tau  \big[\left[\bm{I} + (\bar t - \tau + n P_0) \Lambdat  \right] \tilde{\Phit}^{-1}(\tau)\big]_{\betat  \gammat  } \, \Mt^{\gammat  d}(\tau)  \bm{e}_{dj}(\tau) \bm{a}^j_\mathrm{sig}(\tau)\bigg]. 
\end{split}
\label{Eq:PerturbationIntegral}
\end{equation}
\end{widetext}
We now proceed to calculate the growth on-resonance when $\bm{a}_\mathrm{sig}$ is periodic with period $P_0/k$, with $k\in \mathbb{N}$, and the off-resonance growth when the signal acceleration demonstrates no such periodicity. 

%%%%%%%%%%%%%%%%%%%%%%%%%%%%%%%%%%%%%%%%%%%%%%%%%%%%%%%%%%%%%%%%%%%%%%%%
\subsubsection{On-resonance growth}\label{sec:on_resonance}
%%%%%%%%%%%%%%%%%%%%%%%%%%%%%%%%%%%%%%%%%%%%%%%%%%%%%%%%%%%%%%%%%%%%%%%%

For on-resonance accelerations, \textit{i.e.} $\bm{a}$ with periodicity of $P_0/k$, with $k\in \mathbb{N}^{\neq0}$, the sum over $m$ in Eq.~\eqref{Eq:PerturbationIntegral} yields
\begin{widetext}
\begin{equation}
\begin{split}
\tuple_1^{\alphat }(t) = \tilde{\Phit}_{\alphat \betat  } (\bar t)\bigg[& \int_{\bar t}^{\bar t + P_0} \mathrm{d}\tau  \left[\left[n \bm{I} + \left(n (\bar t - \tau)  + \frac{n(n+1)}{2}P_0\right) \Lambdat \right] \tilde{\Phit}^{-1}(\tau) \right]_{\betat  \gammat  } \Mt^{\gammat  d}(\tau)\bm{e}_{dj}(\tau) \bm{a}^j_\mathrm{sig}(\tau) \\
&+ \int_0^{\bar t} \mathrm{d}\tau  \left[\left[\bm{I} + (\bar t - \tau + n P_0) \Lambdat  \right]\tilde{\Phit}^{-1}(\tau)\right]_{\betat  \gammat  }  \, \Mt^{\gammat  d}(\tau)  \bm{e}_{dj}(\tau) \bm{a}^j_\mathrm{sig}(\tau)\bigg],
\end{split}
\label{Eq:PeriodicIntegral}
\end{equation}
\end{widetext}
The terms that scale as $n^2$ will come to dominate the growth of the perturbation in the large $n$ limit (at long times) so long as they do not vanish. This term is given by
\begin{equation}
\begin{split} 
\frac{n^2 P_0}{2}\Lambdat[f(\bar t)]_{\alphat \gammat}  \int_{\bar t}^{\bar t + P_0} \mathrm{d}\tau 
\Mt^{\gammat  d}(\tau)\bm{e}_{dj}(\tau) \bm{a}^j_\mathrm{sig}(\tau),
\end{split}
\end{equation}
where we used the (surprising) result $\tilde{\Phit} (\bar t)\Lambdat  \tilde{\Phit}^{-1}(\tau)=\Lambdat[f(\bar t)]$. Though this need not be true for more general orbits for different $\bm{a}_\mathrm{bkg}$, it considerably simplifies the evaluation for Keplerian ones. 

From expression~\eqref{eq:Lambdat}, we learn that the only orbital element that can achieve quadratic growth is $f_1$. Furthermore, this also implies that the only integrals which can contribute are
\begin{equation}
\int_{\bar t}^{\bar t + P_0} \mathrm{d}\tau   \Mt^{pd}(\tau)\bm{e}_{dj}(\tau) \bm{a}^j_\mathrm{sig}(\tau),
\label{eq:intP}
\end{equation}
and
\begin{equation}
\int_{\bar t}^{\bar t + P_0} \mathrm{d}\tau   \Mt^{ed}(\tau)\bm{e}_{dj}(\tau) \bm{a}^j_\mathrm{sig}(\tau).
\label{eq:intE}
\end{equation}
This implies that the quadratic growth in $f_1$ is only realized if there is a signal acceleration that generates a response in $p_1$ and $e_1$. From Eq.~\eqref{eq:EoM}, only an acceleration in the $\hat{\bm{\theta}}$ direction will induce such a response in $p_1$, while accelerations in either the $\hat{\bm{r}}$ and $\hat{\bm{\theta}}$ direction will induce this response in $e_1$. More transparently, accelerations that lie within the plane of the orbit, and only those, will generically generate the quadratic growth of $f_1$. 

Now we examine these integrals in detail. Let us consider an acceleration with generic components
in  $\hat{\bm{r}}(\tuple_0)$ and $\hat{\bm{\theta}}(\tuple_0)$,
\begin{equation}
    \bm{a}_\mathrm{sig}(t) = a_r(t) \hat{\bm{r}}(\tuple_0(t)) + a_\theta(t) \hat{\bm{\theta}}(\tuple_0(t))\,.
    \label{eq:ar_at}
\end{equation}
From Eqs.~\eqref{eq:EoM},  the integrals of Eq.~\eqref{eq:intP} and \eqref{eq:intE} are proportional to expressions
\begin{equation}
\begin{split}
 \sqrt{\frac{p_0}{GM}} \int_{\bar t}^{\bar t + P_0}\mathrm{d}\tau \left[ g_{\alphat r}(\tau) a_r(\tau) + g_{\alphat \theta}(\tau) a_\theta(\tau) \right],
\end{split}
\label{eq:IntegralofInterest}
\end{equation}
for $\alphat={p,e}$ and with 
\begin{equation}
\begin{gathered}
g_{p\theta}(t) = \frac{1}{1+e_0 \cos f_0(t)},~~~g_{er}(t) = \sin f_0(t),\\
g_{e\theta}(t) = \frac{2 \cos f_0(t) + e_0 \left[1 + \cos^2 f_0(t)\right]}{1 + e_0 \cos f_0(t)}.
\end{gathered}
\end{equation}
To further examine  Eq.~\eqref{eq:IntegralofInterest}, 
we will consider the Fourier decomposition of Keplerian orbits \cite{Poisson_Will_2014,Maggiore:1900zz,watson1944treatise}. Since $g_{\alphat d}(t)$ are periodic in $P_0$, they must admit a Fourier decomposition into a discrete set of modes 
\begin{equation}
    g_{\alphat d}(t)= \sum_{n} g_{\alphat d, n} \exp[2 \pi i nt /P_0].
\end{equation}
We evaluate these Fourier decompositions in App.~\ref{app:Fourier}, with the general result that $g_{\alphat d, n} \neq 0$ for any $n$. Since in this subsection we assumed that the acceleration is also periodic on $P_0$ (or integer fractions of it), $a_d(t)$ also admits a Fourier decomposition as $a_d(t) =  \sum_{n} a_{d, n} \exp[2 \pi i nt /P_0]$ with at least one $a_{d, n} \neq 0$ for each $d$. We then see that $g_{d\theta}$ and $a_d(t)$ must have some overlapping frequency support, so the integrals of Eq.~\eqref{eq:IntegralofInterest} must be nonvanishing.\footnote{It is possible that $a_d(t)$ and $g_{\alphat d}(t)$ could be precisely out of phase such that even though they have overlapping frequency support, Eq.~\eqref{eq:IntegralofInterest} will still vanish.}
To summarize, we have found that a resonant acceleration in either $\hat{\bm{r}}(\tuple_0)$ or $\hat{\bm{\theta}}(\tuple_0)$ will induce {\it quadratic growth} in time for $f_1$. 

We will also briefly comment on the fate of the remaining perturbations. Since $[\tilde{\Phit}(t) \Lambdat  \tilde{\Phit}^{-1}(\tau)]_{\alphat \gammat}$ is only nonzero for $\alphat =f$,  other perturbations can only receive contributions from terms in Eq.~\eqref{Eq:PeriodicIntegral} which do not depend on $\Lambdat $. Then for $\alphat  \neq f$, we have
\begin{equation}
\begin{split}
\tuple_1^{\alphat }(t) &= \tilde{\Phit}_{\alphat \betat  } (\bar t)\\ 
\times \bigg[&n \int_{\bar t}^{\bar t + P_0} \mathrm{d}\tau   \tilde{\Phit}_{\betat  \gammat  }^{-1}(\tau) \Mt^{\gammat  d}(\tau)\bm{e}_{dj}(\tau) \bm{a}^j_\mathrm{sig}(\tau) \\
&+ \int_0^{\bar t} \mathrm{d}\tau \tilde{\Phit}_{\betat\gammat}^{-1}(\tau) \, \Mt^{\gammat  d}(\tau)  \bm{e}_{dj}(\tau) \bm{a}^j_\mathrm{sig}(\tau)\bigg].
\end{split}
\label{Eq:PeriodicIntegralSpecific}
\end{equation}
Then we see that growth in the perturbation which is linear in $n$, the number of orbits (and hence time), is expected. Without going into detail, one can show from calculations nearly identical to those we performed for $f_1$ that the remaining perturbations will generically grow like $n$ when driven on resonance. 

Moreover, the geometric conditions for the forces to yield the resonant growth in orbital perturbations other than $f_1$ can be read directly off of Eq.~\eqref{eq:EoM}. Indeed, their evolution is more straightforwardly inspected as they evolve following first-order differential equations, whereas $f_1$ evolves following an implicitly second-order in time differential equation, and hence is accelerated. To summarize, resonant acceleration in the $\hat{\bm{r}}$ direction induces linear growth in $p_1$, $e_1$, $\omega_1$, resonant acceleration in the $\hat{\bm{\theta}}$ direction induces linear growth in $e_1$ and $\omega_1$, and resonant acceleration in the $\hat{\bm{z}}$ direction induces linear growth in $\iota_1$ and $\omega_1$.

%%%%%%%%%%%%%%%%%%%%%%%%%%%%%%%%%%%%%%%%%%%%%%%%%%%%%%%%%%%%%%%%%%%%%%%%
\subsubsection{Off-resonance growth}\label{sec:off_resonance}
%%%%%%%%%%%%%%%%%%%%%%%%%%%%%%%%%%%%%%%%%%%%%%%%%%%%%%%%%%%%%%%%%%%%%%%%

Let us briefly consider an acceleration which is off-resonance with respect to the orbital evolution, namely with an oscillating acceleration of angular frequency $\omega$ not periodic on $P_0$,
\begin{equation}
\bm{a}^j_\mathrm{sig}(\tau) = \bm{a}^{j}_\mathrm{sig} \cos (\omega \tau + \phi_j)\,.
\end{equation}
Such an acceleration can be substituted into Eq.~\eqref{Eq:PerturbationIntegral}, and where before we had sums such as $\sum_{m=0}^{n-1}(n-m)$ and $\sum_{m=0}^{n-1} 1$, we now have $\sum_{m=0}^{n-1} (n-m)  \cos[\omega(t + mP_0)+ \phi]$ and $\sum_{m=0}^{n-1} \cos[\omega(t + mP_0)+ \phi]$. These new sums scale as $n^{1}$ and $n^{0}$ where they previously scaled as $n^2$ and $n^1$. The integrals we seek to evaluate are otherwise unchanged. Hence we find that off-resonance accelerations will contribute to linear growth in $f_1$, but no accumulated growth is found in any of the other parameters.

{\color{black}
%%%%%%%%%%%%%%%%%%%%%%%%%%%%%%%%%%%%%%%%%%%%%%%%%%%%%%%%%%%%%%%%%%%%%%%%
\subsubsection{Imperfect periodicity and finite coherent time}
\label{sec:finite_coherent_time}
%%%%%%%%%%%%%%%%%%%%%%%%%%%%%%%%%%%%%%%%%%%%%%%%%%%%%%%%%%%%%%%%%%%%%%%%

The exactly on-resonance and off-resonance cases derived above are idealized limits of a finite-time coherence problem. On resonance, the perturbing acceleration pushes the orbit with the same phase on each cycle, so the contributions from successive orbits add coherently. Off resonance, this phase slips, and cancellations remove the enhanced growth. Realistic systems observed for finite times can lie between these limits: the signal may remain nearly phase-aligned with an orbital harmonic for a finite coherent time, after which phase drift regulates the resonant accumulation. This drift can arise either because the signal frequency varies in time or because the binary is not perfectly periodic, so that the relevant orbital harmonic itself evolves.

To describe this interpolation, let $\nu_{\rm sig}(t)$ be the instantaneous frequency of a component of the perturbing acceleration, and let
\begin{equation}
    \nu_k(t)\equiv \frac{k}{P(t)}
\end{equation}
be the $k$th orbital harmonic, with $P(t)$ the instantaneous period of the binary. The relevant measure of phase coherence is the accumulated mismatch
\begin{equation}
    \Delta\Phi_k(t;t_0)
    =
    2\pi
    \int_{t_0}^{t}\mathrm{d}t'\,
    \left[\nu_{\rm sig}(t')-\nu_k(t')\right].
    \label{eq:phase_mismatch}
\end{equation}
Coherent growth persists only while this phase mismatch remains $\lesssim\mathcal{O}(1)$.

For a signal whose frequency is close to the $k$th harmonic, define the local detuning
\begin{equation}
    \delta\nu_k(t)\equiv \nu_{\rm sig}(t)-\nu_k(t).
\end{equation}
There are two simple ways for the phase mismatch to become large. First, if the signal remains offset from the harmonic by an approximately constant amount, $\delta\nu_k$, then the phase mismatch grows as $\Delta\Phi_k\sim 2\pi\delta\nu_k(t-t_0)$, so the corresponding coherence time is
\begin{equation}
    \tau_{\delta\nu}\sim |\delta\nu_k|^{-1}.
    \label{eq:constant_detuning_coherence_time}
\end{equation}
Second, if the signal or orbital harmonic drifts through resonance, then near the crossing we may write $\delta\nu_k(t)\simeq \dot{\delta\nu}_k(t-t_\star)$. In this case $\Delta\Phi_k\sim \pi\dot{\delta\nu}_k(t-t_\star)^2$, and the corresponding coherent interval around the crossing is
\begin{equation}
    \tau_{\dot\nu}\sim |\dot{\delta\nu}_k|^{-1/2}.
    \label{eq:drifting_detuning_coherence_time}
\end{equation}
More generally, the coherent duration available in a finite observation is set by whichever of these timescales is shortest,
\begin{equation}
    T_{\rm coh}
    \sim
    \min\!\left[
    t_{\rm obs},
    \tau_{\delta\nu},
    \tau_{\dot\nu}
    \right],
    \label{eq:coherent_observing_time}
\end{equation}
where $t_{\rm obs}$ is the finite observing time.

The growth laws derived above then apply over each coherent interval. Within one such interval of duration $T_{\rm coh}$, the directly driven non-phase elements grow linearly, while the true anomaly receives one additional time integration from the accumulated orbital dephasing:
\begin{equation}
    \tuple_1^A\sim T_{\rm coh},
    \quad
    f_1\sim T_{\rm coh}^2,
    \quad
    A\in\{p,e,\iota,\Omega,\omega\}.
    \label{eq:finite_coherence_local_scaling}
\end{equation}
If $T_{\rm coh}=t_{\rm obs}$, this is just the fully coherent on-resonance result. If $T_{\rm coh}<t_{\rm obs}$, the coherent kicks to the directly driven non-phase elements cancel after a time of order $T_{\rm coh}$, so their secular envelope is set by the size accumulated during one coherent interval. The true anomaly behaves differently because it records the accumulated orbital phase. A perturbation to the non-phase elements changes the rate at which $f$ advances; the typical size of this rate perturbation is set by the non-phase response accumulated over a coherent interval, and it is then integrated over the full observing time. Schematically,
\begin{equation}
    \tuple_1^A\sim T_{\rm coh},
    \qquad
    \dot f_1\sim T_{\rm coh},
    \qquad
    f_1\sim t_{\rm obs}T_{\rm coh}.
    \label{eq:finite_coherence_global_scaling}
\end{equation}
Thus finite coherence regulates the fully coherent $f_1\sim t_{\rm obs}^2$ growth by replacing one power of $t_{\rm obs}$ with the coherent time. For nearly constant detuning, $T_{\rm coh}\sim|\delta\nu_k|^{-1}$, giving $f_1\sim t_{\rm obs}/|\delta\nu_k|$, consistent with the linear-in-time off-resonance behavior up to oscillatory factors. These scaling estimates provide qualitative guidance for how more realistic binaries respond to signals whose frequencies are not exactly stationary or exactly resonant. In general, however, the full time-resolved response is obtained by solving Eq.~\eqref{eq:solgen} directly, even in cases where simple analytic scaling arguments are difficult to obtain.}

%%%%%%%%%%%%%%%%%%%%%%%%%%%%%%%%%%%%%%%%%%%%%%%%%%%%%%%%%%%%%%%%%%%%%%%%
\subsection{Numerical examples}\label{sec:comparison}
%%%%%%%%%%%%%%%%%%%%%%%%%%%%%%%%%%%%%%%%%%%%%%%%%%%%%%%%%%%%%%%%%%%%%%%%

\begin{figure*}[!htb]  
    \begin{center}
    \includegraphics[width=\textwidth]{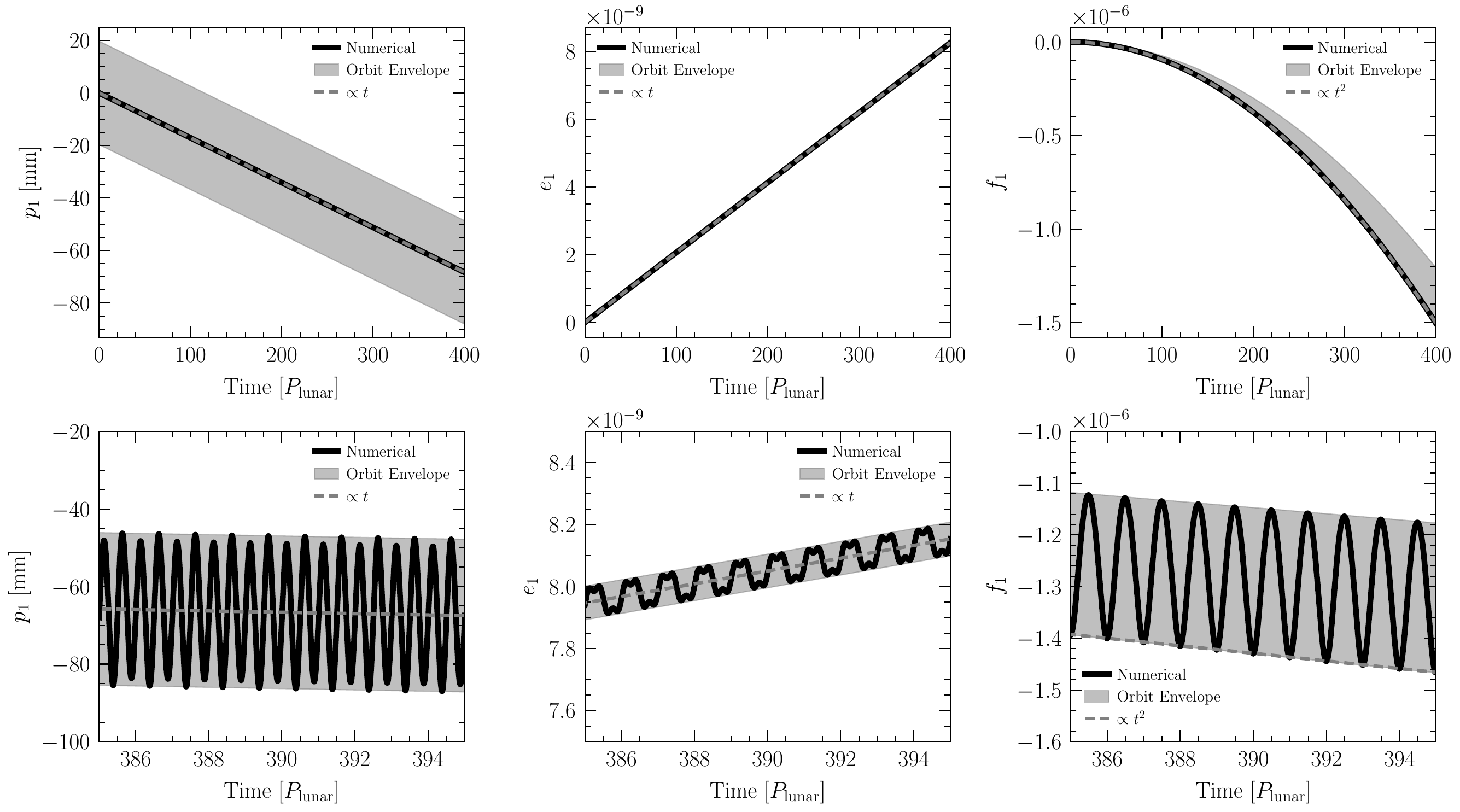}
    \caption{An illustration of the evolution of the $p_1$, $e_1$, and $f_1$ orbital perturbations of the Earth-Moon system subject an on-resonance acceleration in the $\hat{\bm{\theta}}$ direction. See text for more details. In the top panels, we plot the time evolution over a full 400 periods. The heavy black line plots depict the value of the perturbation at $t = \{0, P_0, 2 P_0, \hdots\}$, while the dashed grey line depicts the parametric scaling of the perturbation's evolution. Over the course of a single orbit, the precise value of the perturbation may undergo appreciable oscillations; the grey band indicates the envelope within which those oscillations occur. In the bottom panels, we illustrate the fully time-resolved perturbations as a function of time over 10 orbital periods, which reveals their oscillatory behavior.}
    \label{fig:OnResonanceExample}
    \end{center}
\end{figure*}

As a numerical example that demonstrates the findings of these analytic arguments, we consider the case of the Keplerian orbit of the Earth-Moon system. As a fiducial parametrization, we take $p_0 = 2.54\times10^{-3}\,\mathrm{AU}$, $e_0 = 0.054$, $\iota = 0.36$, $\Omega = 0.21$, $\omega = 1.07$, and $f = 0$ to specify our initial conditions of the unperturbed orbit at time $t = 0$. We also take $G M = 8.9 \times 10^{-10}\,\mathrm{AU}^3/\mathrm{day}^2$. We numerically evolve these initial conditions in the unperturbed evolution, with only $f_0$ being dynamical.

We then go on to consider perturbations to this evolution. We take the acceleration to be given by
\begin{equation*}
    \bm{a}_\mathrm{sig}(t) = \cos(2 \pi \nu t) \hat{\bm{\theta}}(\tuple_0(t)) \, \, \mathrm{mm}\,\mathrm{day}^{-2}\,,
\end{equation*}
with $\nu = 2/P_0$. Since this acceleration is defined to always be in the $\hat{\bm{\theta}}$ direction of the binary system's cylindrical coordinate system, it will induce variations in $p_1$, $e_1$, and $f_1$. The acceleration will also induce perturbations in the $\omega_1$ parameter, but we will not consider this further as it does not undergo a coupled evolution with the $p$, $e$, and $f$ perturbations. We then evolve the perturbed orbital elements for 400 orbits, with the results shown in Fig.~\ref{fig:OnResonanceExample}. The results clearly demonstrate the expected linear growth in $p_1$ and $e_1$ as a function of time as well as the expected quadratic growth of $f_1$ in time for this on-resonance acceleration.

\begin{figure*}[!htb]  
    \begin{center}
    \includegraphics[width=\textwidth]{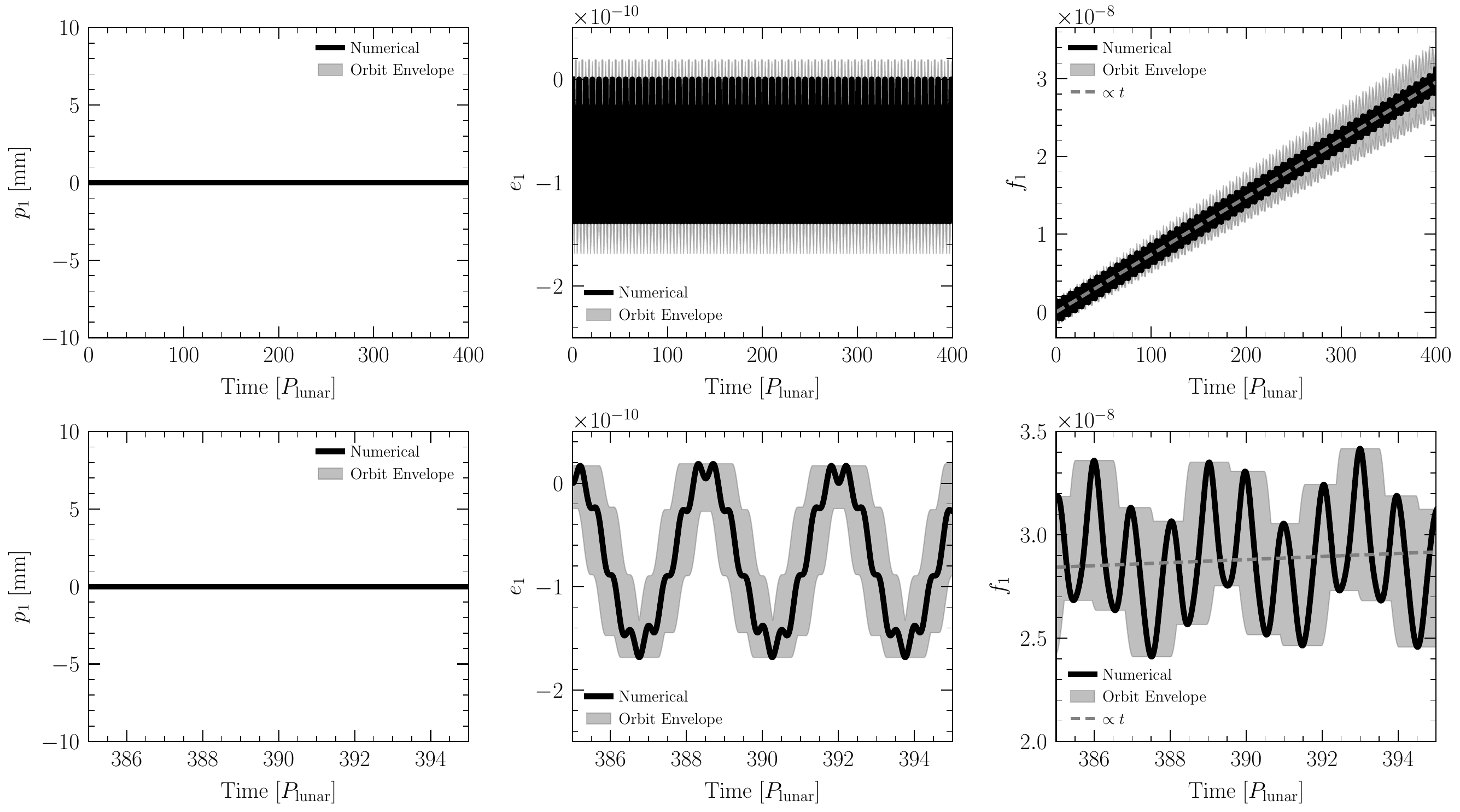}
    \caption{As in Fig.~\ref{fig:OnResonanceExample}, but for an off-resonance acceleration in the $\hat{\bm{r}}$ direction. The radial acceleration produces no response in $p_1$, an oscillatory one in $e_1$ and a linearly growing one in $f_1$ as expected.}
    \label{fig:OffResonanceExample}
    \end{center}
\end{figure*}

As a second example, we take the acceleration to be
\begin{equation*}
    \bm{a}_\mathrm{sig}(t) = \cos(2 \pi \nu t) \hat{\bm{r}}(\tuple_0(t)) \, \, \mathrm{mm}\, \mathrm{day}^{-2}
\end{equation*}
with $\nu = 9/7P_0$ so that it is off-resonance with the binary. Since this acceleration is defined to always be in the $\hat{\bm{r}}$ direction of the binary system's cylindrical coordinate system, it will induce variations in $e_1$, and $f_1$ but not $p_1$. We then evolve the perturbed orbital elements for 400 orbits, with the results shown in Fig.~\ref{fig:OffResonanceExample}. As expected, the $p_1$ perturbation does not experience any dynamics, the $e_1$ perturbation has some time evolution but no growth, and the $f_1$ perturbation realizes linear growth.

%%%%%%%%%%%%%%%%%%%%%%%%%%%%%%%%%%%%%%%%%%%%%%%%%%%%%%%%%%%%%%%%%%%%%%%%
\subsection{Brief comparison to previous methods: the advantage of a time-resolving method}
\label{sec:prev_met_comp}
%%%%%%%%%%%%%%%%%%%%%%%%%%%%%%%%%%%%%%%%%%%%%%%%%%%%%%%%%%%%%%%%%%%%%%%%

The previous results in this section show one of the main advantages of working in a full time-resolved setup:  the dephasing that GWs (or other sources of anomalous acceleration) generate in the perturbed true anomaly of the orbit grows faster than other effects, as shown in Sec.~\ref{sec:on_resonance} and  Sec.~\ref{sec:off_resonance}. The dephasing is ignored when one considers only the secular effects, which do not account for the position of the bodies in the orbit as a function of time. In App.~\ref{app:RecoveringSecular} we show that if these effects are ignored, we recover the previous, less relevant, results of the secular approximation. In brief, the results in this section show that for any observation with a time-resolved tracking of the orbits (such as laser ranging or pulsar timing observations), the analysis of the full time series will be essential to detect or constrain these sources of acceleration. To achieve this aim, the rest of the paper will be devoted to developing the formalism and data analysis framework for this setup.

%%%%%%%%%%%%%%%%%%%%%%%%%%%%%%%%%%%%%%%%%%%%%%%%%%%%%%%%%%%%%%%%%%%%%%%%
\section{Keplerian orbits perturbed by deterministic gravitational waves}
\label{sec:det_GW}
%%%%%%%%%%%%%%%%%%%%%%%%%%%%%%%%%%%%%%%%%%%%%%%%%%%%%%%%%%%%%%%%%%%%%%%%

We begin by specializing to the case of deterministic gravitational waves, characterized by the corresponding components of the metric in the transverse-traceless gauge $\ddot h^{\rm TT}_{ij}(t)$, which are, up to model parameters, perfectly known. While this case can be directly evaluated without the previous perturbative treatment (always working at  $\mathcal{O}(h)$), it provides a useful opportunity to develop the specific formalism for GWs before treating the stochastic case.

The acceleration experienced by the binary in the presence of a gravitational wave of wavelength much larger than the size of the system, as determined in the freely falling center of mass frame, is \cite{Misner:1974qy} 
\begin{equation}
\label{eq:hacc}
    \bm{a}^i_\mathrm{GW} = \frac{1}{2} \ddot h_{ij}^{\rm TT}(t)\bm{r}^j,
\end{equation}
where $\bm{r}$ is the distance vector between the binary companions (we omit the TT label in the following to avoid cluttered expressions). Given this, we have
\begin{equation}
\begin{gathered}
\Ft_1^{\alphat }(t) = \Ft_\mathrm{GW}^{\alphat ij}(t)  \ddot h_{ij}(t),
\end{gathered}
\end{equation}
where
\begin{equation}
\begin{gathered}
\Ft_\mathrm{GW}^{\alphat ij}(t) = \frac{1}{2}\Mt^{\alphat b}(\tuple_0(t)) \bm{e}_{bi}(\tuple_0(t))  \bm{r}^j(\tuple_0(t)),
\end{gathered}
\end{equation}
is the three-index response tensor to the  GW. The first-order solution of Eq.~\eqref{eq:solgen} is then given by
\begin{equation}
\tuple_1^{\alphat }(t) = \Phit_{\alphat \betat  }(t) \int_0^t \mathrm{d}\tau \Phit^{-1}_{\betat\gammat}(\tau) \Ft_\mathrm{GW}^{\gammat  ij}(\tau) \ddot h_{ij}(\tau).
\label{eq:Solution}
\end{equation}
While our calculation is, in principle, now fully complete we note that evaluating Eq.~\eqref{eq:Solution} will require intensive numerical evaluation. In practice, rather than directly integrating Eq.~\eqref{eq:Solution}, we treat the dual ordinary differential equation
\begin{equation}
\begin{gathered}
\tuple_1^\alphat(t) = \Phit_{\alphat \betat}(t) \First_\betat(t),\\
\dot{\First}_1^\betat = \Phit^{-1}_{\betat\gammat}(t) \Ft^{\gammat ij}_\mathrm{GW}(t) \ddot h_{ij}(t),
\label{eq:determ_final}
\end{gathered}
\end{equation}
which is typically more efficient to solve. Moreover, we will find this construction to be particularly useful when evaluating signals and covariance matrices associated with stochastic backgrounds. Notice also that, by using
Eq.~\eqref{eq:FourierTransform} in App.~\ref{app:GWs} we can express our results in terms of responses to the different polarizations
of the GW.

We now make contact with our analytic results for the growth of perturbations of isolated Keplerian binaries. Let us work in a convenient observer coordinate system such at $\Omega_0 = \omega_0 = 0$ and $\iota_0 = \pi/2$ such that the binary orbit lies within the $\hat{\bm{X}}$-$\hat{\bm{Z}}$ plane. Let us also assume a monochromatic gravitational wave at frequency $\nu$; if $\nu = n/P_0$ for integer $n$, then this wave is on resonance. If any of $\ddot h_{11}$, $\ddot h_{13}$ or $\ddot h_{33}$ are nonzero, then there will be an acceleration in the plane of the orbit, leading to quadratic growth of the $f_1$ perturbation in time. Following similar reasoning, we can also determine the components of $\ddot h_{ij}$ which drive linear growth in the remaining orbital elements.

%%%%%%%%%%%%%%%%%%%%%%%%%%%%%%%%%%%%%%%%%%%%%%%%%%%%%%%%%%%%%%%%%%%%%%%%
\section{Stochastic gravitational wave backgrounds}
\label{sec:stochastic_gw}
%%%%%%%%%%%%%%%%%%%%%%%%%%%%%%%%%%%%%%%%%%%%%%%%%%%%%%%%%%%%%%%%%%%%%%%%

We proceed to treat the case of SGWBs, which we characterize in terms of moments of their distribution. In this work, we will assume a SGWB that is isotropic, stationary, unpolarized, Gaussian, and uncorrelated across the sky. Though scenarios in which any or all of these assumptions are violated may appear, we prefer to focus on the simplest scenario.  Note that these backgrounds still represent a large fraction of those relevant for searches of SGWBs \cite{Caprini:2018mtu,Maggiore:1900zz,Maggiore:2018sht}.

%%%%%%%%%%%%%%%%%%%%%%%%%%%%%%%%%%%%%%%%%%%%%%%%%%%%%%%%%%%%%%%%%%%%%%%%
\subsection{Statistics of SGWBs}
%%%%%%%%%%%%%%%%%%%%%%%%%%%%%%%%%%%%%%%%%%%%%%%%%%%%%%%%%%%%%%%%%%%%%%%%

To characterize the statistics of our SGWBs, we consider the time-evolving GW strain at some fixed location, $h_{ij}(t)$. Recalling the definitions from App.~\ref{app:GWs}, under our assumptions of the SGWB, its statistics in Fourier space for the different polarizations $A$ are specified by the first moment
\begin{equation}
 \langle \tilde{h}_A(\nu , \hat{\bm{n}}) \rangle = 0\,,
\end{equation}
and the second moment
\begin{equation}
\begin{split}
    \langle& \tilde{h}_A(\nu , \hat{\bm{n}}) \tilde{h}^*_B(\nu', \hat{\bm{n}}') \rangle\\
    &= \frac{3 H_0^2 \Omega_\mathrm{GW}(\nu)}{32 \pi^3 |\nu|^3} \delta_{AB} \delta(\nu-\nu') \delta(\hat{\bm{n}},\hat{\bm{n}}').
\end{split}
\label{eq:DefiningStatistics}
\end{equation}
Here, $\Omega_\mathrm{GW}(\nu)$ is the one-sided\footnote{Let us also clarify that $\Omega_\mathrm{GW}$ is one-sided in the sense by integrating over positive frequencies, we have $\int \mathrm{d}\ln\nu \Omega_\mathrm{GW} = \rho_\mathrm{GW}/ \rho_c$; however $\Omega_\mathrm{GW}$ is also defined such that $\Omega_\mathrm{GW}(\nu) = \Omega_\mathrm{GW}(-\nu)$. This standard but often unnoticed continuation to negative frequencies is necessary to sensibly work with the two-sided Fourier transformation convention of Eq.~\eqref{eq:FourierTransform} in App.~\ref{app:GWs}. Making this continuation to negative frequencies amounts to a factor of two when performing integrals over $\nu$ which is already fully incorporated in Eq.~\eqref{eq:DefiningStatistics}.} SGWB energy density spectrum defined in terms of the total gravitational wave energy density $\rho_\mathrm{GW}$ and the critical energy density $\rho_c \equiv 3 H_0^2 / (8\pi G)$ (where $H_0$ is the present-day Hubble rate) by 
\begin{equation}
    \Omega_\mathrm{GW}(\nu) \equiv \frac{1}{\rho_c}\frac{\mathrm{d} \rho_\mathrm{GW}}{\mathrm{d} \ln \nu}.
\end{equation}

The first moment of $\ddot h_{ij}$ trivially vanishes:
\begin{equation}
\langle \ddot h_{ij}(t) \rangle \propto \int \mathrm{d}\nu \nu^2 \, e^{2\pi i \nu t} \langle \tilde{h}_A(\nu, \hat{\bm{n}})\rangle = 0.
 \end{equation}
The second moment reads
\begin{widetext}
\begin{equation}
\begin{split}
\langle \ddot h_{ij}(t)  \ddot h_{lm}(t')\rangle &=  16 \pi^4 \int \mathrm{d}^2 \hat{\bm{n}} \mathrm{d}^2 \hat{\bm{n}}' \, e_{ij}^A(\bm{\hat{n}})e_{lm}^B(\bm{\hat{n}}') \int \mathrm{d}\nu \mathrm{d}\nu' \nu^2 \nu'^2 \, e^{2\pi i (\nu t + \nu' t')} \langle \tilde{h}_A(\nu, \hat{\bm{n}}) \tilde{h}_B(\nu', \hat{\bm{n}}')\rangle \\
&=  \frac{3 \pi H_0^2 }{2}  \int \mathrm{d}^2 \hat{\bm{n}}\, e_{ij}^A(\bm{\hat{n}})e_{lm}^B(\bm{\hat{n}}) \delta_{AB}  \int_{-\infty}^{\infty} \mathrm{d}\nu  |\nu| \Omega_{\rm GW}(|\nu|) \, e^{2\pi i \nu (t - t')}  \\
&= \frac{24}{5} \pi^2 H_0^2  C_{ij,lm}  \int_0^{\infty} \mathrm{d}\nu  \nu \Omega_\mathrm{GW}(\nu) \,\cos[ 2\pi\nu (t - t')],
\label{eq:TimeDomainCorrelator}
\end{split}
\end{equation}
\end{widetext}
where in the last step we introduced 
\begin{equation}
    C_{ij,lm} \equiv \frac{5}{8\pi} \int \mathrm{d}^2 \hat{\bm{n}} \, e_{ij}^A(\bm{\hat{n}})e_{lm}^B(\bm{\hat{n}}) \delta_{AB}.
    \label{eq:CovarianceIntegral}
\end{equation}
These integrals are straightforwardly evaluated, and the results are tabulated in Tab.~\ref{tab:covariances}. Since our Gaussian SGWB is fully specified by its first two moments, we are now fully prepared to calculate the defining moments of the orbital perturbations. 

\begin{table}[!htb]
\renewcommand{\arraystretch}{1.5}
\centering
    \begin{tabular}{|c||c|c|c|c|c|c|}
    \hline
    $\{(i,j), (l, m)\}$& $(1, 1)$ & $(2, 2)$ & $(3, 3)$ & $(1, 2)$ & $(1, 3)$ & $(2, 3)$\\ \hline\hline
    $(1, 1)$ & $\frac{4}{3}$ & $-\frac{2}{3}$ & $-\frac{2}{3}$ & 0 & 0 & 0 \\
    \hline
    $(2, 2)$ & $-\frac{2}{3}$ & $\frac{4}{3}$ & $-\frac{2}{3}$ & 0 & 0 & 0 \\
    \hline
    $(3, 3)$ & $-\frac{2}{3}$ & $-\frac{2}{3}$ & $\frac{4}{3}$ & 0 & 0 & 0 \\
    \hline
    $(1, 2)$ & 0 & 0 & 0 & $1$ & 0 & 0 \\
    \hline
    $(1, 3)$ & 0 & 0 & 0 & 0 & $1$ & 0 \\
    \hline
    $(2, 3)$ & 0 & 0 & 0 & 0 & 0 & $1$ \\
    \hline
    \end{tabular}
\caption{The values of $C_{ij,lm}$ as defined by Eq.~\eqref{eq:CovarianceIntegral}. The $C_{ij,lm}$ object is symmetric under $i \leftrightarrow j$, $l \leftrightarrow m$, and $(i,j) \leftrightarrow (l,m)$.}
\label{tab:covariances}
\end{table}

%%%%%%%%%%%%%%%%%%%%%%%%%%%%%%%%%%%%%%%%%%%%%%%%%%%%%%%%%%%%%%%%%%%%%%%%%%%%%%%%%%%%%%%%%%%%%%%%
\subsection{Orbital perturbations from SGWBs}
\label{sec:OrbitalPerturbationsSGWB}
%%%%%%%%%%%%%%%%%%%%%%%%%%%%%%%%%%%%%%%%%%%%%%%%%%%%%%%%%%%%%%%%%%%%%%%%%%%%%%%%%%%%%%%%%%%%%%%%

The first moment of the solution in Eq.~\eqref{eq:Solution} trivially vanishes, as 
\begin{equation}
\begin{split}
&\bm{\mu}_{\alphat }(t) \equiv\langle \tuple_1^{\alphat }(t) \rangle \\&=\Phit_{\alphat \betat  }(t) \int_0^t \mathrm{d}\tau\, \Phit_{\betat  \gammat  }(\tau) \Ft_\mathrm{GW}^{\gammat  ij}(\tau) \langle \ddot h_{ij}(\tau) \rangle  = 0.
\label{eq:FirstMomentFromGWs}
\end{split}
\end{equation}
Hence we see at first order, the mean of $\tuple_1$ vanishes as positive or negative perturbations are equally likely.\footnote{Note that in \cite{Blas:2021mpc,Blas:2021mqw} the non-linear nature of the equations was used to find non-trivial drifts of the averaged values induced by SGWB. However, 
it was found that their impact on the determination of the orbital parameters is subdominant to the growth of the covariance. Hence, the linear approach we follow, based on Eq.~\eqref{eq:general_linear}, suffices for our analysis.}  Proceeding to the second moment we have
\begin{widetext}
\begin{equation}
\begin{split}
\bm{\Sigma}_{\alphat \betat  }(t, t' ) &\equiv \langle \tuple_1^{\alphat }(t) \tuple_1^{\betat  }(t') \rangle
=3\pi H_0^2 C_{ij,lm} 
\Phit_{\alphat\gammat}(t) \Phit_{\betat\kappat}(t') \\
&\times
\int_0^\infty \mathrm{d}\nu \nu \Omega_\mathrm{GW}(\nu) \int_0^{t} \mathrm{d}\tau \int_0^{t'} \mathrm{d}\tau' 
\Phit_{\gammat \lambdat}^{-1}(\tau) \Phit_{\kappat\mut}^{-1}(\tau')\Ft_\mathrm{GW}^{\lambdat ij}(\tau) \Ft_\mathrm{GW}^{\mut lm}(\tau') \cos[2 \pi \nu (\tau-\tau')] .
\end{split}
\end{equation}
\end{widetext}
Taking inspiration from Eqs.~\eqref{eq:determ_final}, we define $\First_{\betat  ij,d}(t|\nu)$ by the differential equation
\begin{equation}\label{eq:XSGWB}
\begin{gathered}
\dot{\First}_{\betat  ij, c}(t |\nu) = \Phit_{\betat\gammat}^{-1}(t) \Ft^{\gammat ij}_\mathrm{GW}(t) \cos(2 \pi \nu t), \\
\dot{\First}_{\betat  ij, s}(t |\nu) = \Phit_{\betat\gammat}^{-1}(t) \Ft^{\gammat ij}_\mathrm{GW}(t) \sin(2 \pi \nu t).
\end{gathered}
\end{equation}
We can then define 
\begin{equation}
\begin{split}
\label{eq:Sigma_GW}
\bm{\Sigma}_{\alphat \betat  }(t, t' | \nu) &= 3 \pi H_0^2 C_{ij,lm} \Phit_{\alphat\gammat}(t) \Phit_{\betat\lambdat}(t')\\
& \times \sum_{d={c,s}}\First_{\gammat ij,d}(t |\nu) \First_{\lambdat lm,d}(t'| \nu),
\end{split}
\end{equation}
so that the two-point correlation function is then given by
\begin{equation}
\begin{split}
\bm{\Sigma}_{\alphat \betat  }(t, t') = \int_0^\infty \mathrm{d} \nu \nu \Omega_\mathrm{GW}(\nu) \bm{\Sigma}_{\alphat \betat  }(t, t' | \nu).
\label{eq:GWOrbitalCovariance}
\end{split}
\end{equation}
This illustrates an additional advantage of working with the dual ordinary differential equation rather than the integral problem: with a single numerical evaluation of $\First_{\betat  ij,d}$ between $0$ and $T$, the full covariance between all times $t, \, t' < T$ may be directly constructed.

%%%%%%%%%%%%%%%%%%%%%%%%%%%%%%%%%%%%%%%%%%%%%%%%%%%%%%%%%%%%%%%%%%%%%%%%%%%%%%%%%%%%%%%%%%%%%%%%
\subsection{Principal components of the SGWB}
\label{sec:PrincipalComponents}
%%%%%%%%%%%%%%%%%%%%%%%%%%%%%%%%%%%%%%%%%%%%%%%%%%%%%%%%%%%%%%%%%%%%%%%%%%%%%%%%%%%%%%%%%%%%%%%%

To relate our construction of the covariance matrix in Sec.~\ref{sec:OrbitalPerturbationsSGWB} with the analytic results of Sec.~\ref{sec:AnalyticResults}, we will make a more detailed consideration of the frequency-domain statistics of our SGWBs. From  Eq.~\eqref{eq:FourierTransform} in App.~\ref{app:GWs} and Eq.~\eqref{eq:DefiningStatistics},
\begin{equation}
\begin{split}
\langle \ddot{\tilde{h}}_{ij}(\nu) \ddot{\tilde{h}}_{lm}^* (\nu') \rangle
&= \frac{12 \pi H_0^2}{5} C_{ij, lm}\delta(\nu - \nu') \Omega_{\rm GW}(|\nu|)|\nu|.
\label{eq:FrequencyDomainCovariance}
\end{split}
\end{equation}
Since $\ddot h_{ij}$ is characterized by Gaussian statistics, through Eq.~\eqref{eq:FrequencyDomainCovariance} we can understand Tab.~\ref{tab:covariances} as specifying the covariance between each of the $\ddot{\tilde{h}}_{ij}(\nu)$. As we see, the $\ddot{\tilde{h}}_{ij}$ at two different frequencies are uncorrelated, so we may consider each frequency at which the SGWB has support independently. 

The covariance matrix between the six components of $\ddot{\tilde{h}}_{ij}$ as defined by Eq.~\eqref{eq:FrequencyDomainCovariance} is singular. The reason for this is that the transverse-traceless condition will eliminate one of these six degrees of freedom in this matrix. The desired statistics can instead be written in terms of five components which are specified by a nonsingular covariance matrix. We will label these five components $\ddot{\tilde{h}}^{i}$. As a definite option for the (underdetermined) mapping between $\ddot{\tilde{h}}^{i}$ and $\ddot{\tilde{h}}_{ij}$, let us choose 
\begin{equation}
\begin{pmatrix}
\ddot{\tilde{h}}_{11} \\
\ddot{\tilde{h}}_{22} \\
\ddot{\tilde{h}}_{33} \\
\ddot{\tilde{h}}_{12} \\
\ddot{\tilde{h}}_{13} \\
\ddot{\tilde{h}}_{23}
\end{pmatrix} =
\begin{pmatrix}
\frac{2}{\sqrt{3}} & 0 & 0 & 0 & 0 \\
-\frac{1}{\sqrt{3}} & 1 & 0 & 0 & 0\\
-\frac{1}{\sqrt{3}} & -1 & 0 & 0 & 0\\
0 & 0 & 1 & 0 & 0 \\
0 & 0 & 0 & 1 & 0 \\
0 & 0 & 0 & 0 & 1 
\end{pmatrix} \times
\begin{pmatrix}
\ddot{\tilde{h}}^1 \\
\ddot{\tilde{h}}^2 \\
\ddot{\tilde{h}}^3 \\
\ddot{\tilde{h}}^4 \\
\ddot{\tilde{h}}^5
\end{pmatrix},
\end{equation}
as it generates uncorrelated variables. The first moment of  $\ddot{\tilde{h}}_i(\nu)$ vanishes, while
\begin{equation}
\begin{split}
   % \langle \ddot{\tilde{h}}^i(\nu) \rangle  &= 0,\\ 
    \langle \ddot{\tilde{h}}^i(\nu) \ddot{\tilde{h}}^{*j}(\nu') \rangle &=  \frac{12 \pi^2 H_0^2}{5} \delta(\nu - \nu') |\nu| \Omega_{\rm GW}(|\nu|) \delta_{ij}\,.
\end{split}
\end{equation}

By Fourier transforming back to the time-domain, we can obtain the principal components \cite{2018arXiv180402502G}. They are defined as 
\begin{equation}
\begin{split}
    \ddot{h}_{ij}^{a,c}(t|\nu) &= \ddot h_{ij}^a \cos(2 \pi \nu t),  \\
    \ddot{h}_{ij}^{a,s}(t|\nu) &= \ddot h_{ij}^a \sin(2 \pi \nu t), \\
\end{split}
\end{equation}
with
\begin{equation}
\begin{gathered}
\ddot h_{ij}^{1} = \frac{1}{\sqrt{3}}\left[2\delta_{i1}\delta_{j1} -\delta_{i2}\delta_{j2}-\delta_{i3}\delta_{j3} \right],\\
\ddot h_{ij}^{2} = \left[\delta_{2i}\delta_{j2}-\delta_{i3}\delta_{j3} \right],~~ 
\ddot h_{ij}^{3} = \delta_{i(1}\delta_{2)j},\\
\ddot h_{ij}^{4} = \delta_{i(1}\delta_{3)j}, ~~
\ddot h^{5}_{ij} = \delta_{i(2}\delta_{3)j},
\end{gathered}
\end{equation}
where indices in parentheses are symmetrized. Hence, any realization of a monochromatic GW be $\ddot h_{ij}$ will be of the form
\begin{equation}
    \ddot h_{ij}(t | \nu) = \sum_{a = 1}^{5} \sum_{d =\{c, s\}} A^\nu_{a, d} \ddot h_{ij}^{a, d}(t|\nu), 
    \label{eq:WaveConstruction}
\end{equation}
where $A^\nu_{a,d}$ are independent, identically distributed Gaussian variables specified by
\begin{equation}
\begin{gathered} 
    \langle A^\nu_{a,d} \rangle = 0,  \\
    \langle A^\nu_{a,d} A^{\nu'}_{p,q}\rangle  = \frac{6 \pi^2 H_0^2}{5} \delta(\nu - \nu') |\nu|\Omega_{\rm GW}(|\nu|) \delta_{aq} \delta_{dq}.
\end{gathered}
\end{equation}
A non-monochromatic stochastic background will be appropriately generated by integrating over frequency. 

%%%%%%%%%%%%%%%%%%%%%%%%%%%%%%%%%%%%%%%%%%%%%%%%%%%%%%%%%%%%%%%%%%%%%%%%%%%%%%
\subsection{The growth of the perturbation covariance}
%%%%%%%%%%%%%%%%%%%%%%%%%%%%%%%%%%%%%%%%%%%%%%%%%%%%%%%%%%%%%%%%%%%%%%%%%%%%%%

Now that we have constructed a useful basis to describe SGWBs, we will use it to understand the scaling properties of the perturbation covariances calculated in Sec.~\ref{sec:OrbitalPerturbationsSGWB}. Let us define 
\begin{equation}
\begin{split}
    \tuple_1^{\alphat i,d}(t | \nu) =  \Phit_{\alphat\betat}(t) \int_0^{t} \mathrm{d}\tau & \, \Phit^{-1}_{\betat  \gammat}(\tau)  \Ft_\mathrm{GW}^{\gammat  lm}(\tau)  \ddot h^{i,d}_{lm}(\tau |\nu).
\end{split}
\end{equation}
Since our system is linear in $\ddot h_{jk}$, then the solution for an arbitrary SGWB constructed from principal components as in Eq.~\eqref{eq:WaveConstruction} is given by
\begin{equation}
\tuple_1^{\alphat}(t | \nu) = \sum_{i=1}^{5}\sum_{d=\{c,s\}}A_{i, d} \tuple_1^{\alphat i,d}(t |\nu),
\label{eq:SolutionSuperposition}
\end{equation}
where $A_{i, d}$ are the same Gaussian variables from  Eq.~\eqref{eq:WaveConstruction}.

From Eq.~\eqref{eq:WaveConstruction}, the solution of \eqref{eq:SolutionSuperposition} 
in the SGWB case can be interpreted from those of the deterministic case, Eq.~\eqref{eq:Solution},
driven by the principal components of the SGWB. For instance, if the monochromatic wave is on-resonance, we will induce quadratic growth in $f_1$ if we have an acceleration in either the $\hat{\bm{r}}$ or $\hat{\bm{\theta}}$ directions. Working in our convenient $\omega = \Omega = 0$ and $\iota = \pi/2$ coordinates, acceleration in these directions are generated by nonzero $\ddot h_{11}(t|\nu)$, $\ddot h_{13}(t|\nu)$, or $\ddot h_{33}(t|\nu)$. This condition is realized by $\ddot h_{ij}^{1, d}(t|\nu)$, $\ddot h_{ij}^{2, d}(t|\nu)$, and $\ddot h_{ij}^{4, d}(t|\nu)$ for either $d = c$ or $d = s$. This suggests that six of our ten principal components could generate quadratic growth of $f_1$. Similar reasoning can be applied to identify that linear growth of each of the orbital elements is generically induced by some subset of the principal components of $\ddot h_{ij}(t|\nu)$.

\begin{figure*}[!htb]  
    \begin{center}
    \includegraphics[width=\textwidth]{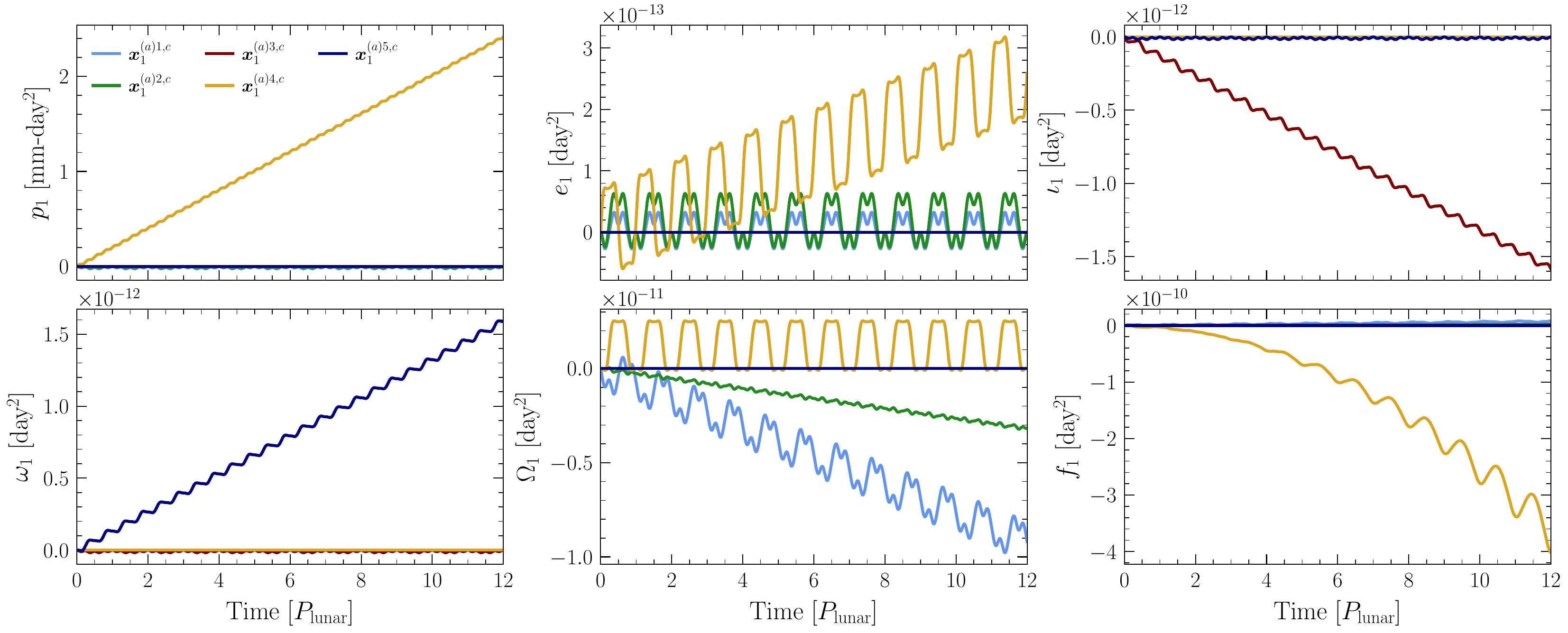}
    \caption{The numerical solutions for $\tuple_1^{\alphat i,c}$ for an on resonance GW over 12 orbits of the Earth-Moon system. The predicted quadratic growth in $f_1$ is realized in $\tuple_1^{\alphat 4,c}$ but not in $\tuple_1^{\alphat 1,c}$ or $\tuple_1^{\alphat 2,c}$ due to the phase mismatch between the GW and the unperturbed orbit. The predicted linear growth is realized for some $\tuple_1^{\alphat ic}$ in the remaining elements.}
    \label{fig:CosineComponents}
    \end{center}
\end{figure*}

\begin{figure*}[!htb]  
    \begin{center}
    \includegraphics[width=\textwidth]{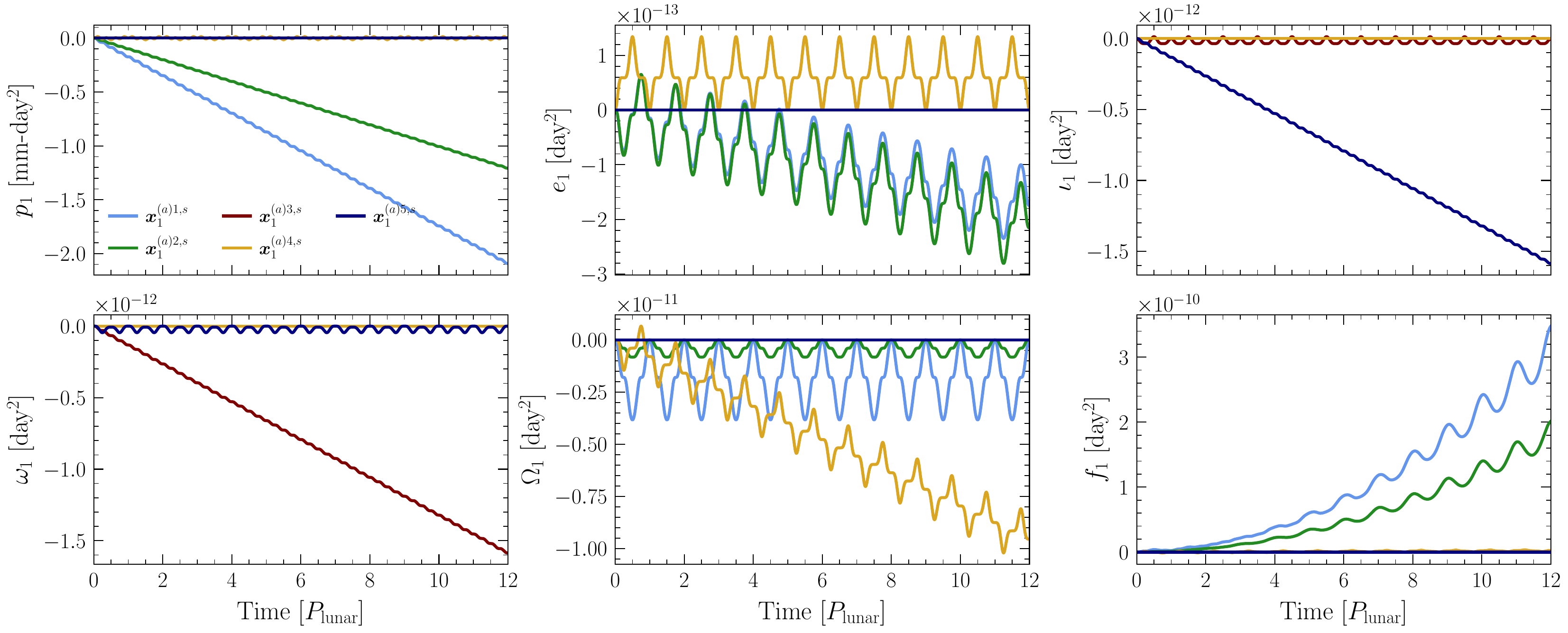}
    \caption{As in Fig.~\ref{fig:CosineComponents}, but for the $\tuple_1^{\alphat i,s}$. The predicted quadratic growth in $f_1$ is realized in $\tuple_1^{\alphat 1,s}$ and $\tuple_1^{\alphat 2,s}$ but not in $\tuple_1^{\alphat 4,s}$ due to the phase mismatch between the GW and the unperturbed orbit. The predicted linear growth is realized for some $\tuple_1^{\alphat i,s}$ in the remaining elements.}
    \label{fig:SinePrincipalComponents}
    \end{center}
\end{figure*}

As a demonstration of these results, we consider the Earth-Moon system in our convenient coordinates and evaluate the $\tuple_1^{\alphat i,d}(t|\nu)$ for each $i, d$ with $\nu = 2/P_0$ so that the GW is on-resonance. These results for the evolution of all six orbital elements for $d = c$ are shown in Fig.~\ref{fig:CosineComponents}, while the results for $d = s$ are shown in Fig.~\ref{fig:SinePrincipalComponents}. Quadratic growth of $f_1$ and linear growth in the remaining parameters as induced by some of the $\ddot h_{ij}^{a,d}$ is clearly demonstrated.

Collecting these results, we can then calculate the covariance $\bm\Sigma_{\alphat \betat  }$by 
\begin{equation}
\begin{split}
\bm{\Sigma}_{\alphat\betat}&(t,t' | \nu) = \langle \tuple_1^{\alphat }(t | \nu )\tuple_1^{\betat  }(t' | \nu) \rangle \\
&= \sum_{i=1}^{5} \sum_{d=\{c, s\}} \langle A_{i, d}^2\rangle \tuple_1^{\alphat i,d}(t | \nu) \tuple_1^{\betat  i,d}(t' |\nu). 
\end{split}
\end{equation}
{\color{black}
This expression makes clear how the deterministic growth laws translate into covariance growth. Consider first the idealized monochromatic limits. If the Fourier component lies exactly on an orbital harmonic and the binary is perfectly periodic over the full observing time, then inserting the deterministic scaling results of Sec.~\ref{sec:AnalyticResults} gives
\begin{equation}
\begin{gathered}
    \bm{\Sigma}_{ff}(t,t'|\nu)\sim (t t')^2,
    \qquad
    \bm{\Sigma}_{Af}(t,t'|\nu)\sim t\,t'^2, \\
    \bm{\Sigma}_{fA}(t,t'|\nu)\sim t^2\,t',
    \qquad
    \bm{\Sigma}_{AB}(t,t'|\nu)\sim t t',
    \label{eq:monochromatic_on_resonance_covariance_hierarchy}
\end{gathered}
\end{equation}
where $A,B\in\{p,e,\iota,\Omega,\omega\}$. The true-anomaly block contains two powers of the accumulated orbital dephasing, the mixed blocks contain one, and the purely non-phase block grows slowest. By contrast, for a monochromatic component far from resonance, the non-phase elements do not accumulate secularly while the true anomaly grows linearly, giving
\begin{equation}
\begin{gathered}
    \bm{\Sigma}_{ff}(t,t'|\nu)\sim t t',
    \qquad
    \bm{\Sigma}_{Af}(t,t'|\nu)\sim t', \\
    \bm{\Sigma}_{fA}(t,t'|\nu)\sim t,
    \qquad
    \bm{\Sigma}_{AB}(t,t'|\nu)\sim 1 .
    \label{eq:monochromatic_off_resonance_covariance_hierarchy}
\end{gathered}
\end{equation}
While these results are relevant for perfectly periodic binaries and narrowband backgrounds, a realistic broadband SGWB requires integrating over a continuum Fourier components.
}

{\color{black}
Since the background is stationary, $\Omega_{\rm GW}(\nu)$ is time independent, and we do not need to track any intrinsic time dependence of the signal frequencies themselves. Instead, the binary selects a finite resonant band around each orbital harmonic. Using Eq.~\eqref{eq:GWOrbitalCovariance}, the broadband scaling is obtained by inserting the monochromatic covariance scalings into the frequency integral. For a perfectly periodic binary observed for a time $t_{\rm obs}$, the resonant part of $\bm{\Sigma}_{\alpha\beta}(t,t'|\nu)$ is coherently enhanced only within a finite-time bandwidth $\Delta\nu_{\rm res}\sim t_{\rm obs}^{-1}$ around each orbital harmonic $\nu_k$. If $\Omega_{\rm GW}(\nu)$ varies slowly across this band, then at equal times $t=t'=t_{\rm obs}$,
\begin{equation}
\begin{gathered}
    \bm{\Sigma}_{ff}
    \sim
    \nu_k\Omega_{\rm GW}(\nu_k)\,
    t_{\rm obs}^4\,\Delta\nu_{\rm res}
    \sim
    \nu_k\Omega_{\rm GW}(\nu_k)\,
    t_{\rm obs}^3, \\
    \bm{\Sigma}_{Af},
    \bm{\Sigma}_{fA}
    \sim
    \nu_k\Omega_{\rm GW}(\nu_k)\,t_{\rm obs}^2,
    \\
    \bm{\Sigma}_{AB}
    \sim
    \nu_k\Omega_{\rm GW}(\nu_k)\,t_{\rm obs}.
    \label{eq:broadband_periodic_covariance_hierarchy}
\end{gathered}
\end{equation}
Thus the broadband frequency integral softens the coherent monochromatic scaling by one power of $t_{\rm obs}$, because only a bandwidth $\Delta\nu_{\rm res}\sim t_{\rm obs}^{-1}$ contributes coherently.
}

{\color{black}
The same reasoning applies when the binary is not perfectly periodic. In that case, the orbital harmonic $\nu_k(t)$ may drift through the fixed-frequency SGWB spectrum, so the response should be assembled over coherent intervals rather than over the full observing time at once. Let $\tau_j$ be the duration of the $j$th coherent interval, as defined by Eq.~\ref{eq:coherent_observing_time}, centered near $t_j$, over which the harmonic is approximately constant. The local resonant bandwidth is then $\Delta\nu_j\sim \tau_j^{-1}$, and Eq.~\eqref{eq:GWOrbitalCovariance} gives
\begin{equation}
\begin{split}
    \bm{\Sigma}_{ff}^{(j)}
    &\sim
    \nu_k(t_j)\Omega_{\rm GW}(\nu_k(t_j))\,
    \tau_j^4\,\Delta\nu_j \\
    &\sim
    \nu_k(t_j)\Omega_{\rm GW}(\nu_k(t_j))\,
    \tau_j^3 .
\end{split}
\end{equation}
The mixed and non-phase blocks similarly scale as
\begin{equation}
\begin{gathered}
    \bm{\Sigma}_{Af}^{(j)},
    \bm{\Sigma}_{fA}^{(j)}
    \sim
    \nu_k(t_j)\Omega_{\rm GW}(\nu_k(t_j))\,\tau_j^2, \\
    \bm{\Sigma}_{AB}^{(j)}
    \sim
    \nu_k(t_j)\Omega_{\rm GW}(\nu_k(t_j))\,\tau_j .
    \label{eq:local_broadband_covariance_hierarchy}
\end{gathered}
\end{equation}
If the spectrum varies slowly over the relevant resonant bands and the coherent intervals have a characteristic duration $\tau$ over a total observing time $t_{\rm obs}$, summing over $N_{\rm coh}\sim t_{\rm obs}/\tau$ intervals gives
\begin{equation}
\begin{gathered}
    \bm{\Sigma}_{ff}
    \sim
    \nu_k\Omega_{\rm GW}(\nu_k)\,t_{\rm obs}\tau^2,
    \\
    \bm{\Sigma}_{Af},
    \bm{\Sigma}_{fA}
    \sim
    \nu_k\Omega_{\rm GW}(\nu_k)\,t_{\rm obs}\tau,
    \\
    \bm{\Sigma}_{AB}
    \sim
    \nu_k\Omega_{\rm GW}(\nu_k)\,t_{\rm obs}.
    \label{eq:broadband_finite_coherence_covariance_hierarchy}
\end{gathered}
\end{equation}
The perfectly periodic broadband result is recovered by setting $\tau=t_{\rm obs}$. Thus imperfect periodicity regulates the resonant enhancement by replacing powers of the full observing time with powers of the coherent interval, while preserving the hierarchy among covariance blocks.
}

%%%%%%%%%%%%%%%%%%%%%%%%%%%%%%%%%%%%%%%%%%%%%%%%%%%%%%%%%%%%%%%%%%%%%%%%%%%%%%
\section{Fluctuations of ultra-light dark matter}
\label{sec:stochastic_uldm}
%%%%%%%%%%%%%%%%%%%%%%%%%%%%%%%%%%%%%%%%%%%%%%%%%%%%%%%%%%%%%%%%%%%%%%%%%%%%%%

In dark matter candidates with masses $m_{\rm DM}\lesssim 1\,$eV, the typical distance between particles in the Milky Way, $\sim n^{-1/3}$ (with $n$ the number density), is smaller than the typical size that one can associate to these particles, related to the average momentum in wavepackets  $\bar p\approx m \sigma_0$, where $\sigma_0\sim 10^{-3}$ is the dispersion velocity in the dark matter halo of the Milky Way. Models with masses $m_{\rm DM}\lesssim$\,eV are hence better described with classical wave equations, and they are known as ultra-light dark matter (ULDM) of fuzzy 
dark matter models~\cite{Hui:2016ltb,Ferreira:2020fam}. In this section, we will consider the case of a scalar ULDM candidate, represented by the field $\phi$.

In this work, we will consider how this field \textit{universally} affects orbital motion through purely gravitational effects. Note, however, that
more model-dependent cases of direct coupling may also generate relevant effects that can be readily accommodated within our framework, see \cite{Blas:2016ddr,Blas:2019hxz,Kus:2024vpa} for previous work. Now, let us consider the metric of the galactic halo,
\begin{equation}
\mathrm{d} s^2=-(1+2 \phi) \mathrm{d}  t^2+(1-2 \psi) \delta_{i j} \mathrm{d}  x^i \mathrm{d}  x^j.
\label{eq:MWmetric}
\end{equation}
These potentials are sourced by the density $\rho_{\rm DM}$ and pressure $p_{\rm DM}$ of the ULDM field as \cite{Flanagan:2005yc,Blas:2024duy,Khmelnitsky:2013lxt}
\begin{equation}
    \Delta \psi=4 \pi G \rho_\mathrm{DM}, ~~6 \ddot{\psi}-2 \Delta(\psi-\phi)=24 \pi G p_{\rm DM}.
    \label{eq:MWpotentials}
\end{equation}
In these expressions, the energy density $\rho$ and pressure $p$ are
\begin{equation}
    \begin{split}
        \rho_{\rm DM}=\frac{1}{2}\left[\dot{\phi}^2+(\nabla \phi)^2+m_{\rm DM}^2 \phi^2\right],\\
        p_{\rm DM} =\frac{1}{2}\left[\dot{\phi}^2-(\nabla \phi)^2-m_{\rm DM}^2 \phi^2\right].
        \label{eq:rho_p}
    \end{split}
\end{equation}
The gravitational impact of ULDM on a binary system, as seen in coordinates 
defined in the center of mass, is given by the acceleration induced from metric in Eq.~\eqref{eq:MWmetric}
\begin{equation}
  \bm{a}_\mathrm{ULDM}  = -\ddot{\psi}\,\bm{r}.
  \label{eq:ULDMa}
\end{equation}
From this, the relevant quantity in Eq.~\eqref{eq:general_linear} reads
\begin{equation}
\label{eq:FaULDM}
\Ft_1^{\alphat }(t) = \Ft^{\alphat}_\mathrm{ULDM}(\tuple, t)\ddot \psi(t),
\end{equation}
with 
\begin{equation}
    \Ft^{\alphat }_\mathrm{ULDM}(t) = \Mt^{\alphat b}(\tuple_0(t), t) \bm{e}_{bc}(\tuple_0(t)) \bm{r}^c(\tuple_0(t)).
\end{equation}
We can now follow the same steps as in Sec.~\ref{sec:det_GW} and Sec.~\ref{sec:stochastic_gw} to find the effects from deterministic (or highly coherent) or stochastic $\psi$ configurations.
As we show in the next sections, the DM configurations of $\phi$ in a virialized halo generate both situations. As before, we consider here only the case where the unperturbed orbit is Keplerian, while the inclusion of other external forces is discussed in Sec.~\ref{sec:beyond_isolation}.

Before moving to the study of the gravitational interaction of Eq.~\eqref{eq:ULDMa}, it is important to notice that the
direct coupling of ULDM to the members of the binary will also generate other accelerations which perturb orbital
motion \cite{Blas:2016ddr,Blas:2019hxz}. Their study can follow a similar logic as the one we will undertake in the following, but, in general, requires a separate study. However, for the quadratic and universal coupling, this is not the case, and it can be shown that its phenomenology corresponds to simply modifying  the metric to an effective metric, with a coupling $\beta$ to matter, $\psi\mapsto \beta\, \psi$ \cite{Blas:2016ddr,Blas:2019hxz}. This coupling can be understood as the ratio between the fundamental coupling of ULDM to matter $\Lambda$, to the gravitational one, $\beta\equiv \pi \,G_N\Lambda^2$. With this simple prescription, all the formalism developed for gravitational coupling, directly applies to the case of universal quadratic direct coupling, see also \cite{Blas:2024PRLForward}.

%%%%%%%%%%%%%%%%%%%%%%%%%%%%%%%%%%%%%%%%%%%%%%%%%%%%%%%%%%%%%%%%%%%%%%%%%%%%%%%%%%%%
\subsection{Statistics of the stochastic scalar potential}
%%%%%%%%%%%%%%%%%%%%%%%%%%%%%%%%%%%%%%%%%%%%%%%%%%%%%%%%%%%%%%%%%%%%%%%%%%%%%%%%%%%%

To find $\ddot\psi$, one requires the values of $\rho_{\rm DM}$ and $p_{\rm DM}$, and hence the knowledge of the properties of different quantities quadratic in $\phi$. These are determined by the dynamics of the DM halo and admit a stochastic description in terms of fluctuations of $\phi$ with a distribution generated by the
virialization of the halo. The precise form of this distribution has been clarified in 
\cite{Foster:2017hbq} in terms of the distribution function of particles per phase-space volume (see also \cite{Bar-Or:2018pxz,Foster:2020fln,Kim:2023pkx,Cheong:2024ose}),
\begin{equation}
f(\vec{v}) = \frac{\rho_0/m_{\rm DM}}{[2\pi \sigma_0^2]^{3/2}} \exp\bigg[-\frac{|\vec{v}+\vec{v}_{\rm loc}|^2}{2\sigma_0^2}\bigg],
\end{equation}
where $\sigma_0\sim 10^{-3}$ is the velocity dispersion of the ULDM in the halo \cite{Lisanti:2016jxe}, $\vec{v}_{\rm loc}$ the local velocity of the center of mass of the binary with respect to the halo, and $\vec{v}$ is the ULDM velocity as measured in this frame.

As the local field $\psi$ is sourced by the density and pressure from Eq.~\eqref{eq:rho_p}, we can compute its defining statistics from the defining statistics of the DM field $\phi$. Following \cite{Kim:2023pkx}, we first calculate the expected value of $\ddot \psi(t)$. We have 
\begin{equation}
    \ddot \psi(\omega,  \vec{k}) = \frac{4 \pi G \omega^2}{k^2} \delta\rho_\phi(\omega, \vec{k}),
    \label{eq:psiddoteq}
\end{equation}
where $\delta \rho_\phi$ is the density contrast in the DM distribution. By definition, $\langle \delta \rho_\phi\rangle = 0$, so it is clear that we have
\begin{equation}
    \langle \ddot \psi(t) \rangle = 0.
\end{equation}
We now proceed to the second moment. Our interest is in the time-time correlation function 
\begin{equation}
\langle \ddot \psi(t) \ddot \psi(t') \rangle = \int \frac{\mathrm{d}\omega \mathrm{d}\omega'}{(2\pi)^2} e^{\mathrm{i} (\omega t + \omega' t')} \langle \ddot \psi(\omega) \ddot \psi(\omega' ) \rangle,
\end{equation}
which we have already expressed in terms of the correlation in Fourier space (recall Eq.~\eqref{eq:FourierTransform} for our Fourier transform conventions). Following \cite{Kim:2023pkx}, the Fourier-space correlation function can itself be written in terms of the power spectrum as\footnote{From the statistics of the ULDM field, it follows that $\delta \rho$, and hence $\ddot\psi$, are non-Gaussian, while our projections assume they are gaussian and characterized by the one- and two-point functions for $\psi$. While consistent with existing ULDM analysis procedures \cite{Porayko:2018sfa, EPTA:2023sfo, EuropeanPulsarTimingArray:2023egv, NANOGrav:2023hvm}, the intrinsic non-gaussianity introduce some corrections to the anticipated sensitivity. We leave study of this effect to future work, see also App.~\ref{app:MonochromaticCoherence}\label{foot:nongauss}.}
\begin{equation}
\begin{split}
\langle \ddot \psi(\omega) \ddot \psi(\omega' ) \rangle &= (2 \pi) \delta(\omega + \omega') \int \frac{\mathrm{d}^3 \vec{k}}{(2 \pi)^3} P_{\ddot \psi}(\omega, \vec{k}), 
\end{split}
\end{equation}
where the power-spectrum integral can be split into low- and high-frequency correlation functions. Specifically, we can define 
\begin{equation}
2\pi\int \frac{\mathrm{d}^3 \vec{k}}{(2 \pi)^3} P_{\ddot \psi}(\omega, \vec{k})=A(|\omega|) + B(|\omega|),
\end{equation}
with 
\begin{widetext}
\begin{equation}
\begin{gathered}
A(\omega) \equiv 32 \pi ^4 G^2 m^2 \omega^4 \int \frac{\mathrm{d}^3 \vec{v}_1 \mathrm{d}^3\vec{v}_2 \mathrm{d}^3\vec{k}}{|\vec{k}|^4} f(\vec{v}_1) f(\vec{v}_2) \left[\delta^{(4)}(\bm{k} + \bm{p}_1 - \bm{p}_2) + \delta^{(4)}(\bm{k} - \bm{p}_1 + \bm{p}_2)\right], \\
B(\omega) \equiv 2 \pi ^4 G^2 m^2 \omega^4  \int \frac{\mathrm{d}^3 \vec{v}_1 \mathrm{d}^3\vec{v}_2\mathrm{d}^3\vec{k}}{|\vec{k}|^4} f(\vec{v}_1) f(\vec{v}_2) |\vec{v}_1 + \vec{v}_2|^{4} \left[ \delta^{(4)}(\bm{k} - \bm{p}_1 - \bm{p}_2)  + \delta^{(4)}(\bm{k} + \bm{p}_1 + \bm{p}_2)\right].
\label{eq:ULDMPowerComponents}
\end{gathered}
\end{equation}
\end{widetext}
The four-vectors $\bm{k}$, $\bm{p}_1$, and $\bm{p}_2$ are defined by $\bm{k} =(\omega, \vec{k})$ and $\bm{p}_{i} = m \left( 1+|\vec{v}_{i}|^2/2, \vec{v}_{i}\right)$. 

For simplicity, we will consider the effects of the ULDM fluctuations on a binary system which are at rest with respect to the DM rest frame, $\vec{v}_{\rm loc}=0$, though accounting for possible boosts of the binary with respect to the DM may yield slightly modified predictions associated with boosting the DM velocity distribution to the binary frame. For our assumed isotropic velocity distribution, these integrals can be analytically evaluated, 
\begin{equation}
\begin{split}
A(\omega) &= \frac{16 \pi ^3 G^2 \omega^3 \rho_0^2}{m_{\rm DM}^4 \sigma_0^4}  K_1\left(\frac{\omega}{m_{\rm DM} \sigma_0^2}\right),\\
B(\omega)& =  \frac{\pi ^4 G^2  \rho_0^2}{m_{\rm DM}^7 \sigma_0^6}  \omega^4 (\omega- 2 m_\mathrm{DM})^2 \\
&~~~\times e^{-\frac{\omega -2 m_{\rm DM}}{m_{\rm DM} \sigma_0^2}} \theta(\omega -2 m_{\rm DM}), 
\end{split}
\end{equation}
where $K_1$ is a modified Bessel function of the second kind with an index of 1.

Finally, the time-domain correlator of $\ddot\psi$ reads
\begin{equation}
\begin{split}
    \langle \ddot \psi(t)\ddot \psi(t') \rangle 
    =  \int_0^{\infty} \frac{\mathrm{d}\omega}{2\pi^2}
        \cos(\omega\tau)(A(\omega)+ B(\omega)),
\label{eq:PotentialCorrelator}
\end{split}
\end{equation}
with $\tau\equiv t-t'$. These two terms have a clear interpretation \cite{Kim:2023pkx}: the first one has support over a wide range of low frequencies less than roughly 
\begin{equation}
\tau_{\rm coh}^{-1}\equiv m_{\rm DM}\sigma_0^2\sim 10^{-6} m_{\rm DM},
\end{equation}
while the second one is localized at $\omega\approx 2 m_{\rm DM}$, with width of order $\tau_{\rm coh}^{-1}$. Since in both cases the width of the distributions around a certain frequency are similar,
both $A(\omega)$ and $B(\omega)$ terms have a coherence time of order $\tau_{\rm coh}$.

When an observation spans a time less than the coherence time, the fluctuations of the scalar potential $\ddot \psi$ are often treated as oscillating at a single frequency mode with unknown amplitude and phase, 
\begin{equation}
\label{eq:coh_naive}
    \ddot \psi(t)=4\pi  \hat\psi \, G\rho_{\rm DM} \cos(2 m_{\rm DM}t+\varphi),
\end{equation}
see, \textit{e.g.} \cite{Khmelnitsky:2013lxt}. The amplitude $\hat\psi$ is Rayleigh distributed with scale parameter equal to one, and the phase has a flat distribution form $0$ to $2\pi$ \cite{Foster:2017hbq}. 
For completeness, in App.~\ref{app:MonochromaticCoherence} we show how this single frequency mode picture emerges from the Gaussian statistics described in Eq.~\eqref{eq:PotentialCorrelator}, cf. Eq.~\eqref{eq:coh_psi}. As both times, shorter and longer in duration than the coherence time, are of observational relevance for this work, to develop a unified approach, we elect to work solely with the covariance matrix for subsequent calculations, able to address both cases. Moreover, working with the covariance matrix description will enable a treatment of perturbations induced by ULDM fluctuations that is similar to that of perturbations induced by gravitational waves. The expression of Eq.~\eqref{eq:coh_naive} will be modified if we boost the binary with respect to the ULDM rest frame \cite{Blas:2019hxz,Kim:2023kyy}. However, because the expected values of $\vec{v}_{\rm loc}$ are always non-relativistic (e.g. $\vec{v}_{\rm loc}\sim 10^{-3}$ for the binaries in the Solar System), the non-zero $\vec{v}_{\rm loc}$  results in suppressed effects with respect to those we describe in this section, and we ignore them for this reason. These boosts may yield significant effects in the case of direct couplings that violate the equivalence principle, as studied in \cite{Blas:2019hxz}. We do not consider these models in this work.

%%%%%%%%%%%%%%%%%%%%%%%%%%%%%%%%%%%%%%%%%%%%%%%%%%%%%%%%%%%%%%%%%%%%%%%%%%%
\subsection{Binary dynamics from coherent ULDM perturbations}
\label{sec:coh_ULDM}
%%%%%%%%%%%%%%%%%%%%%%%%%%%%%%%%%%%%%%%%%%%%%%%%%%%%%%%%%%%%%%%%%%%%%%%%%%%

The coherent ULDM perturbations, characterized by Eq.~\eqref{eq:coh_naive},
are the source of many of the tests of this dark matter paradigm~\cite{Hui:2016ltb,Ferreira:2020fam}. For
the case of orbital motion, their secular effects have been used in \cite{Blas:2016ddr,LopezNacir:2018epg,Blas:2019hxz,Armaleo:2020yml,Kus:2024vpa} to find first constraints on gravitational and direct coupling from binary pulsars, while \cite{Rozner:2019gba} advanced some conclusions related to non-secular effects. Note that some of them, \cite{Blas:2019hxz,Kus:2024vpa} in particular, not only described the resonant case, but also the cases away from resonances.

In Eq.~\eqref{eq:FaULDM} we already found $F^\alphat_{1,\rm ULDM}$, to be used in  Eq.~\eqref{eq:solgen}. 
As done in Sec.~\ref{sec:det_GW}, rather than directly integrating this solution, it is more convenient to treat the dual ordinary differential equation
\begin{equation}
\begin{gathered}
\tuple_1^\alphat(t) = \Phit_{\alphat \betat}(t) \First_\betat(t),\\
\dot{\First}_1^\betat = \Phit^{-1}_{\betat\gammat}(t)\Ft^{\gammat}_\mathrm{ULDM}(\tuple, t)\ddot \psi(t).
\label{eq:determ_finalULDM}
\end{gathered}
\end{equation}
As in Sec.~\ref{sec:det_GW}, we now make contact with our analytic results for the growth of perturbations of isolated Keplerian binaries of Sec.~\ref{sec:AnalyticResults}. Since the acceleration due to the ULDM always acts in the radial direction, ULDM perturbations can affect the $e$, $\omega$, and $f$ perturbations. Similarly to before, on-resonance, we will have quadratic growth of $f_1$ and linear growth in $e_1$ and $\omega_1$, while off-resonance we will have linear growth in $f_1$ and no accumulation in either $e_1$ or $\omega_1$. 

For ULDM, we do not anticipate any scenarios in which the amplitude and phase parameters of the ULDM are known, even in the sub-coherence time regime, motivating the stochastic description we subsequently develop. However, in App.~\ref{app:MonochromaticCoherence}, we develop a statistically equivalent treatment to that which follows this discussion by calculating the induced perturbations from a coherent fluctuation of the form in Eq.~\eqref{eq:coh_naive} as a function of amplitude and phase using Eq.~\eqref{eq:determ_finalULDM}. In that context, the amplitude and phase represent unknown parameters drawn from known distributions (Rayleigh and uniform, respectively) which follow from the approxiate Gaussian statistics of $\ddot \psi$. These values are valid for regions of the size of the order of the de Broglie wavelength characteristic of the distribution, $\lambda_{\rm dB}\sim \frac{1}{m_{\rm DM}\sigma_0}$.

%%%%%%%%%%%%%%%%%%%%%%%%%%%%%%%%%%%%%%%%%%%%%%%%%%%%%%%%%%%%%%%%%%%%%%%%%%%
\subsection{Binary dynamics from stochastic scalar perturbations}
%%%%%%%%%%%%%%%%%%%%%%%%%%%%%%%%%%%%%%%%%%%%%%%%%%%%%%%%%%%%%%%%%%%%%%%%%%%

Next, we proceed to calculate the two-point correlator
\begin{widetext}
\begin{equation}
\begin{split}
\bm{\Sigma}_{\alphat \kappat}(t, t') =\langle \tuple_1^{\alphat }(t) \tuple_1^{\kappat}(t')  \rangle 
&= \Phit_{\alphat \betat  }(t)\Phit_{\kappat\lambdat}(t')\int_0^t \mathrm{d}\tau \int_0^{t'} \mathrm{d}\tau' \Phit_{\betat  \gammat  }^{-1}(\tau) \Phit_{\lambdat\mut}^{-1}(\tau')\Ft_\mathrm{ULDM}^{\gammat  }(\tau)  \Ft_\mathrm{ULDM}^{\mut}(\tau') \\
&~~~~~~~~~~~~~~~~~~~~~~~~~~~~~~~~~~~~\times \int \frac{\mathrm{d}\omega}{2\pi^2} \cos[\omega(\tau-\tau')][A(\omega) + B(\omega)].
\end{split}
\end{equation}
\end{widetext}
In analogy to our treatment of an SGWB (cf. Eq.~\eqref{eq:XSGWB}), we define $\First_{\betat  , c}(t|\omega)$ and $\First_{\betat  , s}(t|\omega)$ as the solutions to the differential equations  
\begin{equation}
\begin{gathered}
\dot{\First}_{\betat,c}(t|\omega) = \Phit^{-1}_{\betat \gammat}(t) \Ft_\mathrm{ULDM}^{\gammat}(t)\cos(\omega t), \\
\dot{\First}_{\betat,s}(t|\omega) = \Phit^{-1}_{\betat  \gammat}(t) \Ft_\mathrm{ULDM}^{\gammat}(t)\sin(\omega t),
\end{gathered}
\end{equation}
so that the correlation function is given by
\begin{equation}
\begin{split}
    \bm{\Sigma}_{\alphat \kappat}(t, t') &= \Phit_{\alphat \betat  }(t) \Phit_{\kappat\lambdat}(t') \int_0^\infty \frac{\mathrm{d}\omega}{2 \pi^2}   [A(\omega) +B(\omega) ] \\
   &~~~~~~~~~~~~ \times \First_{\betat,d}(t |\omega) \First_{\lambdat,d}(t' |\omega).
\end{split}
\end{equation}
As mentioned above, and will be clear when we describe the analysis method, the knowledge of this correlation function is also enough to search for the coherent ULDM oscillations of Sec.~\ref{sec:coh_ULDM} in the data, and it will be our main object of interest for the ULDM case.

%%%%%%%%%%%%%%%%%%%%%%%%%%%%%%%%%%%%%%%%%%%%%%%%%%%%%%%%%%%%%%%%%%%%%%%%%%%%%%%%
\subsection{The growth of the perturbation covariance}
%%%%%%%%%%%%%%%%%%%%%%%%%%%%%%%%%%%%%%%%%%%%%%%%%%%%%%%%%%%%%%%%%%%%%%%%%%%%%%%%

{\color{black}
The time evolution of the covariance of the orbital perturbations induced by ULDM can be understood in close analogy with the SGWB case, but with an important simplification: at fixed frequency, the scalar perturbation has only two quadratures, $\cos(\omega t)$ and $\sin(\omega t)$, and the acceleration in Eq.~\eqref{eq:ULDMa} is always radial. As a result, ULDM directly drives only the $e_1$ and $\omega_1$ non-phase perturbations. The faster growth of $f_1$ arises because the radial acceleration makes $e_1$ dynamical, which in turn changes the rate of orbital phase advance.

For a fixed-frequency component exactly on an orbital harmonic, $e_1$ and $\omega_1$ grow linearly while $f_1$ grows quadratically. The corresponding covariance scales as
\begin{equation}
\bm{\Sigma}_{\alphat\betat}(t, t'|\omega) \propto 
\begin{cases} 
(t t')^2 & \text{if } \alphat = \betat = f, \\
t^2 t' & \text{if } \alphat = f,\; \betat \in \{e, \omega\}, \\
t t'^2 & \text{if } \alphat \in \{e, \omega\},\; \betat = f,\\
t t' & \text{if  } \alphat,\betat \in \{e, \omega\}, \\
0 & \text{else}.
\end{cases}
\label{eq:uldm_on_resonance_covariance_scaling}
\end{equation}
For a fixed-frequency component far from resonance, $e_1$ and $\omega_1$ do not accumulate secularly, while $f_1$ grows only linearly. Thus
\begin{equation}
\bm{\Sigma}_{\alphat\betat}(t, t'|\omega) \propto 
\begin{cases} 
t t' & \text{if } \alphat = \betat = f, \\
t  & \text{if } \alphat = f,\; \betat \in \{e, \omega\}, \\
t' & \text{if } \alphat \in \{e, \omega\},\; \betat = f,\\
1 & \text{if  } \alphat,\betat \in \{e, \omega\}, \\
0 & \text{else}.
\end{cases}
\label{eq:uldm_off_resonance_covariance_scaling}
\end{equation}
}

{\color{black}
The stochastic ULDM covariance is obtained by integrating these fixed-frequency responses against the spectral weight $A(\omega)+B(\omega)$ appearing in Eq.~\eqref{eq:PotentialCorrelator}. However, the two terms should be interpreted differently. The low-frequency component $A(\omega)$ is broadband over $\omega\lesssim m_{\rm DM}\sigma_0^2$. If an orbital harmonic lies within this support and $A(\omega)$ varies slowly across the finite-time resonant bandwidth, then the SGWB-like bandwidth argument applies. For a perfectly periodic binary observed for a time $t_{\rm obs}$, the resonant bandwidth is $\Delta\omega_{\rm res}\sim t_{\rm obs}^{-1}$, and at equal times
\begin{equation}
\begin{gathered}
    \bm{\Sigma}_{ff}^{A}
    \sim
    A(\omega_k)\,t_{\rm obs}^4\,\Delta\omega_{\rm res}
    \sim
    A(\omega_k)\,t_{\rm obs}^3,\\
    \bm{\Sigma}_{ef}^{A},\bm{\Sigma}_{fe}^{A},
    \bm{\Sigma}_{\omega f}^{A},\bm{\Sigma}_{f\omega}^{A}
    \sim
    A(\omega_k)\,t_{\rm obs}^2,\\
    \bm{\Sigma}_{ee}^{A},\bm{\Sigma}_{e\omega}^{A},
    \bm{\Sigma}_{\omega e}^{A},\bm{\Sigma}_{\omega\omega}^{A}
    \sim
    A(\omega_k)\,t_{\rm obs}.
\end{gathered}
\label{eq:uldm_A_broadband_periodic_covariance_scaling}
\end{equation}
Here $\omega_k$ denotes the relevant orbital harmonic, and we have suppressed numerical constants and orbit-dependent transfer functions. Components involving $p$, $\iota$, or $\Omega$ vanish at this order for the purely radial gravitational coupling considered here.
}

{\color{black}
The narrow component $B(\omega)$ is qualitatively different. It is localized near $\omega\simeq 2m_{\rm DM}$ with width $\Delta\omega_{\rm ULDM}\sim m_{\rm DM}\sigma_0^2$, and therefore behaves as an effectively monochromatic feature whenever its linewidth is not resolved over the observation. In that case, the fixed-frequency scalings in Eqs.~\eqref{eq:uldm_on_resonance_covariance_scaling} and \eqref{eq:uldm_off_resonance_covariance_scaling} apply directly: if $2m_{\rm DM}$ lies within the finite-time resonant band of an orbital harmonic, the response follows the on-resonance hierarchy, while if the harmonic drifts away from the narrow feature, the response becomes off resonance. Thus, unlike the smooth $A(\omega)$ contribution, drifting off the $B(\omega)$ feature does not produce a continued broadband resonant response; it simply removes the resonant enhancement.
}

{\color{black}
For imperfectly periodic binaries, this distinction remains important. If the harmonic $\omega_k(t)$ drifts through the smooth low-frequency support of $A(\omega)$, the response can be assembled over coherent intervals as in the SGWB case. Let $\tau_j$ be the duration of the $j$th interval, centered near $t_j$, over which the harmonic is approximately constant, with local bandwidth $\Delta\omega_j\sim \tau_j^{-1}$. For slowly varying $A(\omega)$,
\begin{equation}
\begin{gathered}
    \bm{\Sigma}_{ff}^{A,(j)}
    \sim
    A(\omega_k(t_j))\,\tau_j^4\,\Delta\omega_j
    \sim
    A(\omega_k(t_j))\,\tau_j^3,\\
    \bm{\Sigma}_{ef}^{A,(j)},\bm{\Sigma}_{fe}^{A,(j)},
    \bm{\Sigma}_{\omega f}^{A,(j)},\bm{\Sigma}_{f\omega}^{A,(j)}
    \sim
    A(\omega_k(t_j))\,\tau_j^2,\\
    \bm{\Sigma}_{ee}^{A,(j)},\bm{\Sigma}_{e\omega}^{A,(j)},
    \bm{\Sigma}_{\omega e}^{A,(j)},\bm{\Sigma}_{\omega\omega}^{A,(j)}
    \sim
    A(\omega_k(t_j))\,\tau_j.
\end{gathered}
\label{eq:uldm_A_local_broadband_covariance_scaling}
\end{equation}
If the coherent intervals have a characteristic duration $\tau$ over a total observing time $t_{\rm obs}$, summing over $N_{\rm coh}\sim t_{\rm obs}/\tau$ intervals gives
\begin{equation}
\begin{gathered}
    \bm{\Sigma}_{ff}^{A}
    \sim
    A(\omega_k)\,t_{\rm obs}\tau^2,\\
    \bm{\Sigma}_{ef}^{A},\bm{\Sigma}_{fe}^{A},
    \bm{\Sigma}_{\omega f}^{A},\bm{\Sigma}_{f\omega}^{A}
    \sim
    A(\omega_k)\,t_{\rm obs}\tau,\\
    \bm{\Sigma}_{ee}^{A},\bm{\Sigma}_{e\omega}^{A},
    \bm{\Sigma}_{\omega e}^{A},\bm{\Sigma}_{\omega\omega}^{A}
    \sim
    A(\omega_k)\,t_{\rm obs}.
\end{gathered}
\label{eq:uldm_A_finite_coherence_covariance_scaling}
\end{equation}
The perfectly periodic broadband result is recovered by setting $\tau=t_{\rm obs}$.
}

{\color{black}
For the narrow $B(\omega)$ feature, an imperfectly periodic binary should instead be treated by tracking the overlap between the evolving harmonic $\omega_k(t)$ and the narrow spectral line near $2m_{\rm DM}$. While the harmonic remains within the overlap of the ULDM linewidth and the finite-time resonant band, the response is effectively coherent and follows the fixed-frequency resonant scaling over that interval. Once the harmonic drifts out of this overlap, the perturbation behaves as an off-resonance fixed-frequency component. The full covariance used in our forecasts is computed from the frequency integral above, so both the smooth low-frequency component and the narrow feature are treated without imposing these limiting estimates by hand.
}

%%%%%%%%%%%%%%%%%%%%%%%%%%%%%%%%%%%%%%%%%%%%%%%%%%%%%%%%%%%%%%%
\section{Observational sensitivities}
\label{sec:ObservationalSensitivities}
%%%%%%%%%%%%%%%%%%%%%%%%%%%%%%%%%%%%%%%%%%%%%%%%%%%%%%%%%%%%%%%

To derive our prediction for current and future sensitivities, we will follow 
a relatively standard approach, based on a statistical analysis of the time series
of the observable, evolving in an environment with noise sources (which we assume to be of the order of those reported in current analysis or forecasts of future missions) and a signal background of GWs or ULDM.  This approach is similar to the one followed in \cite{Blas:2021mpc,Blas:2021mqw}. \textcolor{black}{In the scaling estimates below, we focus on the case of a perfectly periodic binary, for which the resonant response remains coherent over the full observing time. Similar Fisher-scaling estimates can be obtained for imperfectly periodic binaries by using the finite-coherence covariance scalings derived in Sec.~\ref{sec:finite_coherent_time} and in the covariance-growth discussions of Secs.~\ref{sec:stochastic_gw} and~\ref{sec:stochastic_uldm}. In practice, the most accurate sensitivities are obtained by evaluating the full time-resolved covariance numerically, rather than by imposing these asymptotic scaling limits.}

%%%%%%%%%%%%%%%%%%%%%%%%%%%%%%%%%%%%%%%%%%%%%%%%%%%%%%%%%%%%%%%
\subsection{Perturbation mapping}
%%%%%%%%%%%%%%%%%%%%%%%%%%%%%%%%%%%%%%%%%%%%%%%%%%%%%%%%%%%%%%%

The data of interest will be a time-series of time of arrivals of photons, $t_i$, acquired with a certain accuracy $\delta t_i$. The two possibilities we will study are the data related to \emph{ranging}, corresponding to a two-way light travel time, where we measure the difference between the receiving time and the emitting time for the signals, and to \emph{timing}, where we measure the difference in time of arrival between two pulses.
To convert these observations into a forecast for ULDM or GWs, we now connect them to the determination of orbital parameters and use this information to seek the effects of the fundamental background. Some first steps towards a more complex situation with other backgrounds are described in Sec.~\ref{sec:beyond_isolation}.

Regarding ranging, the Keplerian elements of a binary system can be 
connected to the two-way light travel time as
\begin{equation}
    \Delta_{||}(\tuple) =  \frac{2 p}{(1+e\cos f)} +\Delta_{||}^0,
    \label{eq:Delta||}
\end{equation}
assuming no noise sources in the orbital parameters, and emission and reception at the center of mass. In $\Delta_{||}^0$, we include all other quantities of the ranging model included in standard LLR/SLR data analysis (atmospheric delay, relativistic correction, \dots)

As far as pulsar timing is concerned, we can adopt a simplified timing model summing R\"omer, Shapiro, and Einstein time delays. The time of arrival of the pulsar signal is then parametrized as
\begin{equation}
\label{eq:Delta|}
    \Delta_{|}(\tuple)=\Delta_{R}+\Delta_{S}+\Delta_{E} +\Delta_{|}^0,
\end{equation}
given by
\begin{equation}
\begin{aligned}
\Delta_{R} &= \frac{M_2 p \sin \iota}{\cc (1 - e^2) (M_1 + M_2)} 
\big[ (\cos E - e) \sin \omega  \\
&\quad  + \sqrt{1 - e^2} \sin E \cos \omega \big], \\
\Delta_{S} &= - 2 G M_2  
\log \big[ 1 - e \cos E - \sin \iota
\\
&\times\big( \sin \omega (\cos E - e)  
 + \sqrt{1 - e^2} \cos \omega \sin E \big) \big], \\
\Delta_{E} &= \left( \frac{G e^2 p}{1 - e^2} \right)^{1/2} 
\frac{M_2 (M_1 + 2 M_2)}{(M_1 + M_2)^{3/2}} \sin E,
\end{aligned}
\end{equation}
where $M_1$ is the mass of the pulsar, $M_2$ is the mass of the companion, and $E$ is the eccentric anomaly calculable from the true anomaly $f$ \cite{Maggiore:1900zz}. Note that the effect of GWs and ULDM is only considered in the way they modify $\tuple$, and not in the way they modify the propagation of the signal from the source (as considered in PTA or other analysis, see \textit{e.g.} \cite{Blas:2024duy,Khmelnitsky:2013lxt}). As the latter is not secularly affected, and is mainly sensitive to other frequencies, we ignore this effect in the following.

In both cases, the first-order perturbation to observable timing data can be evaluated as
\begin{equation}
  \label{eq:signal}
    \Delta_{I1}\approx \bm{T}_{I\alphat}(t) \tuple^{\alphat}_1, \qquad \bm{T}_{I\alphat}(t) \equiv \frac{\partial (\Delta_I)}{\partial \tuple^{\alphat }}   \bigg|_{\tuple=\tuple_0(t)},
\end{equation}
where $I=\{|,||\}$. Since  $\bm{T}_{I\alphat}$ is a linear mapping, it may also be used to suitably transform the covariance of orbital perturbations to the covariance of the observable timing data.

Note that both the light-travel-time and the pulse-time-of-arrival depend on the true anomaly $f_1$, \textit{i.e.} $\Delta_{I} \supset \bm{T}_{If} f_1$. Since $f_1$ grows more rapidly than any other perturbation, $\Delta_I \approx \bm{T}_{If} f_1$ in the large time limit and so will inherit the time-evolving behavior of $f_1$.

%%%%%%%%%%%%%%%%%%%%%%%%%%%%%%%%%%%%%%%%%%%%%%%%%%%%%%%%%%%%%%%%%%%%%%%%%%%%%
\subsection{Deterministic GW and coherent ULDM signals}\label{sec:detGWcohULDM}
%%%%%%%%%%%%%%%%%%%%%%%%%%%%%%%%%%%%%%%%%%%%%%%%%%%%%%%%%%%%%%%%%%%%%%%%%%%%%

We will start by considering our sensitivity to the amplitude of a deterministic gravitational wave. Let us take a GW of the form
\begin{equation}
\ddot{h}_{ij}(t) =  \omega^2 h_c \hat{h}_{ij}(t),
\end{equation}
and assume we observe its imprint upon a binary system for a total duration of $t_{\rm obs}$. From Eq.~\eqref{eq:signal}, we can compute the expected signal $\bm{\mu}_\mathrm{\rm sig}$ defined as 
\begin{equation}
\begin{gathered}  
\label{eq:signal_mGW}
\Delta^{\rm GWs}_{I}(t) = h_c \bm{\mu}_I (t), \\
    \bm{\mu}_I(t) = \omega^2 \bm{T}_{I\alphat }(t) \Phit_{\alphat \betat  }(t) \int_0^t \mathrm{d}\tau \Phit^{-1}_{\betat  \gammat  }(\tau) \Ft_\mathrm{GW}^{\gammat  ij}(\tau) \hat{h}_{ij}(\tau).
\end{gathered}
\end{equation}
Our Fisher information for $h_c$ (see App.~\ref{app:FisherInformation}) is then given by
\begin{equation}
    \bm{I}_{h_c h_c}= \bm{\mu}_I^T \bm{\Sigma}^{-1} \bm{\mu}_I,
\end{equation}
where $\bm{\Sigma}$ is the covariance matrix of the measurements.
Assuming that our observational uncertainties are independent and identically distributed (\textit{i.i.d.}) Gaussian with measurement variance $\sigma^2$ at each data point, we have
\begin{equation}
\begin{split}
    \bm{I}_{h_c h_c} &= \frac{1}{\sigma^2 }\bm{\mu}_I^T \bm{\mu}_I = \frac{1}{\sigma^2\Delta t }\sum_i \Delta t\bm{\mu}_I(t_i)\bm{\mu}_I(t_i)\\
    & \approx \int_0^{t_{\rm obs}} \frac{\mathrm{d}t}{\sigma^2\Delta t} \bm{\mu}_I(t)^2,
\end{split}
\end{equation}
where $\Delta t$ is the time between measurements and we are in the limit of dense resolution of $\bm{\mu}_I$.

Supposing that the deterministic GW is resonant with the zeroth-order orbit and that the GW drives quadratic growth of $f_1$, then we have to good approximation in the large time limit, 
\begin{equation}
\label{eq:Fisherh}
\bm{I}_{h_c h_c} \propto \int_0^{t_{\rm obs}} \frac{\mathrm{d}t}{\sigma^2} t^4 \propto\frac{t_{\rm obs}^5}{\sigma^2}.
\end{equation}
As a result, we see that our sensitivity to $h_c$ scales like $t_{\rm obs}^{5/2}/\sigma$. When off-resonance, we would instead have linear growth of $f_1$, resulting in 
\begin{equation}
\bm{I}_{h_c h_c} \propto \int_0^{t_{\rm obs}}  \frac{\mathrm d t}{\sigma^2} t^2 \propto \frac{t_{\rm obs}^3}{\sigma^2},
\end{equation}
and sensitivity to $h_c$ scaling like $t_{\rm obs}^{3/2}/\sigma$.

The previous analysis follows almost identically for the ULDM case. To get an estimate of the 
reach in sensitivity, it is enough to consider that $\omega^2 h_c$ is of the order 
of $A\,\rho_{\rm DM}$ and $\Ft^{\alphat}_\mathrm{ULDM}$ substitutes $\Ft_\mathrm{GW}^{\gammat  ij}$ in Eq.~\eqref{eq:signal_mGW}.

{\color{black}
%%%%%%%%%%%%%%%%%%%%%%%%%%%%%%%%%%%%%%%%%%%%%%%%%%%%%%%%%%%%%%%%%
\subsection{Stochastic GW and ULDM signals}
%%%%%%%%%%%%%%%%%%%%%%%%%%%%%%%%%%%%%%%%%%%%%%%%%%%%%%%%%%%%%%%%%

Now let us consider a stochastic signal whose amplitude is controlled by a single normalization parameter. For an SGWB we write
\begin{equation}
    \Omega_{\rm GW}(\nu)=\Omega_{\rm GW}^0 \hat\Omega_{\rm GW}(\nu),
\end{equation}
while for ULDM the analogous normalization is set by $\rho_{\rm ULDM}^2$. In both cases, the stochastic signal enters the observable time series through the signal covariance. For an SGWB, suppressing the observable index $I$, this covariance is
\begin{equation}
    \bm{\Sigma}_{\rm sig}(t,t')
    =
    \Omega_{\rm GW}^0\,
    \bm{T}_{\alphat}(t)\bm{T}_{\betat}(t')
    \int_0^\infty \mathrm{d}\nu\,
    \nu \hat\Omega_{\rm GW}(\nu)
    \bm{\Sigma}_{\alphat\betat}(t,t'|\nu),
    \label{eq:observable_sgwb_signal_covariance}
\end{equation}
with the analogous ULDM expression obtained by replacing the SGWB spectral weight by the scalar-potential spectrum $A(\omega)+B(\omega)$ described in Sec.~\ref{sec:stochastic_uldm}.

Assuming \textit{i.i.d.} Gaussian measurement noise with variance $\sigma^2$, the total covariance is
\begin{equation}
    \bm{\Sigma}(t,t')=\sigma^2\bm{I}+\bm{\Sigma}_{\rm sig}(t,t').
\end{equation}
The Fisher information for the signal normalization, evaluated under the null hypothesis, is
\begin{equation}
    \bm{I}
    =
    \frac{1}{2\sigma^4}
    \mathrm{Tr}\!\left[\bm{\Sigma}_{\rm sig}^2\right]
    \approx
    \frac{1}{2\sigma^4(\Delta t)^2}
    \int_0^{t_{\rm obs}}\mathrm{d}t\,\mathrm{d}t'\,
    \bm{\Sigma}_{\rm sig}(t,t')^2 ,
    \label{eq:stochastic_signal_fisher_scaling}
\end{equation}
where $\Delta t$ is the cadence. Thus the sensitivity follows directly from the time scaling of the observable covariance.

For a monochromatic stochastic signal exactly on resonance, the dominant true-anomaly contribution scales as
$\bm{\Sigma}_{\rm sig}(t,t')\sim (tt')^2$, giving
\begin{equation}
    \bm{I}
    \propto
    \frac{t_{\rm obs}^{10}}{\sigma^4(\Delta t)^2}.
\end{equation}
The minimum detectable stochastic amplitude therefore scales as $\sigma^2\Delta t/t_{\rm obs}^{5}$, or equivalently the reach scales as $t_{\rm obs}^{5}/\sigma^2$ at fixed cadence. Off resonance, $\bm{\Sigma}_{\rm sig}(t,t')\sim tt'$, so
\begin{equation}
    \bm{I}
    \propto
    \frac{t_{\rm obs}^{6}}{\sigma^4(\Delta t)^2},
\end{equation}
corresponding to a weaker reach scaling as $t_{\rm obs}^{3}/\sigma^2$ at fixed cadence. These are the covariance-level analogues of the deterministic on- and off-resonance sensitivities discussed above.

For a broadband SGWB with support near the resonant harmonics, the frequency integral softens the coherent monochromatic covariance by one power of the observing time, as discussed in Sec.~\ref{sec:stochastic_gw}. Equivalently, the dominant equal-time signal covariance scales as
\begin{equation}
    \bm{\Sigma}_{\rm sig}\sim t_{\rm obs}^{3},
\end{equation}
rather than $t_{\rm obs}^{4}$. Inserting this scaling into Eq.~\eqref{eq:stochastic_signal_fisher_scaling} gives
\begin{equation}
    \bm{I}_{\Omega_{\rm GW}^0\Omega_{\rm GW}^0}
    \propto
    \frac{t_{\rm obs}^{8}}{\sigma^4(\Delta t)^2}.
\end{equation}
Thus the minimum detectable SGWB normalization scales as $\sigma^2\Delta t/t_{\rm obs}^{4}$, or equivalently the reach scales as $t_{\rm obs}^{4}/\sigma^2$ at fixed cadence.

The ULDM case follows the same Fisher logic, but with the spectral distinction described in Sec.~\ref{sec:stochastic_uldm}. The low-frequency component $A(\omega)$ is broadband over $\omega\lesssim m_{\rm DM}\sigma_0^2$, so when it supports a resonant harmonic its normalization sensitivity follows the broadband covariance scaling. Since the ULDM covariance is proportional to $\rho_{\rm ULDM}^2$, this corresponds to sensitivity to $\rho_{\rm ULDM}$ scaling as $t_{\rm obs}^{2}/\sigma$ at fixed cadence. By contrast, the high-frequency component $B(\omega)$ is a narrow feature near $\omega\simeq 2m_{\rm DM}$ with quality factor $Q\sim\sigma_0^{-2}\sim10^6$. For observational baselines shorter than this coherence time, the feature is effectively monochromatic and is not suppressed by the finite resonant bandwidth. Its sensitivity therefore follows the coherent monochromatic scaling, giving sensitivity to $\rho_{\rm ULDM}$ proportional to $t_{\rm obs}^{5/2}/\sigma$, as in the coherent-signal estimate of Sec.~\ref{sec:detGWcohULDM}.
}

%%%%%%%%%%%%%%%%%%%%%%%%%%%%%%%%%%%%%%%%%%%%%%%%%%%%%%%%%%%%
\section{Sensitivity case studies}\label{sec:CaseStudies}
%%%%%%%%%%%%%%%%%%%%%%%%%%%%%%%%%%%%%%%%%%%%%%%%%%%%%%%%%%%%

With the entirety of our formalism for predicting GW or ULDM-induced signals and projecting observational sensitivities now specified, we proceed to consider a representative set of sensitivity case studies in the context of two systems: Earth-Moon binary and the binary pulsar J1829+2456.  

For the Moon, we assume an initial condition for $\tuple^{\alphat}\equiv \{p,e,\iota,\Omega,\omega,f\}$ of 
\begin{equation*}
    \tuple_0^{\alphat } = 
    \begin{pmatrix}
    .025 \, \mathrm{AU}, & 0.054, & 0.089, & 2.18, & 5.55, & 0  
    \end{pmatrix} ,
\end{equation*}
in ICRS coordinates in which the inclination is measured with respect to the ecliptic,
see \textit{e.g.} \cite{nasaMoonFact, Simon1994}, associated with an unperturbed orbital period of $P_0 \approx 27$ days. For J1829+2456, we take
\begin{equation*}
    \tuple_0^{\alphat } = 
    \begin{pmatrix}
    .0304 \, \mathrm{AU}, & 0.13, & 1.32, & 0.8, & 4,013, & 0  
    \end{pmatrix} 
\end{equation*}
corresponding to an unperturbed orbital period of $P_0 \approx 1.2$ days.  The mass of the MSP is $M_1 \approx 1.31\, M_\odot$ while its companion has mass $M_2 \approx 1.57 M_\odot$~\cite{Haniewicz:2020jro}. For both systems, we take an initial condition of $\tuple_1^{\alphat } = 0$ for the first-order perturbations. Note that the inclination is taken to be $\iota = \pi / 3 \approx 1.32$. In practice, the companion mass $M_2$ and the inclination realize strong degeneracies which are broken only by relativistic effects (see the form of of the R\"omer time delay which depends on the degenerate combination of $M_2 \sin \iota$ as compared to the Shapiro and Einstein time delays, which do not and then break the degeneracy). Here we choose fiducial values of $M_2$ and $\iota$, matching the treatment of \cite{Blas:2021mqw}, but varying these parameters over reasonable prior ranges may be necessary in a comprehensive search. 

Throughout these sections, we will project sensitivities for an observational program measuring perturbations to these binary systems with fiducial sensitivities adopted from the ``2021 sensitivity" scenario of \cite{Blas:2021mpc, Blas:2021mqw}. For observations of the Earth-Moon system, we assume 15 years of measurement, with 260 light-travel-time measurements per year each at $20\,\mathrm{ps}$ precision, corresponding to a distance sensitivity of $3\,\mathrm{mm}$. For the binary pulsar J1829+2456, we assume a similar 15-year observing campaign, making 168 time-of-arrival measurements per year, each with $1\,\mu\mathrm{s}$ precision.

%%%%%%%%%%%%%%%%%%%%%%%%%%%%%%%%%%%%%%%%%%%%%%%%%%%%%%%%%%%%%%%%%%%%%%%
\subsection{Monochromatic GW sensitivities}
\label{sec:MonochromaticSensitivity}
%%%%%%%%%%%%%%%%%%%%%%%%%%%%%%%%%%%%%%%%%%%%%%%%%%%%%%%%%%%%%%%%%%%%%%%

\begin{figure}[!htb]
\begin{center}
\includegraphics[width=0.5\textwidth]{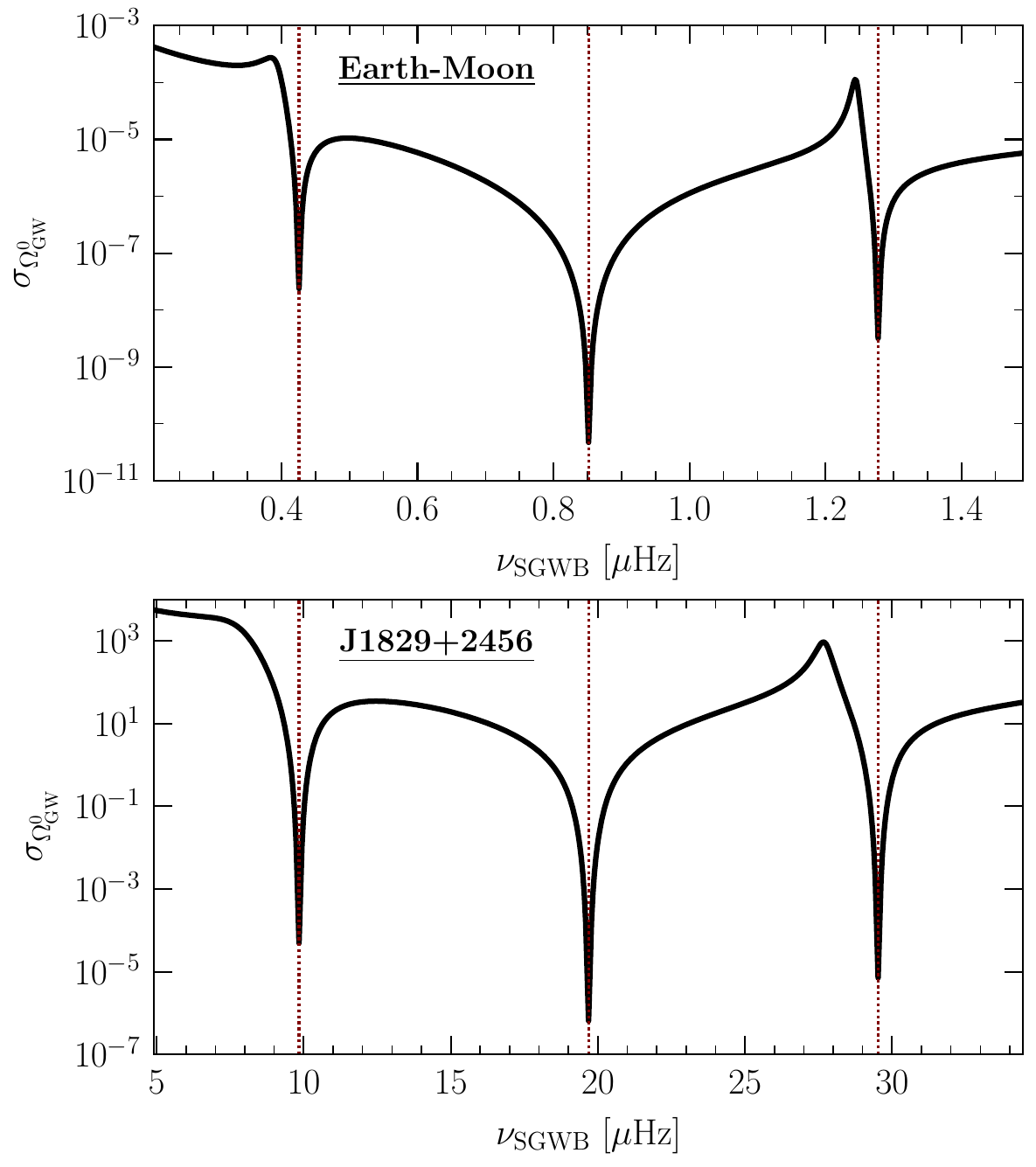}
\end{center}
\caption{The sensitivity to a monochromatic SGWB amplitude $\Omega^0_\mathrm{GW}$ as a function of the SGWB frequency $\nu_\mathrm{SGWB}$. In the top panel, we illustrate the sensitivity derived from a 15-year-long observation of the Earth-Moon system, while in the bottom panel, we illustrate the sensitivity derived from an equal-duration observation of the J1829+2456 binary MSP. Integer multiples of the fundamental frequency $1/P_0$ for the two systems are highlighted in dashed red in both panels. See text for more details.}
\label{fig:MonochromaticSGWBSensitivity}
\end{figure}

We begin our case studies by demonstrating the sensitivity of our two binary systems to a monochromatic SGWB of the form
\begin{equation}
\Omega_\mathrm{GW}(\nu) = \Omega_\mathrm{GW}^0 \delta(\ln\nu - \ln\nu_\mathrm{SGWB}).
\label{eq:MonochromaticSGWB}
\end{equation}
and project our sensitivity to $\Omega_\mathrm{GW}^0$ using our Fisher information treatment. In Fig.~\ref{fig:MonochromaticSGWBSensitivity}, we demonstrate the estimated sensitivity to $\Omega_\mathrm{GW}$ as a function of $\nu_\mathrm{SGWB}$ for frequencies between $0.5 / P_0$ and $3.5 /P_0$ where $P_0$ is the unperturbed period of the binary. The sharp resonant responses at integer multiples of the fundamental frequency $1/P_0$ are clearly visible. For an otherwise unperturbed binary orbit, the quality factor of the response is regulated by the number of orbital periods. For the Earth-Moon system and the binary MSP J1829+2456, we have assumed an equal duration observing campaign, leading to a visibly higher quality factor resonant response in the shorter period J1829+2456 system. Though we do not expect any monochromatic SGWB signals in reality, they are a useful test case for the formalism we have developed as the response to an arbitrary SGWB spectrum can be treated as an integral over the monochromatic responses in our linearized treatment.

%%%%%%%%%%%%%%%%%%%%%%%%%%%%%%%%%%%%%%%%%%%%%%%%%%%%%%%%%%%%%%%%%%%%%%%
\subsection{Time-dependence of sensitivities}
\label{sec:SensitivityTimeDependence}
%%%%%%%%%%%%%%%%%%%%%%%%%%%%%%%%%%%%%%%%%%%%%%%%%%%%%%%%%%%%%%%%%%%%%%%

After having inspected the frequency-dependence of our sensitivities, we now examine the time-dependence of our sensitivities, testing the analytic results developed in Sec.~\ref{sec:ObservationalSensitivities}. For this, we focus on the simplest cases: we first consider two cases of the monochromatic sensitivity, with $\nu_\mathrm{0} = 1.5 / P_0$ and $\nu_\mathrm{0} = 2/P_0$ where $P_0$ is, like before, the unperturbed period of the system. We have chosen to examine frequencies in the vicinity of the second resonance as we have seen in Sec.~\ref{sec:MonochromaticSensitivity} that this is the one that provides the greatest sensitivity. For $\nu_\mathrm{0} = 1.5 / P_0$, the SGWB is maximally off-resonance while for $\nu_\mathrm{0} = 2/ P_0$, the SGWB is exactly on-resonance. As a result, we expect our sensitivity to $\Omega_\mathrm{GW}^0$ to scale as $t_{\rm obs}^3$ and $t_{\rm obs}^5$, respectively, where $t_{\rm obs}$ is the total observation time.

We additionally consider a broadband spectrum following a Gaussian envelope centered on the second resonance given
\begin{equation}
\Omega_\mathrm{GW}(\nu) = \frac{\Omega^0_\mathrm{GW}}{\sqrt{2 \pi \sigma_\mathrm{0}^2}} \exp \left[-\frac{(\nu - \nu_\mathrm{0})^2}{2 \sigma_\mathrm{0}^2} \right]
\label{eq:BroadbandSpec}
\end{equation}
with $\nu_\mathrm{0} = 2/P_0$ and $\sigma_\mathrm{0} = \nu_\mathrm{0}/4$. This SGWB is broadband compared to the resonant response of the binary systems of interest, so we expect the sensitivity to asymptotically scale as $t_{\rm obs}^4$.

\begin{figure}[!htb]
\begin{center}
\includegraphics[width=0.48\textwidth]{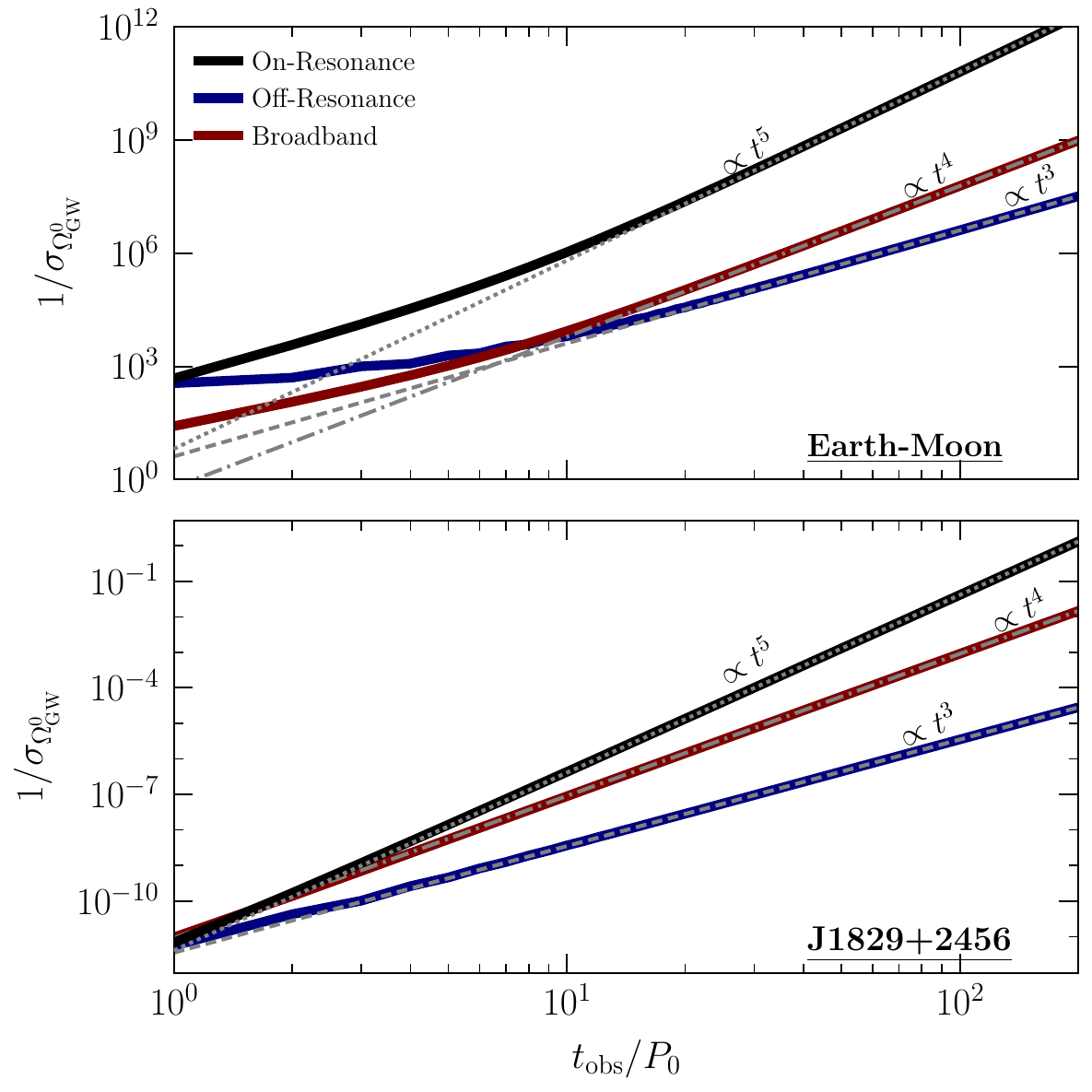}
\end{center}
\caption{(\textit{Above}) The sensitivity to the $\Omega_\mathrm{GW}^0$ parameter for the Earth-Moon system in each of our three SGWB scenarios. In black, we show the growing sensitivity for a monochromatic SGWB that is exactly on-resonance with $\nu_\mathrm{SGWB} = 2/P_0$, while in blue, we show the growing sensitivity for a monochromatic SGWB that is exactly off-resonance with $\nu_\mathrm{SGWB} = 1.5/P_0$. As expected, in these scenarios, the asymptotic sensitivity grows as $t_{\rm obs}^5$ and $t_{\rm obs}^3$, respectively; these power-law scalings are indicated in dotted grey and dashed grey lines. In red, we illustrate the growth of sensitivity to a broadband spectrum as parametrized in Eq.~\eqref{eq:BroadbandSpec}, which realizes the expected asymptotic sensitivity scaling of $t_{\rm obs}^4$, shown in dot-dashed grey. (\textit{Below}) As in the above, but for the J1829+2456 binary MSP. Like before, the expected $t_{\rm obs}^3$, $t_{\rm obs}^4$, and $t_{\rm obs}^5$ scalings are observed for the off-resonance, broadband, and on-resonance scenarios, respectively. }
\label{fig:TimeDependencePlot}
\end{figure}

For each of these three scenarios, we evaluate the time-dependence of the sensitivity to $\Omega_\mathrm{GW}^0$ parameter as a function of time over the first 200 orbits of the binary system. The results are provided in Fig.~\ref{fig:TimeDependencePlot} and realize the expected scaling. Note that the two systems we examine here have different unperturbed periods; as a result, the SGWB spectra used in these examples for the Earth-Moon system differ in frequency parameters (which are defined in terms of $P_0$) from those used for J1829+2456. In summary, these results validate the analytic arguments made in Sec.~\ref{sec:ObservationalSensitivities}.

%%%%%%%%%%%%%%%%%%%%%%%%%%%%%%%%%%%%%%%%%%%%%%%%%%%%%%%%%%%%%%%%%%%%%%%
\subsection{Eccentricity-dependence of sensitivities}
%%%%%%%%%%%%%%%%%%%%%%%%%%%%%%%%%%%%%%%%%%%%%%%%%%%%%%%%%%%%%%%%%%%%%%%

While we have chosen the Earth-Moon system and J1829+2456 as representative examples that demonstrate the relevant calculations for laser-ranging and pulsar-timing sensitivity to binary resonances, there exist several other possibilities for similar binary systems that might also be used to interrogate low-frequency GWs. For instance, any well-timed binary MSP might serve as a viable observational target; additionally, artificial satellites orbiting, \textit{e.g.}, the Earth might have their distance laser-ranged.\footnote{Position measurements via a global navigation satellite system might also provide useful, albeit lower precision, distance measurements.} While the orbital period $P_0$ sets the frequencies to which a given system provides sensitivity, as we demonstrate in this section, the eccentricity $e_0$ also plays an important role in determining the on-resonance sensitivity (see also \cite{Blas:2021mqw}).

\begin{figure*}[!htb]  
    \begin{center}
    \includegraphics[width=\textwidth]{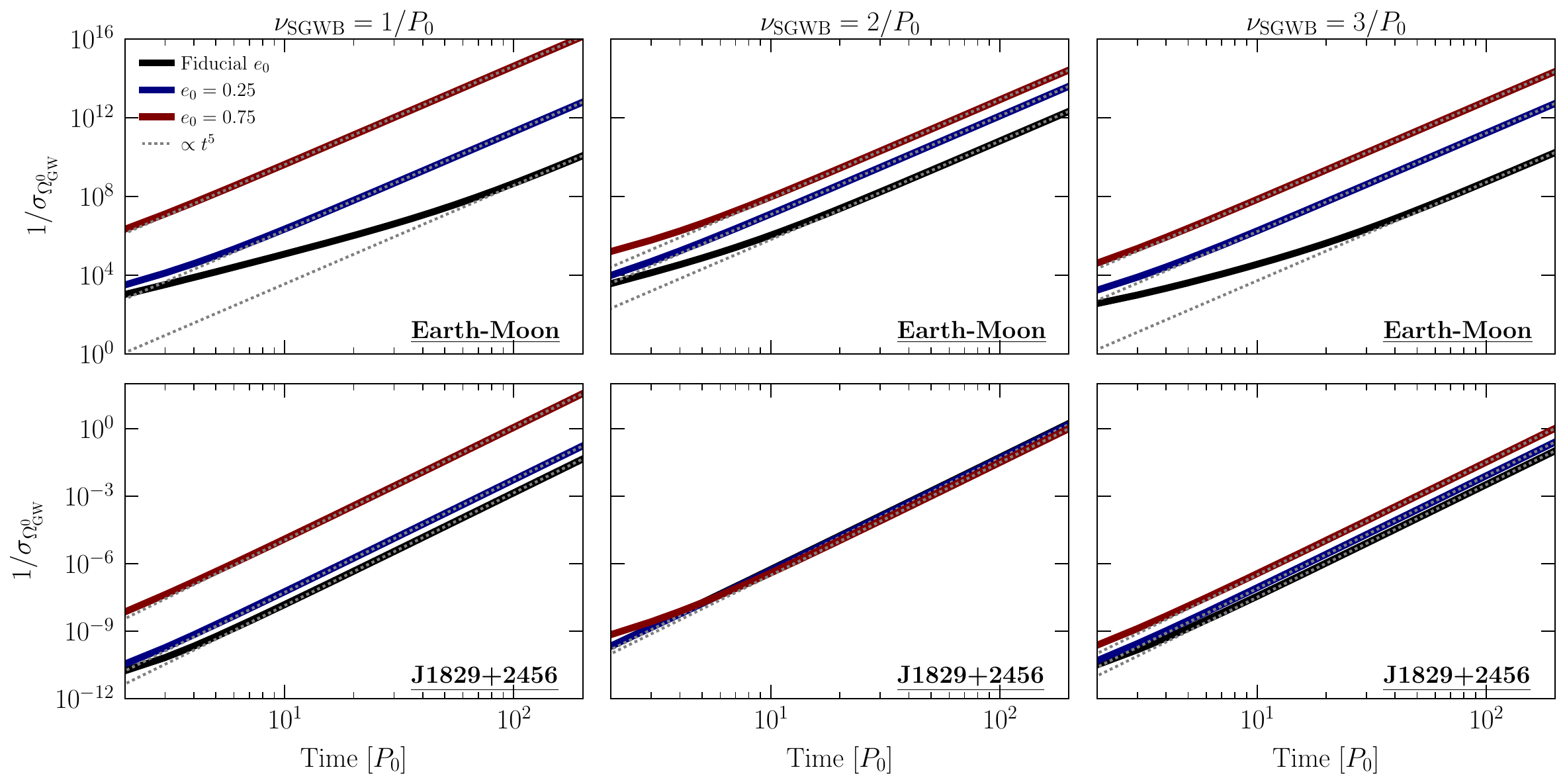}
    \caption{(\textit{Top panels}) The time-evolution of the sensitivity of laser-ranging of the Earth-Moon system over 200 orbits to a monochromatic SGWB which is on-resonance at one of the first three resonances. We illustrate the sensitivity for the true value of $e_0 = 0.054$ (black), as well as the sensitivity that would be realized if $e_0 = 0.25$ (blue) or $e_0 = 0.75$ (red). We also illustrate the asymptotic scaling of $t^5$ for each eccentricity scenario. (\textit{Bottom panels}) As in the top panels, but for the J1829+2456 binary MSP.}
    \label{fig:EccentricityTimeScaling}
    \end{center}
\end{figure*}

To examine the impact of the eccentricity on the sensitivity, we reconsider our sensitivity for a monochromatic SGWB which is on-resonance at $\nu_\mathrm{0} \in \{1, 2, 3\}/P_0$ for the Earth-Moon system and for J1829+2456. We hold all orbital parameters fixed, except for $e_0$, which we vary between its fiducial value, $e_0 = 0.25$ and $e_0 =0.75$. We then evaluate the observational sensitivity as a function of observing time as done in Sec.~\ref{sec:SensitivityTimeDependence}. 

\begin{figure}[!htb]
\begin{center}
\includegraphics[width=0.48\textwidth]{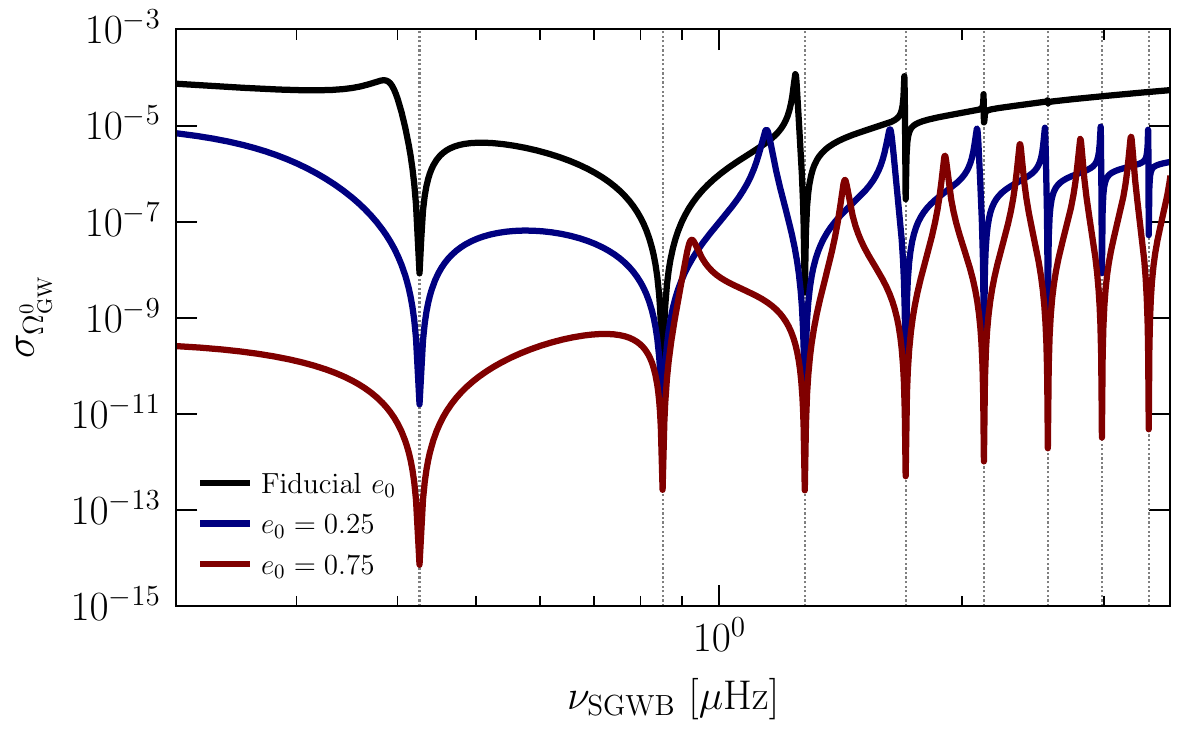}
\end{center}
\caption{The sensitivity to a monochromatic SGWB through laser-ranging of the Earth-Moon distance, as in the top panel Fig.~\ref{fig:MonochromaticSGWBSensitivity}, but varying the unperturbed eccentricity of the system. Dotted grey lines indicate the frequencies which are harmonic with the orbital period of the moon. At the fiducial eccentricity of $e_0 = 0.054$ shown in black, there is not a strong resonant response above the fifth harmonic of the orbital period. As the unperturbed eccentricity increases to $e_0 = 0.25$ (shown in blue) and then $e_0 = 0.75$ (shown in red), a strong resonant response is realized at increasingly large harmonics of the orbital period.}
\label{fig:EccentricityResonance}
\end{figure}

From the results of this calculation, presented in Fig.~\ref{fig:EccentricityTimeScaling}, a general trend that increasing $e_0$ results in a stronger resonant response and thus greater sensitivity can be observed. It is particularly important to note that at lower $e_0$, a weaker resonant response may result in a later time at which the asymptotic sensitivity scaling of $t_{\rm obs}^5$ is realized, though in all cases we consider, this scaling behavior is achieved within 200 orbits. Moreover, as we discuss in App.~\ref{app:Fourier}, higher eccentricity orbits will provide more support at higher harmonics than lower eccentricity orbits, which we illustrate in Fig.~\ref{fig:EccentricityResonance} where we project the sensitivity of the Earth-Moon system to monochromatic SGWB at frequencies between $0.5/P_0$ and $12.5/P_0$ for the three eccentricity scenarios considered in this section.

In general, these results motivate the selection of high-eccentricity systems and the consideration of only those low-eccentricity systems that have been observed for a sufficiently long duration such that the resonant response has come to dominate. For instance, in the case of pulsar timing a population of MSP binaries, these results motivate dedicating more observing time to the most eccentric systems (and with higher periods). In the case of laser ranging, while the lunar eccentricity is as we observe it, the laser-ranging artificial satellites with optimized orbits provide additional opportunities to probe low-frequency GWs. We consider both of these possibilities further in~\cite{Blas:2024PRLForward}.

%%%%%%%%%%%%%%%%%%%%%%%%%%%%%%%%%%%%%%%%%%%%%%%%%%%%%%%%%%%%%%%%%%%%%%%
\subsection{On-sky sensitivity map}
\label{sec:sky}
%%%%%%%%%%%%%%%%%%%%%%%%%%%%%%%%%%%%%%%%%%%%%%%%%%%%%%%%%%%%%%%%%%%%%%%

\begin{figure}[!htb]
\begin{center}
\includegraphics[width=0.5\textwidth]{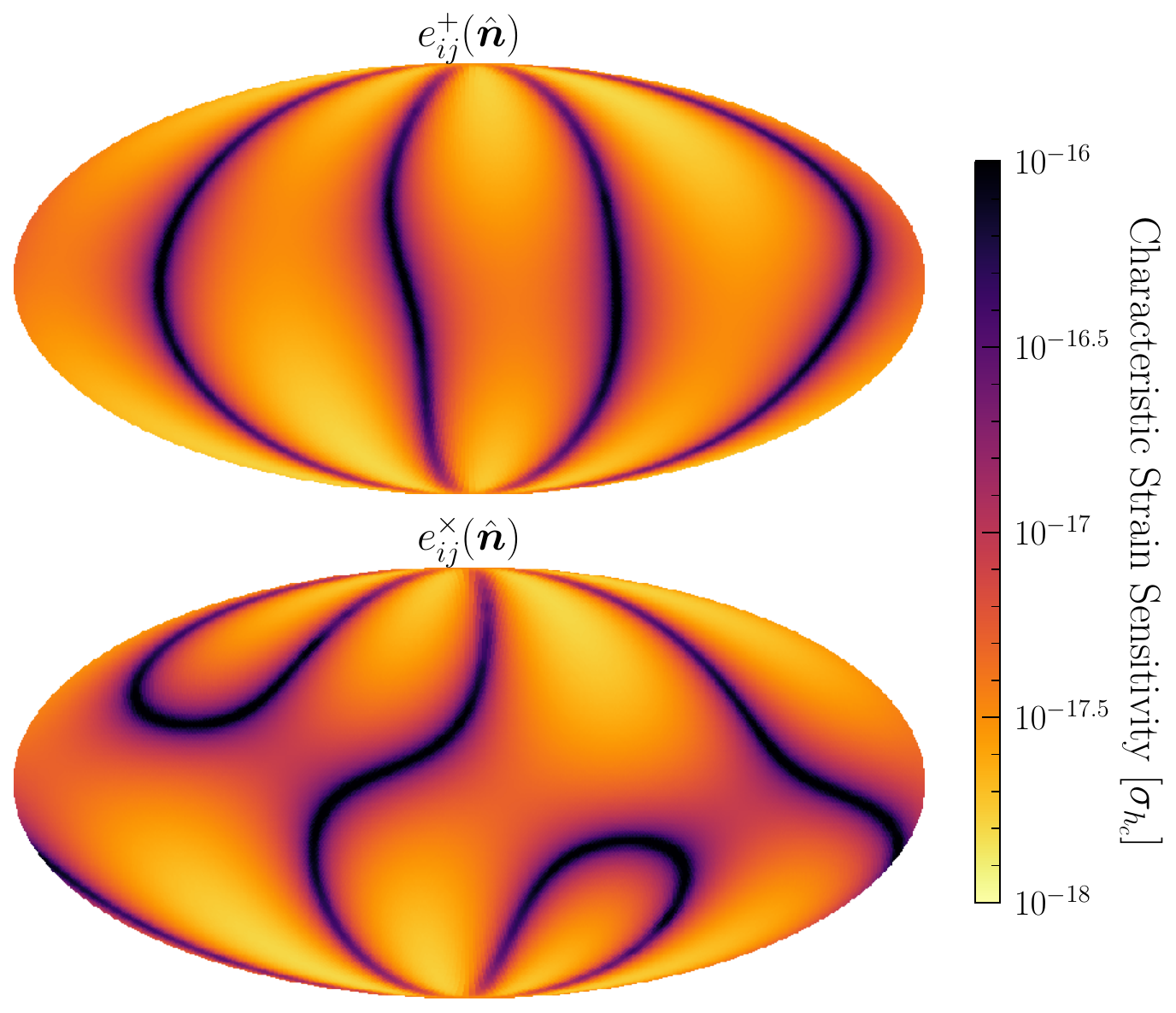}
\end{center}
\caption{Sky-maps for the sensitivity to a monochromatic GW of $+$ (top) and $\times$ polarization through laser-ranging of the Earth-Moon distance. The celestial sphere is shown in ICRS coordinates. With the exception of relatively small-area contours along the celestial sphere at which the binary response is degraded, good sensitivity is achieved across most of the sky for both plus and cross-polarization.}
\label{fig:LLROnSkySensitivity}
\end{figure}

\begin{figure}[!htb]
\begin{center}
\includegraphics[width=0.49\textwidth]{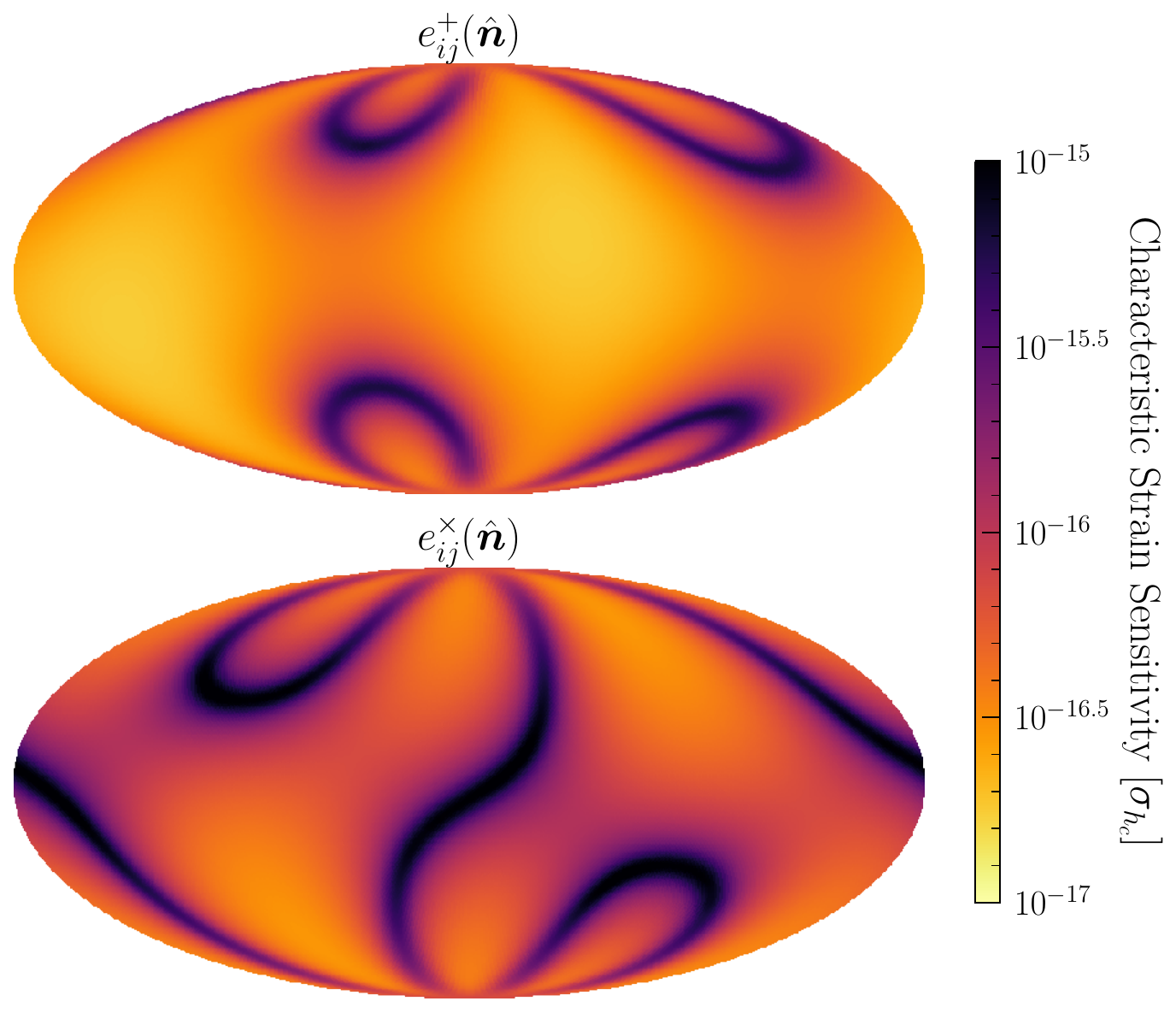}
\end{center}
\caption{As in Fig.~\ref{fig:LLROnSkySensitivity}, but the on-sky sensitivity to a monochromatic GW of each polarization from pulsar timing of the binary pulsar J1829+2456.}
\label{fig:PTAPolarizationOnSky}
\end{figure}

When considering deterministic GW sources rather than isotropic SGWBs, it is useful to examine the sensitivity of the binary resonant response as a function of the propagation direction of the incident GW. To do so, we consider the effect of a gravitational wave with frequency $\nu = 2/P_0$ of the form
\begin{equation}
    h_{ij}(t, \hat{\bm{n}}) = h_c \cos(2 \pi \nu t+\varphi_0)  \bm{e}_{ij}^{A}(\hat{\bm{n}})
\label{eq:MonopolarMonochromatic}
\end{equation}
where $A$ is either the $+$ or $\times$ polarization. It is most instructive to examine the sensitivities at this frequency as we have seen in Sec.~\ref{sec:MonochromaticSensitivity} that the binary response is most observable at this frequency. In Fig.~\ref{fig:LLROnSkySensitivity}, we illustrate the characteristic strain sensitivity $\sigma_{h_c}$ for the Earth-Moon system as a function of $\hat{\bm{n}}$ while in Fig.~\ref{fig:PTAPolarizationOnSky} we show the analogous result for J1829+2456. For simplicity, we set $\varphi_0=0$, though a nonzero phase could change the on-sky sensitivity pattern somewhat.

%%%%%%%%%%%%%%%%%%%%%%%%%%%%%%%%%%%%%%%%%%%%%%%%%%%%%%%%%%%%%%%%%%%%%%%%%%%%%%%%%%%%
\subsection{Chirping black hole inspirals}
%%%%%%%%%%%%%%%%%%%%%%%%%%%%%%%%%%%%%%%%%%%%%%%%%%%%%%%%%%%%%%%%%%%%%%%%%%%%%%%%%%%%

While a comprehensive investigation of the detectability of individual supermassive black hole binaries is beyond the scope of this work, we examine aspects of the problem and point to directions for future inquiry. This is partially motivated by the results in Figs.~\ref{fig:LLROnSkySensitivity} and~\ref{fig:PTAPolarizationOnSky}, which show that our method may be sensitive to the amplitude of GWs expected in the Milky Way by some of these binaries, cf.~\cite{Sesana:2019vho}.

The key opportunity associated with a supermassive black hole binary search is that the binary itself may be time-varying through the inspiral process on observationally relevant timescales. As the binary black hole system evolves, its orbital period shrinks, causing it to emit gravitational waves at increasingly high frequencies. In doing so, GW emission from the binary black hole system may become resonant with the observed binary system, \textit{e.g.}, the Earth-Moon system This can lead to potentially detectable perturbations in the observed orbit without requiring an impossible tuning of the period of the observed binary system. 

As a representative example, we consider a supermassive black hole binary with $M_1 = M_2 = 5 \times 10^8 \, M_\odot$. We place the binary at a distance of 100 Mpc from the Earth and on the celestial sphere at $\theta = \phi = 0$ in the ICRS coordinates of our Earth-Moon system. Using the \texttt{TaylorT4} Post-Newtonian approximant implemented in \texttt{PyCBC}\cite{alex_nitz_2024_10473621}, we generate the GW waveform for the black hole binary assuming spinless black holes and that their orbit has $e = \iota = \Omega = \omega = 0$ for simplicity. In a real search in data, these parameters would be unknown for an as-of-yet undetected black hole binary, and a realistic analysis would necessarily vary them over physically motivated ranges. However, for this toy example, we assume all parameters of the black hole binary are perfectly known.

We proceed to evaluate the orbital perturbations to the Earth-Moon system associated with the GW emission from the black hole binary assuming our 15 years of observation coincide with the final 15 years of the inspiral process. From the orbital perturbations and associated distance perturbation measured by light-travel-time, we evaluate the cumulative $\chi^2$ associated with the perturbations that appear above the null model of no black hole binary which is defined by
\begin{equation}
    \chi^2(t) = \sum_{t_i< t} \left[ \frac{\delta r(t_i)}{\sigma} \right]^2
    \label{eq:Chi2}
\end{equation}
where $\delta r(t_i)$ is the distance perturbation at time $t_i$, $\sigma$ is the measurement precision, and the sum is performed over all measurements. Recall that this $\chi^2$ test statistic follows directly from the Gaussian likelihood specified in App.~\ref{app:FisherInformation} and that $\chi^2 = 25$ corresponds to a so-called ``5$\sigma$ detection".

\begin{figure}[!htb]
\begin{center}
\includegraphics[width=0.5\textwidth]{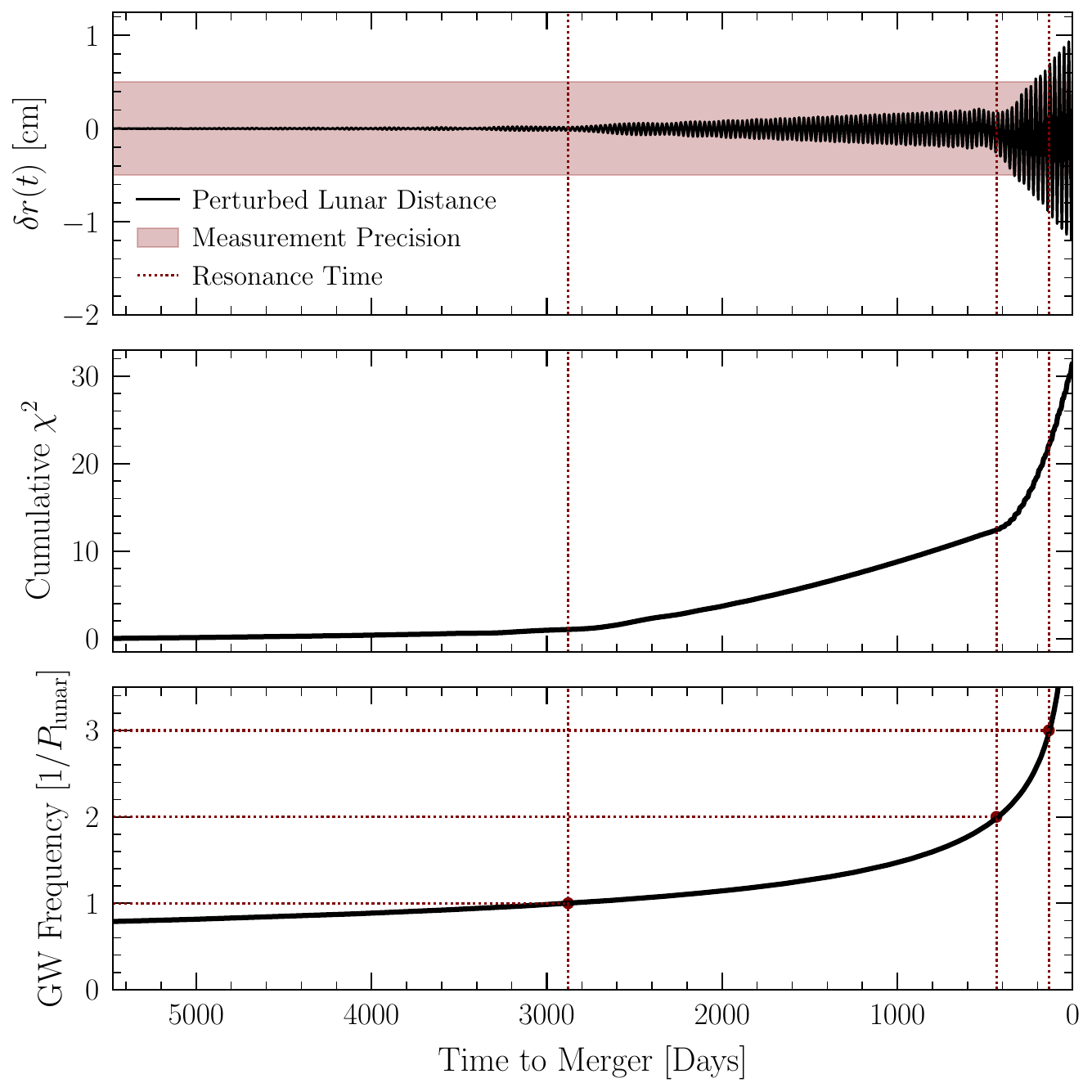}
\end{center}
\caption{(\textit{Top panel}) In black, we show the time-evolving perturbation to the binary separation of the Earth-Moon system as a function of time as measured by time until the merger of the black hole binary. The red band indicates the precision of a single distance measurement $\sigma = 0.3 \,\mathrm{cm}$, and, as before, we assumed a measurement cadence of 260 measurements per year. Vertical dotted red lines indicate the time times at which exact resonance between the Earth-Moon system and the GW emission from the binary black hole system is achieved. The distance perturbation can be seen to grow appreciably around these times, associated with the $t^2$ growth of the distance perturbation when driven on resonance demonstrated in Sec.~\ref{sec:AnalyticResults}. (\textit{Middle panel}) The cumulative $\chi^2$ as a function of time to merger as defined Eq.~\eqref{eq:Chi2}. The growth in the distance perturbation results in the growth of the $\chi^2$ that is clearly visible around and after the resonance times. (\textit{Bottom panel}) The frequency of GW emission as measured in units of $1/P_\mathrm{lunar}$. When the frequency is equal to $1$ in these units, then resonance is achieved. Note that zero-eccentricity inspirals, like that considered here, are characterized by instantaneous monochromatic emission \cite{Maggiore:1900zz, Maggiore:2018sht}, so this single parameter fully characterizes the GW signal.}
\label{fig:InspiralExample}
\end{figure}

The results of these calculations are shown in Fig.~\ref{fig:InspiralExample}, which demonstrate the magnitude of the orbital perturbations and the cumulative $\chi^2$ as a function of measurement time. We also depict the characteristic frequency of the GW emission as a function of the time to merger. As is expected, the orbital perturbations do not grow appreciably until roughly 3000 days before the merger, which is when the emission frequency of the GWs is equal to $1/P_\mathrm{lunar}$ and resonance at the first harmonic is achieved. At this tie, the orbital perturbations (and the associated cumulative $\chi^2$) begin to grow, with even larger growth as the black hole binary continues to undergo the inspiral process and the resonances at the second and third harmonic are achieved. By the end of the inspiral process, the total $\chi^2$ has grown to approximately $30$, indicating that this inspiral process is expected to be discovered in this projected LLR analysis.

%%%%%%%%%%%%%%%%%%%%%%%%%%%%%%%%%%%%%%%%%%%%%%%%%%%%%%%%%%%%%%%%%%%%%%%%%%%%%%%%%%%%
\subsection{Ultra-light dark matter}
%%%%%%%%%%%%%%%%%%%%%%%%%%%%%%%%%%%%%%%%%%%%%%%%%%%%%%%%%%%%%%%%%%%%%%%%%%%%%%%%%%%%

Projecting sensitivity to ULDM follows nearly identically to the procedure to project sensitivity to GWs. However, some care must be taken as the scalar potential fluctuations associated with ULDM support oscillations in two very different frequency regimes: at $\omega \approx 2 \,m_\mathrm{DM}$, which we refer to as coherent fluctuations, and at $\omega \lesssim 10^{-6} m_\mathrm{DM}$, which we refer to as incoherent fluctuations. Recall that we consider the fluctuations on $\ddot \psi$ as Gaussian as a first approximation, see footnote~\ref{foot:nongauss}.

\begin{figure*}[!htb]
\begin{center}
\includegraphics[width=0.9\textwidth]{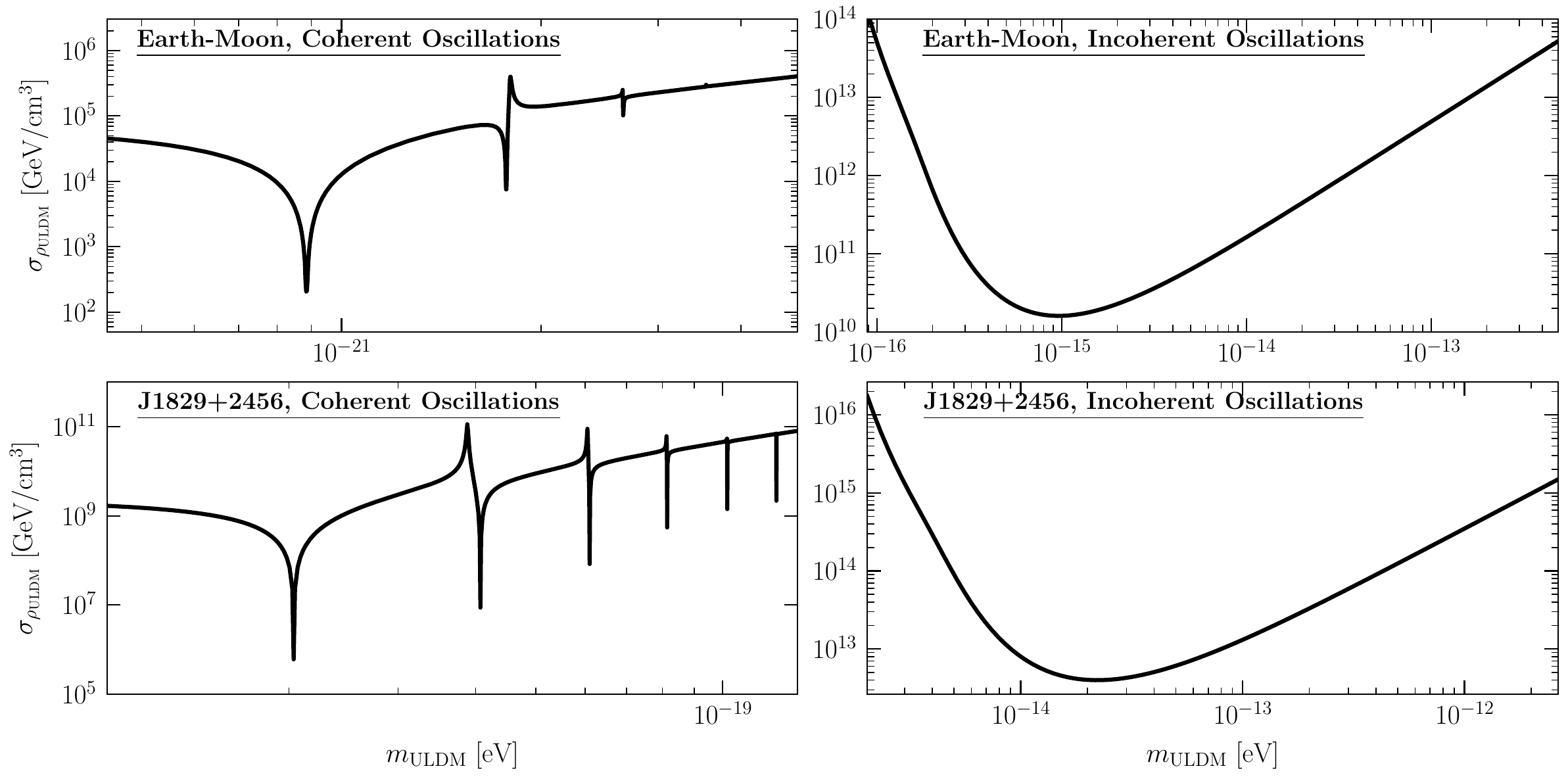}
\end{center}
\caption{The sensitivity to the local density of ULDM, $\rho_\mathrm{ULDM}$ as a function of the ULDM mass. In the top left panel, we illustrate the sensitivity derived from a 15-year-long observation of the Earth-Moon system in search of the coherent oscillations of ULDM at frequencies  $\omega  \approx 2 m_\mathrm{DM}$ described by $B(\omega)$ in Eq.~\eqref{eq:ULDMPowerComponents}. In the top right panel, we illustrate the sensitivity derived from that same projected lunar observation, but in search of the incoherent oscillations of ULDM at frequencies $\omega \approx m v^2$ described by $A(\omega)$ in Eq.~\eqref{eq:ULDMPowerComponents}. In the bottom left and right panels, we illustrate the projected sensitivities of an equal duration observation of the J1829+2456 binary MSP. Recall that the expected DM energy density in the Solar System is $\rho_\odot\sim 0.3 \rm\, GeV/cm^3$. See text for more details.}
\label{fig:ULDMSensitivity}
\end{figure*}

In Fig.~\ref{fig:ULDMSensitivity}, we develop compelling projected sensitivities to the local ULDM abundance at the location of the perturbed binary system for the Earth-Moon binary and the J1829+2456 system. We realize sensitivity at very low masses through the coherent fluctuations and larger masses through the incoherent fluctuations. In both cases, we observe that the sensitivity to $\rho_\mathrm{ULDM}$ is appreciably better through the coherent fluctuations than through the incoherent fluctuations, though they support sensitivity to appreciably different mass ranges. For interferometric searches, the incoherent fluctuations generate better sensitivity at larger masses than the coherent one does at lower masses  \cite{Kim:2023pkx}.

\begin{figure*}[!htb]
\begin{center}
\includegraphics[width=0.95\textwidth]{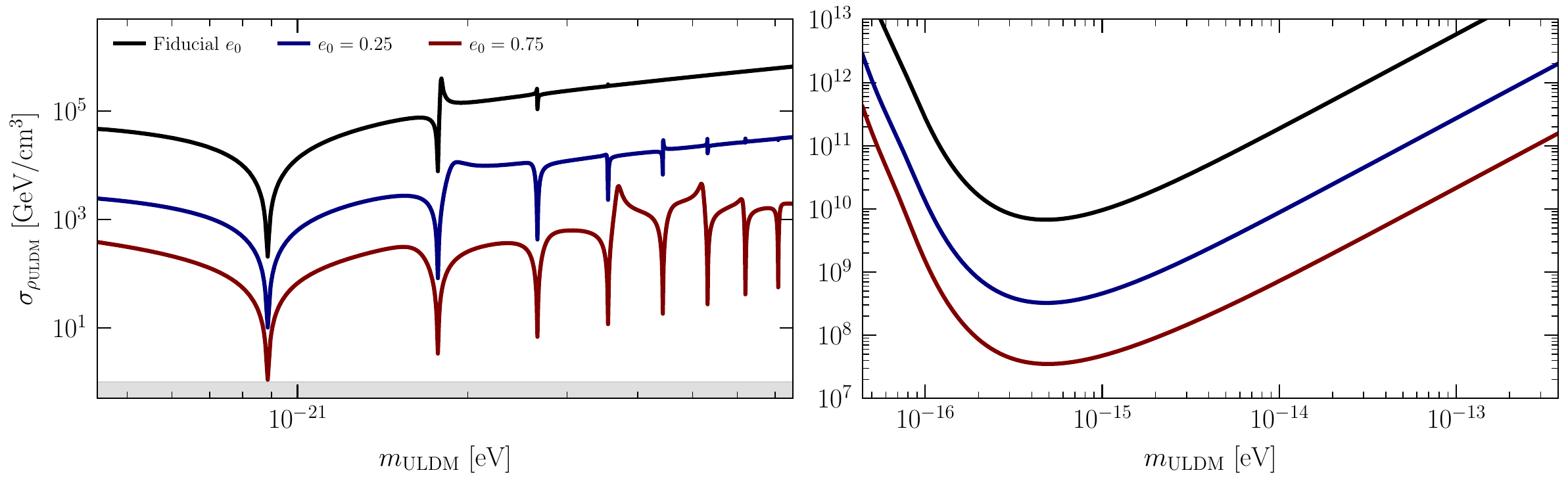}
\end{center}
\caption{As in Fig.~\ref{fig:EccentricityResonance}, but comparing the sensitivity to the local ULDM density via lunar laser ranging as a function of eccentricity. Like in Fig.~\ref{fig:ULDMSensitivity}, in the left (right) panel, we depict the projected sensitivities from searches from the coherent (incoherent) oscillations of the ULDM and induced scalar potential described by $B(\omega)$ ($A(\omega)$) in Eq.~\eqref{eq:ULDMPowerComponents}. As in the case of GW sensitivities, we find our ULDM sensitivity is highly eccentricity dependent, while larger eccentricities result in resonance features with higher harmonics and generally stronger resonant responses. The grey-shaded region shows the expected ULDM abundance in the Solar System.}
\label{fig:EccentricityResonanceULDM}
\end{figure*}

Since we have already demonstrated in Sec.~\ref{sec:SensitivityTimeDependence} that the time dependence of
the sensitivities corresponds to the one derived analytically, we do not repeat it for the ULDM case. Furthermore, for the ULDM the effect is not directional, simplifying the analysis and conclusions. Regarding the dependence on eccentricity, in Fig.~\ref{fig:EccentricityResonanceULDM} we see that more eccentric orbits are preferred for the same reasons as those for GWs. 

Overall, as in the case of GW sensitivities, our results for the coherent piece are considerably more optimistic than those that appeared in 
\cite{Blas:2016ddr,Blas:2019hxz,LopezNacir:2018epg,Kus:2024vpa,Rozner:2019gba}, based on binary pulsars. In \cite{Blas:2024PRLForward}, we develop further the prospects for LLR, SLR, and pulsar timing to detect ULDM using several systems as observational targets. In general, however, sensitivity to the expected DM energy density would only be achieved if the binary systems of interest resided within overdense dark matter environments, which may be realized in some dark matter models or for some particularly fortuitously located pulsar binaries. We also note that ULDM may have direct couplings to baryonic matter and that the effective strength of the external acceleration applied to binary systems may exceed that of the external acceleration generated by the purely gravitational interaction of the oscillating ULDM field with a binary system. The external acceleration will take a more model-dependent and less universal form than that of the one induced by the gravitational interaction, but building upon the original results of \cite{Blas:2016ddr, Blas:2019hxz} with the new formalism we present here represents an interesting direction for future study. The simplest possibility is the universal quadratic coupling, which can be studied by trivially transforming $\psi\mapsto \beta\, \psi$, which, in our formalism, maps $\rho_{\rm ULDM}\mapsto \beta\, \rho_{\rm ULDM}$. This is exploited in \cite{Blas:2024PRLForward} to project costraints on $\beta$.

%%%%%%%%%%%%%%%%%%%%%%%%%%%%%%%%%%%%%%%%%%%%%%%%%%%%%%%%%%%%%%%%%%%%%%%%%%%%%%%%%%%%
\section{Beyond isolated Keplerian binaries}
\label{sec:beyond_isolation}
%%%%%%%%%%%%%%%%%%%%%%%%%%%%%%%%%%%%%%%%%%%%%%%%%%%%%%%%%%%%%%%%%%%%%%%%%%%%%%%%%%%%

Even absent perturbations by GWs or ULDM, binary systems do not exist in isolation, and high-precision modeling of these binary systems requires incorporating both environment and post-Newtonian effects in describing their dynamical evolution. A key advantage of the computational framework we have developed in this work is that it is highly flexible in terms of incorporating arbitrary accelerations in the fully nonlinear evolution of $\tuple_0$. 

It is important to assess the role of additional accelerations which determine the dynamics of the two-body system in specifying its resonant response to signals of interest. In this Section, as illustrative examples of the Earth-Moon system, two additional effects are discussed. We will first consider the impact of the perturbation by the Sun, which, after the Earth-Moon gravitational interaction, results in the largest acceleration acting on the system. As we will show, the presence of this new acceleration does not spoil the strong response of the Earth-Moon orbit to a resonant acceleration from a SGWB. Afterward, we will consider the effect of an additional background acceleration, the one sourced by the Earth's tides, on the Earth-Moon orbit, assessing its degeneracy with the SGWB-induced acceleration of interest in this work.

%%%%%%%%%%%%%%%%%%%%%%%%%%%%%%%%%%%%%%%%%%%%%%%%%%%%%%%%%%%%%%%%%%%%%%%%%%%%%%%%%%%%
\subsection{Non-resonant perturbations from the Solar Potential}
\label{sec:nonres_strong_background}
%%%%%%%%%%%%%%%%%%%%%%%%%%%%%%%%%%%%%%%%%%%%%%%%%%%%%%%%%%%%%%%%%%%%%%%%%%%%%%%%%%%%

We begin with our binary system experiencing an environmental perturbation acceleration induced by the Sun. This tidal acceleration is given by
\begin{equation}
            \bm{a}_\odot = -\frac{GM_\odot}{\left|\bm R_\odot\right|^3} \left(\bm{r}-  3 (\bm{r} \cdot \hat{\bm{R}}_\odot) \hat{\bm{R}}_\odot\right)\, ,
\end{equation}
where $\bm{R}_\odot$ is the vector from the binary system to the perturbing third body (the Sun) and $M_\odot$ is the Sun mass. We use planetary ephemerides from \cite{fienga:2021} to evaluate $\bm{R}_\odot$ as a function of time.

We go on to characterize the impact of this tidal acceleration on GW sensitivities for the Earth-Moon system in the context of a monochromatic SGWB as in Sec.~\ref{sec:MonochromaticSensitivity}. Concretely, we follow the procedure described in the main text to evaluate the zeroth-order solution 
\begin{equation}
\begin{split}
    \dot{\tuple}^\alpha_0  = \Ft_0^{\alphat }(\tuple_0, t) 
    \equiv& \Mt^{\alphat b}(\tuple_0)\bm{e}_{bc}(\tuple_0)\bm{a}_\odot^c(\tuple_0, t) \\
    &+ \sqrt{\frac{G M}{p_0^3}}(1+e_0\cos(f_0))^2 \delta^{\alpha 6},
\end{split}
\end{equation}
which now includes the effect of the third-body perturbation through the inclusion of $\bm{a}_\odot$. Next, we calculate the distance perturbation covariance $\bm{\Sigma}(t, t')$, accounting for the effect of the three-body disruption here through its inclusion in $\Ft_0^{\alphat \betat}$. Otherwise, our sensitivity projections make identical assumptions and are developed with identical procedures. 

\begin{figure}[!t]
\begin{center}
\includegraphics[width=0.48\textwidth]{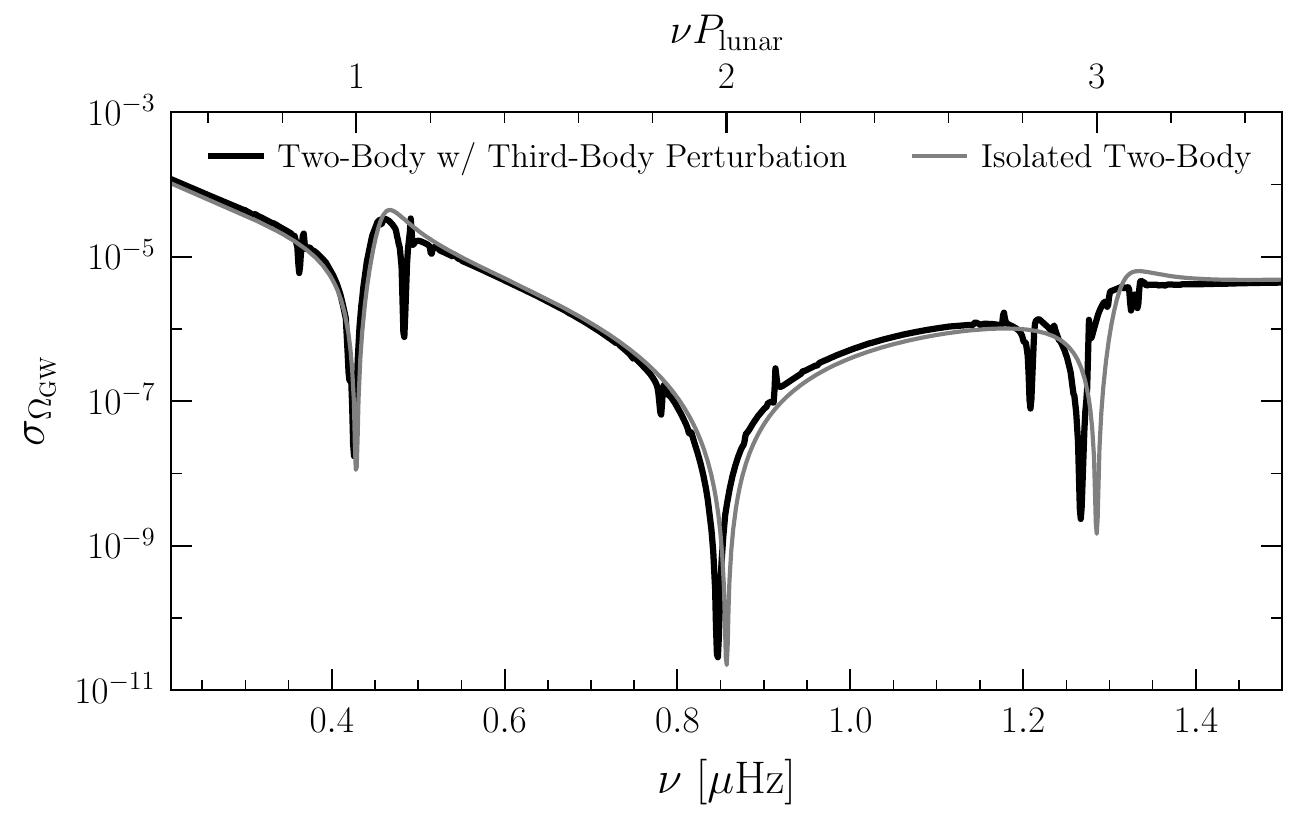}
\end{center}
\caption{The projected sensitivity of the binary response of the Earth-Moon system to a monochromatic SGWB as a function of frequency when accounting for third-body disruption by the sun (\textit{black}) as compared to when this effect is neglected (\textit{grey}). The precise resonance frequencies are somewhat shifted due to solar perturbation, which also generates weaker resonant responses in the sidebands of dominant binary resonances. The general sensitivity of the binary response, particularly on resonance, is essentially unchanged.}
\label{fig:TidalDisruption}
\end{figure}

The results of these sensitivity projections, compared with those developed absent the solar tidal acceleration, are shown in Fig.~\ref{fig:TidalDisruption}. They demonstrate that the on-resonance sensitivity is left essentially unchanged while new features are generated in the response that may provide subleading contributions to integrated sensitivities. More generally, this example shows how the formalism developed in this work can be extended to accommodate detailed modeling of effects that determine the fine-grained details of orbital motion. 

%%%%%%%%%%%%%%%%%%%%%%%%%%%%%%%%%%%%%%%%%%%%%%%%%%%%%%%%%%%%%%%%%%%%%%%%%%%%%%%%%%%%%%%%%%%%%%%%%%
\subsection{Degenerate resonant backgrounds}
\label{sec:ConfoundingBackgrounds}
%%%%%%%%%%%%%%%%%%%%%%%%%%%%%%%%%%%%%%%%%%%%%%%%%%%%%%%%%%%%%%%%%%%%%%%%%%%%%%%%%%%%%%%%%%%%%%%%%%

In Sec.~\ref{sec:nonres_strong_background}, we considered how the impact of additional background accelerations impacts the signal of interest, concluding that the resonant response of our binary systems was robust. A different question, which we begin to assess here, is the degree to which potentially confounding backgrounds can be distinguished from signals. Indeed, many parameters are fitted when performing the orbit determination for pulsars, the Moon, or for a spacecraft orbiting the Earth like \textit{e.g.},  initial conditions, some gravity potential coefficients, tidal interactions, etc. 

In this section, we consider the dissipative effect of the Earth's tides, which results in roughly a $3\,\mathrm{cm}/\mathrm{year}$ recession rate of the lunar semilatus rectum \cite{2014IPNPR.196C...1F}. We follow~\cite{2014IPNPR.196C...1F} to develop a simplified model of tidal acceleration. In particular, we consider only tidal dissipation associated with the second-degree Love numbers $k_{2i}$, and we approximate $k_{20} = k_{21} = k_{22} = k= 0.33$ \cite{petit:2010fk}. Similarly, we assume that the effects at each order are characterized by a single phase lag term $\delta = 4.02 \times 10^{-2}$ \cite{2014IPNPR.196C...1F}. Subject to these assumptions, the radial, azimuthal, and vertically oriented accelerations that act on the Earth-Moon distance are
\begin{widetext}
\begin{equation}\label{eq:tide}
\begin{gathered}
\bm{a}_{\mathrm{tide}} \cdot \hat{\bm{r}} = -3 k G M_\mathrm{moon}\left(1 + \frac{M_\mathrm{moon}}{M_\mathrm{earth}}\right) \frac{R_E^5}{r^7}, \\ 
\bm{a}_{\mathrm{tide}} \cdot \hat{\bm{\theta}} = -3 k G M_\mathrm{moon}\left(1 + \frac{M_\mathrm{moon}}{M_\mathrm{earth}}\right) \frac{R_E^5}{r^7}\left( \sin \varepsilon \sin \iota \cos\Omega - \cos\varepsilon\cos\iota\right)\delta, \\ 
\bm{a}_{\mathrm{tide}} \cdot \hat{\bm{z}} = -3 k G M_\mathrm{moon}\left(1 + \frac{M_\mathrm{moon}}{M_\mathrm{earth}}\right) \frac{R_E^5}{r^7}\left[\sin \varepsilon \cos \iota \cos\Omega - \cos\varepsilon\sin\iota\cos(\omega+f) -\sin\varepsilon\sin\Omega\sin(\omega + f)\right]\delta, \\ 
\end{gathered}
\end{equation}
\end{widetext}
with $\varepsilon \approx 0.41$ in ICRS coordinates where $\iota \approx 0.09$. Notably, these accelerations are,  \textit{e.g.}  through their dependence on the Earth-Moon distance $r$, harmonic with the Earth-Moon period. As a result, we expect them to generate a resonant response which might be degenerate with the GW signal.

\begin{figure}[!t]  
    \begin{center}
    \includegraphics[width=0.49\textwidth]{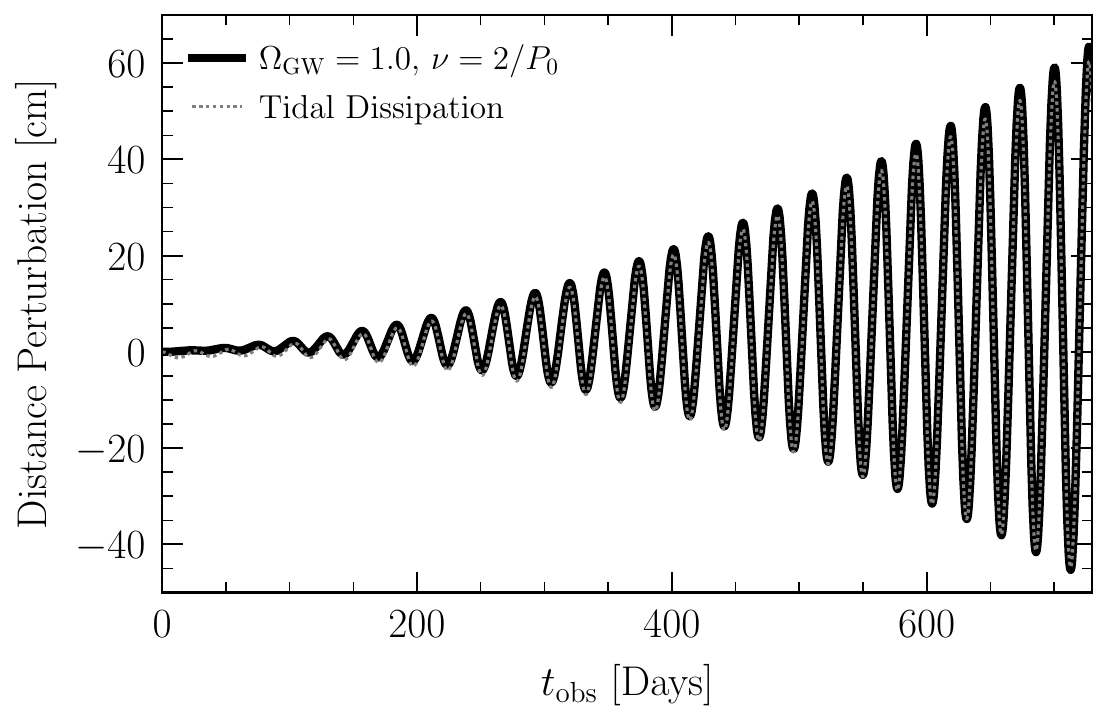}
    \caption{A comparison of the resonant effect of tidal dissipation on the perturbation to the Earth-Moon distance (dotted grey) as compared to the ``typical" effect of a monochromatic SGWB at $\nu_\mathrm{SGWB} = 2/P_0$ with $\Omega_\mathrm{GW} = 1.0$, which very nearly matches the observed lunar recession rate. See text for more details.}
    \label{fig:DegeneracyComparison}
    \end{center}
\end{figure}

As an initial assessment of the significance of this effect, we evolve over a two-year-long period the effect of the dissipative acceleration by the tides, and we compare this to the ``typical" effect of a monochromatic SGWB at $\nu_\mathrm{SGWB} = 2/P_0$ (with $P_0$ the Moon orbital period) and with $\Omega_\mathrm{GW} = 1.0$. We examine this typical effect by performing a principal component analysis of the covariance matrix for the GW-induced distance perturbations. Recall that from a degree-of-freedom counting, there are statistically 10 independent components of $\ddot h_{ij}$, see Sec.~\ref{sec:PrincipalComponents} and hence at most 10 nontrivial principal components of the covariance matrix. In fact, the first principal component is associated with an eigenvalue (variance) that is larger by a factor of nearly $10^5$ than that of the second principal component. As a result, any monochromatic GW signal will be, to good approximation, described as this first principal component with an amplitude drawn from a Gaussian distribution with variance corresponding to the associated eigenvalue of the principal component.

The comparison of these effects is shown in Fig.~\ref{fig:DegeneracyComparison}, where we plot the time-evolving distance perturbation associated with the tidal dissipation and associated with the principal component of the GW-induced perturbations with root-mean-square amplitude expected under the choice of $\Omega_\mathrm{GW} = 1.0$ in this monochromatic scenario. These perturbations are visibly highly degenerate and coincidentally agree at the percent level for $\Omega_\mathrm{GW}=1.0$. As a result, if one infers simultaneously the amplitude of the GW and the Earth love number, these two parameters will be highly correlated. For this reason, achieving sensitivity to $\Omega_\mathrm{GW} < 1$ will require some knowledge regarding the tides with some level of precision, either using other observations that are not sensitive to the GW searched for in the LLR dataset, or using some theoretical Earth interior modeling that is independent of a possible GW signal. 
We additionally note that including the subdominant principal components of the SGWB at $\nu=2/P_0$, as well as signal contributions from other frequencies, may also help to break this degeneracy.

The $k$ Love number for a body is always between 0 (rigid body) and approximately 19 (homogeneous perfect fluid). Total ignorance on the Earth Love number, therefore, sets a sensitivity floor at $\Omega_\mathrm{GW} \approx 19$ \footnote{There is a quadratic relationship between $\Omega_\mathrm{GW}$ and $k$.}, while modeling the tidal dissipation at $0.1\%$($0.01\%$) precision, will lower this sensitivity floor to the level of $\Omega_\mathrm{GW} = 10^{-6}$($10^{-8}$) \cite{2014IPNPR.196C...1F}. We leave a complete study of these effects and current measurement/modeling capabilities to future work, but given this initial investigation, it is likely that high-precision measurements of the Earth's Love numbers which are independent of lunar laser ranging are of the most critical complementary measurements necessary to reach our projected sensitivities.

%%%%%%%%%%%%%%%%%%%%%%%%%%%%%%%%%%%%%%%%%%%%%%%%%%%%%%%%%%%%%%%%%%%%%%%%%%%%%%%%%%%%%%%%%%%%%%%%%%
\subsection{Parameter degeneracies in LLR data}
\label{sec:LLR}
%%%%%%%%%%%%%%%%%%%%%%%%%%%%%%%%%%%%%%%%%%%%%%%%%%%%%%%%%%%%%%%%%%%%%%%%%%%%%%%%%%%%%%%%%%%%%%%%%%
In Sec.~\ref{sec:ConfoundingBackgrounds} we considered a single nuisance parameter (the Earth's love number) that is particularly likely to give rise to degeneracies when searching for a SGWB signal in LLR data.
However, realistic LLR searches will include $\mathcal{O}(100)$ nuisance parameters, whose degeneracy (or lack thereof) with a given GW signal template must be characterized using state-of-the-art models of the Lunar orbit and of the ranging observations.
This question also depends sensitively on the GW signal morphology, so that a search for, e.g., an individual compact binary signal may suffer from completely different parameter degeneracies compared to the SGWB case considered above.

\begin{figure}[!b]  
    \begin{center}
    \includegraphics{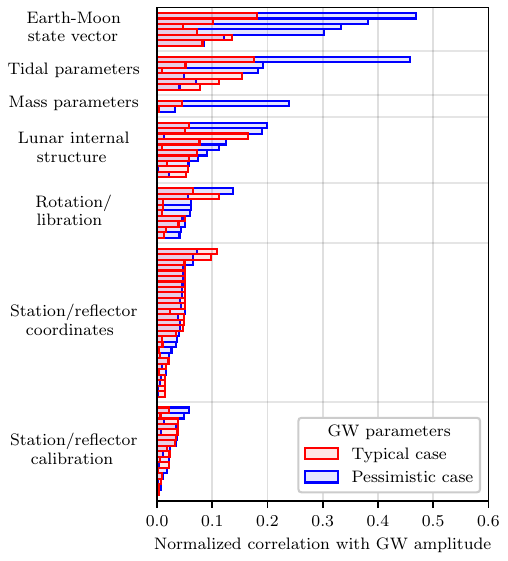}
    \caption{Estimated correlation between the GW amplitude and other parameters fitted in LLR data analysis for two choices of GW parameters (sky location, polarization, etc.): one randomly selected case, and one in which we have attempted to maximize the correlation. These coefficients have been obtained using a Fisher analysis considering a realistic full-force model and the real distribution and uncertainties of the existing LLR data.}
    \label{fig:param_correlation}
    \end{center}
\end{figure}

This is illustrated in Fig.~\ref{fig:param_correlation}, where we show the correlation between the amplitude of a deterministic and monochromatic GW point source (such as that emitted by a slowly-inspiralling compact binary) with 76 parameters used in a realistic LLR orbit determination model~\cite{Bourgoin:2016ynf, bourgoin:2017aa}.
We compute these correlations using a Fisher forecast, where the covariance matrix $\Sigma_{ij}$ for the 77 parameters (76 LLR parameters $+$ 1 GW amplitude) is obtained by propagating the derivatives of the ranging data with respect to each parameter.
The normalized absolute correlation between parameters $i$ and $j$ is then given by
\begin{equation}
    \mathrm{Corr}(i,j)=\frac{|\Sigma_{ij}|}{\sqrt{\Sigma_{ii}\Sigma_{jj}}},
\end{equation}
such that $\mathrm{Corr}(i,j)=1$ represents perfect degeneracy and $\mathrm{Corr}(i,j)=0$ represents zero degeneracy.

We consider two choices of the GW signal parameters (sky location, polarization, etc.), one in which these are chosen randomly, and one in which we attempt to maximize the degeneracy with the LLR parameters.
As shown in Fig.~\ref{fig:param_correlation}, the GW amplitude generally decorrelates relatively well from most of the other fitted parameters.
However, this statement depends strongly on the GW signal parameters, and there are some situations where the correlation can be as large as $\sim60\%$---these cases would correspond to approximate ``blind spots'' in a full-sky search for GW point sources.
The main source of degeneracy appears to be the masses and initial orbital parameters of the Earth and Moon, as well as the tidal parameters.

Further investigation is required in order to build a comprehensive understanding of these issues.
However, our preliminary findings are an encouraging indication that (some) GW signals can indeed be decorrelated from the large number of nuisance parameters involved in a realistic analysis of LLR data.

%%%%%%%%%%%%%%%%%%%%%%%%%%%%%%%%%%%%%%%%%%%%%%%%%%%%%%%%%%%%%%%%%%%%%%%%%%%%%%%%%%%%
\section{Conclusion}
\label{sec:Conclusion}
%%%%%%%%%%%%%%%%%%%%%%%%%%%%%%%%%%%%%%%%%%%%%%%%%%%%%%%%%%%%%%%%%%%%%%%%%%%%%%%%%%%%
Though GWs have thus far only been observed in a region of frequencies around 100~Hz, with an additional tentative detection in the nHz band, their detection represents one of the most transformative results of modern physics. The potential of GWs to be used as a probe of the fundamental physics of our Universe is only just now being realized, and the possibility of extending GW sensitivities to as-of-yet unprobed frequency regions is highly compelling. In particular, the $\mu$Hz range has been largely overlooked due to intrinsic technical difficulties, even though this band has great potential for astrophysics and fundamental physics. 

In this work, we have considerably strengthened the case for using \textit{resonant binary systems,} such as Earth-satellite systems or binary pulsars, to detect $\mu$Hz GWs~\cite{Blas:2021mqw,Blas:2021mpc}. Critically, we have developed a detailed modeling framework for describing the resonant (or non-resonant) transfer of energy between GWs and the binary system. These results confirm and extend early efforts in~\cite{Blas:2021mqw,Blas:2021mpc} which first demonstrated that resonant GWs would drive secular growth of the orbital elements. Critically, however, the calculations presented here go beyond time-averaged secular approximations for the evolution of orbital elements to make time-resolved predictions for timing residuals of pulsar timing and laser ranging measurements.

A key result of these calculations is that observational sensitivities to GWs through binary resonances are parametrically stronger than previously appreciated. The origin of this enhanced sensitivity is the rapid dephasing that GWs (or other sources of anomalous acceleration) drive in the perturbed true anomaly of the orbit. As we demonstrate, a resonant (non-resonant) acceleration will drive growth in the true anomaly perturbation that is then imparted on the laser ranging or pulsar timing observable that grows as $t_\mathrm{obs}^2$ ($t_\mathrm{obs}^1$) while all other perturbations grow only as  $t_\mathrm{obs}^1$ ($t_\mathrm{obs}^0$). In the secular approximation, the perturbations to the true anomaly are unresolved (they are not traced when one averages over orbits), and so prior calculations neglected the most rapidly growing perturbations which lead in turn to the greatest observational sensitivities. We demonstrate these scaling behaviors in detail using a Floquet-theoretic approach in Sec.~\ref{sec:AnalyticResults}, while in App.~\ref{app:RecoveringSecular} we demonstrate in detail a comparison between our methodology and the more limited secular approximation approach.

To put our findings in context, we have applied them to estimate possible sensitivities of current and future searches. While a dedicated discussion of our observational forecasts will appear elsewhere \cite{Blas:2024PRLForward}, here we have used the orbital parameters of the Earth-Moon system and the binary pulsar J1829+2456 to determine some properties of the expected sensitivity for laser ranging and pulsar timing data, respectively. We confirm the scaling laws derived analytically and also the key role played by eccentricity in the final sensitivity. The existence of a sky pattern for GWs, clarified in Sec.~\ref{sec:sky}, hints towards the existence of new possibilities to determine if an effect measured in enough binaries is indeed generated by GWs via a cross-correlation analysis of multiple satellites. We present the necessary formalism for such a computation in App.~\ref{app:NetworkCorrelations} and plan to study its prospects at greater length as part of a program to design optimal artificial satellite laser ranging schemes. Finally, we have also made an initial study of background effects that may bias or weaken our sensitivities. These motivate complementary independent measurements, such as studies of the Earth's tidal deformability, that will be necessary to reach the measurement-precision-limited sensitivities projected here and in \cite{Blas:2024PRLForward}.

In parallel, we have also extended the analysis of the impact of ultra-light dark matter (ULDM) in the evolution of binary systems from \cite{Hui:2012yp,LopezNacir:2018epg,Rozner:2019gba,Blas:2019hxz,Armaleo:2020yml,Kus:2024vpa} to include the new insights we attained in the GW analysis, and also extend them to the treatment of the stochastic effect \cite{Bar-Or:2018pxz,Kim:2023pkx} of the virialized distribution relevant in the Milky Way. Our conclusion parallels the one for GWs: our treatment opens up the parameter space of ULDM models that may impact resonant binary systems.

In summary, our findings are highly promising: precision timing or ranging of binary system offers a unique opportunity to cover the gap of searches of GWs between the nHz and the mHz and ULDM of masses in the range $m_{\rm DM}\in\{10^{-21},10^{-18})\,{\rm eV}$. The most natural direction forward is to combine our computational framework with existing methods for the modeling and analysis of laser-ranging data and pulsar timing. This will allow for even more realistic sensitivity projections, optimized observational strategies, and, eventually, real data analyses. Given the prospects from the present paper and its companion article~\cite{Blas:2024PRLForward}, there is a realistic possibility that our methods will open a new band to detect GWs and/or ULDM in the near future.

%%%%%%%%%%%%%%%%%%%%%%%%%%%%%%%%%%%%%%%%%%%%%%%%%%%%%%%%%%%%%%
\section*{Acknowledgments}
%%%%%%%%%%%%%%%%%%%%%%%%%%%%%%%%%%%%%%%%%%%%%%%%%%%%%%%%%%%%%%

{\it We thank D.~Hackett, Y.~Kahn, N.~Rodd, B.~Safdi, T.~Trickle, and J.~Urban for helpful discussions. This research used resources from the Lawrencium computational cluster provided by the IT Division at the Lawrence Berkeley National Laboratory, supported by the Director, Office of Science, and Office of Basic Energy Sciences, of the U.S. Department of Energy under Contract No. DE-AC02-05CH11231.
Part of this work was carried out at the Munich Institute for Astro-, Particle and BioPhysics (MIAPbP), which is funded by the Deutsche Forschungsgemeinschaft (DFG, German Research Foundation) under Germany's Excellence Strategy – EXC-2094 – 390783311. This manuscript has been authored in part by Fermi Forward Discovery Group, LLC under Contract No. 89243024CSC000002 with the U.S. Department of Energy, Office of Science, Office of High Energy Physics.

D.B. acknowledges the support from the Departament de Recerca i Universitats from Generalitat de Catalunya to the Grup de Recerca 00649 (Codi: 2021 SGR 00649).
The research leading to these results has received funding from the Spanish Ministry of Science and Innovation (PID2020-115845GB-I00/AEI/10.13039/501100011033). This publication is part of the grant PID2023-146686NB-C31 funded by MICIU/AEI/10.13039/501100011033/ and by FEDER, UE.
IFAE is partially funded by the CERCA program of the Generalitat de Catalunya. Project supported by a 2024 Leonardo Grant for Scientific Research and Cultural Creation from the BBVA Foundation. The BBVA Foundation accepts no responsibility for the opinions, statements and contents included in the project and/or the results thereof, which are entirely the responsibility of the authors.   This work is supported by ERC grant ERC-2024-SYG 101167211. Funded by the European Union.  Views and opinions expressed are however those of the author(s) only and do not necessarily reflect those of the European Union or the European Research Council Executive Agency. Neither the European Union nor the granting authority can be held responsible for them.  D.B. acknowledges the support from the European Research Area (ERA)
via the UNDARK project of the Widening participation
and spreading excellence programme (project number 101159929). 

The work of M. H-V. has been partially supported by the Spanish State Research Agency MCIN/AEI/10.13039/501100011033 and the EU NextGenerationEU/PRTR funds, under grant IJC2020-045126-I.

A.C.J. was supported by the Gavin Boyle Fellowship at the Kavli Institute for Cosmology, Cambridge, and by the Science and Technology Facilities Council through the UKRI Quantum Technologies for Fundamental Physics Programme (Grant No. ST/T005904/1).

X.X. is funded by the grant CNS2023-143767. 
Grant CNS2023-143767 funded by MICIU/AEI/10.13039/501100011033 and by European Union NextGenerationEU/PRTR.
}

\appendix

%%%%%%%%%%%%%%%%%%%%%%%%%%%%%%%%%%%%%%%%%%%%%%%%%%%%%%%%%%%%%%%%%%%%%%%%%%%%%%%%%%%%%%%%%%%%%%%%%%
\section{Conventions for GWs}
\label{app:GWs}
%%%%%%%%%%%%%%%%%%%%%%%%%%%%%%%%%%%%%%%%%%%%%%%%%%%%%%%%%%%%%%%%%%%%%%%%%%%%%%%%%%%%%%%%%%%%%%%%%%

For the acceleration in Eq.~\eqref{eq:hacc}, we need to work in the TT gauge. We will follow closely the conventions of \cite{Maggiore:1900zz}. We first define the standard GW polarization tensors for a wave propagating in the $\hat{\bm{n}}$ direction given by
\begin{equation}
\bm{e}_{ij}^+(\hat{\bm{n}}) = \hat{\bm{u}}_i \hat{\bm u}_j - \hat{\bm v}_i \hat{\bm v}_j, \qquad \bm{e}_{ij}^\times = \hat{\bm u}_i \hat{\bm v}_j + \hat{\bm v}_i \hat{\bm u}_j.
\end{equation}
Working in spherical coordinates, we take 
\begin{equation}\begin{split}
    \bm{\hat{n}} &= [\sin\theta\cos\phi, \sin\theta\sin\phi, \cos\theta], \\
    \bm{\hat{u}} &= [\cos\theta\cos\phi, \cos\theta\sin\phi, -\sin\theta],\\
    \bm{\hat{v}} &= [-\sin\phi, \cos\phi, 0].
\end{split}\end{equation}
Defining $h_A(t, \hat{\bm{n}})$ to be the differential GW strain propagating in the $\hat{\bm{n}}$ direction, we can write the total GW strain as
\begin{equation}
    h_{ij}(t) = \int \mathrm{d}^2 \hat{\bm{n}} \, \bm{e}_{ij}^A(\bm{\hat{n}}) h_A(t, \hat{\bm{n}}).
\end{equation}
We can also Fourier transform these components to go to the plane-wave basis so that we have
\begin{equation}
    h_{ij}(t) =  \int \mathrm{d}^2 \hat{\bm{n}} \, \bm{e}_{ij}^A(\bm{\hat{n}}) \int \mathrm{d}\nu\, e^{2\pi i \nu t} \tilde{h}_A(\nu, \hat{\bm{n}}).
    \label{eq:FourierTransform}
\end{equation}
This also implicitly defines our Fourier-transform convention. 

%%%%%%%%%%%%%%%%%%%%%%%%%%%%%%%%%%%%%%%%%%%%%%%%%%%%%%%%%%%%%%%%%%%%%%%%%%%%%%%%%%%%%%%%%%%%%%%%%%
\section{Validity of the perturbative approach}
\label{app:PerturbativeDemonstration}
%%%%%%%%%%%%%%%%%%%%%%%%%%%%%%%%%%%%%%%%%%%%%%%%%%%%%%%%%%%%%%%%%%%%%%%%%%%%%%%%%%%%%%%%%%%%%%%%%%

\begin{figure}[!htb]  
    \begin{center}
    \includegraphics[width=0.49\textwidth]{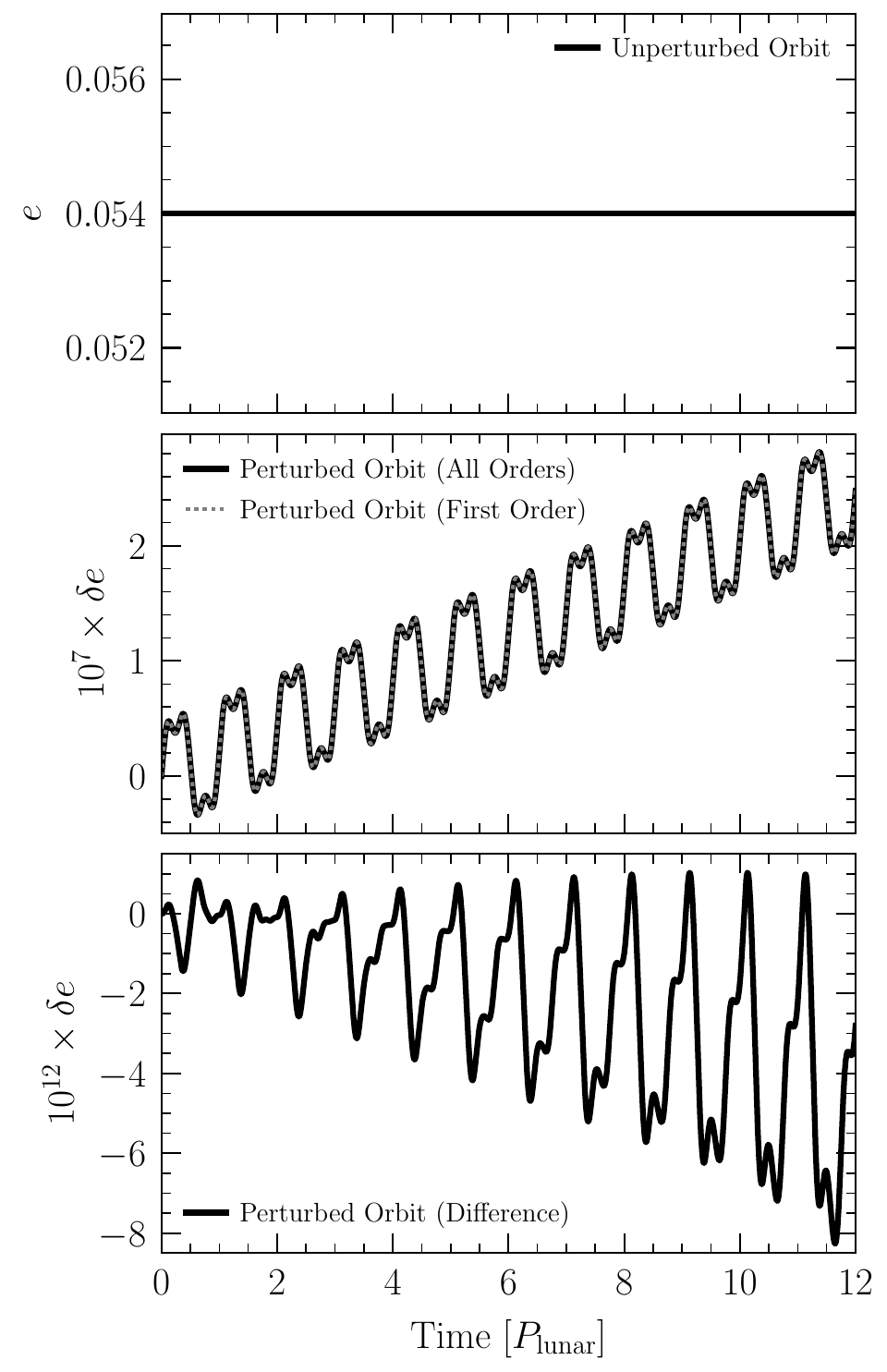}
    \caption{An illustration of the time-evolution of the eccentricity and its perturbations for the scenario discussed in Sec.~\ref{app:PerturbativeDemonstration}. (\textit{Top panel}) The time-evolution of the eccentricity in the unperturbed orbit scenario. Here, the eccentricity is constant. (\textit{Middle panel}) In black, we depict the perturbations to the otherwise constant eccentricity calculated by including the perturbing acceleration in the fully nonlinear dynamics of the binary system. In dotted grey, we show the same perturbations, but calculated at only first-order using our perturbative formalism. (\textit{Bottom panel}) The small difference between the all-orders perturbation and the first-order perturbation to $e$. See text for more details.}
    \label{fig:PerturbativeDemonstration}
    \end{center}
\end{figure}

The calculations in this work depend heavily on a perturbative approach in which the effects of perturbing accelerations are considered at first-order but no higher. As a numerical example demonstrating that, at least in the case of the Earth-Moon system, this is satisfactory, we consider an acceleration of the form 
\begin{equation*}
    \bm{a}_\mathrm{sig}(t) = \cos(2 \pi \nu t) \hat{\bm{\theta}}(\tuple_0(t)) \, \, \mathrm{m}\, \mathrm{day}^{-2}\,,
\end{equation*}
with $\nu = 2/P_0$. We point out that this acceleration is $10^6$ larger but otherwise identical to that considered in Sec.~\ref{app:RecoveringSecular}. We then perform several calculations of the orbital dynamics, illustrating the results for the evolution of $e$ in Fig.~\ref{fig:PerturbativeDemonstration}. In the top panel, we evaluate the zeroth-order orbit absent the perturbing acceleration; in this scenario, the eccentricity is constant. We then evaluate the perturbed dynamics in two ways. First, we solve the orbital dynamics with a brute-force approach by including $\bm{a}_\mathrm{sig}(t)$ in the ``zeroth-order" evolution of $\tuple_0$, then defining the perturbation as the difference between the solution $\tuple_0$ with and without the inclusion of this perturbing acceleration. This approach captures the effect of the perturbing acceleration at all orders. We also compute the perturbation to the solution restricted only to first-order effects using Eq.~\eqref{eq:solgen}. Finally, in the bottom panel, we illustrate the difference between the perturbations calculated at first order as compared to those calculated at all orders. 

From these examples, we can see that the first-order perturbation is parametrically smaller than the zeroth-order value $e_0 = 0.054$, and the contribution of higher-order terms is parametrically smaller than the first-order perturbations. We see then that for $\mathcal{O}(\mathrm{cm}/\mathrm{day}^2)$ accelerations the perturbative approach more than suffices, while even smaller accelerations may well within detectability. 

%%%%%%%%%%%%%%%%%%%%%%%%%%%%%%%%%%%%%%%%%%%%%%%%%%%%%%%%%%%%%%%%%%%%%%
\section{Fourier decompositions of functions of orbital elements}
%%%%%%%%%%%%%%%%%%%%%%%%%%%%%%%%%%%%%%%%%%%%%%%%%%%%%%%%%%%%%%%%%%%%%%
\label{app:Fourier}

Here we discuss the Fourier decomposition of periodic functions of orbital elements as relevant for the calculations in Sec.~\ref{sec:AnalyticResults}. Consider a function $g$ which satisfies $g(t+P_0) = g(t)$ such that it admits a Fourier decomposition as $g(t) = \sum_n g_n \exp(2 \pi i n t/P_0)$. The coefficients $g_n$ may be evaluated as
\begin{equation}
g_n = \frac{1}{P_0}\int_0^{P_0} \mathrm{d}t \, g(t)e^{- 2\pi i n t/P_0}.
\end{equation}
To proceed, we first change variable time to the zeroth-order eccentric anomaly, which satisfies
\begin{equation}
    E_0 - e \sin E_0 = \frac{2\pi}{P_0}t - M_0(0)
\end{equation}
where $M_0(0)$ is the zeroth-order mean anomaly at time $t = 0$. We then have
\begin{equation}
\begin{split}
g_n = \frac{e^{i n M_0(0)}}{2\pi}\int_{E_0(0)}^{E_0(P_0)} &\mathrm{d}E_0 (1 - e \cos E_0) g(t) \\
&\times e^{-i n (E_0-e_0\sin E_0)},
\end{split}
\end{equation}
where $M_0(0)$ is the zeroth-order mean anomaly. From the Jacobi-Anger identity,
\begin{equation}
\begin{split}
g_n = \frac{e^{i n M_0(0)}}{2\pi}&  \int_{E_0(0)}^{E_0(0)+2\pi} \mathrm{d}E_0 (1 - e \cos E_0)  \  g(t) \\
&\times \sum_{m=-\infty}^{\infty} J_m(ne_0) e^{i(m-n)E_0},
\end{split}
\end{equation}
where $J_m$ is a Bessel function of the first kind. We now proceed to consider the specific $g_{p\theta}$,  $g_{e\theta}$, and $g_{r\theta}$ defined in Sec.~\ref{sec:AnalyticResults}. Each of these functions is only a function of time through $f_0$, which may in turn be related to $E_0$ by 
\begin{equation}
\begin{gathered}
\cos f_0 = \frac{\cos E_0 -e}{1- e_0 \cos E_0},\\
\sin f_0 = \frac{\sqrt{1-e^2}\sin E_0}{1 - e_0 \cos E_0}.
\end{gathered}
\end{equation}

First, let us consider $g_{p\theta}$. For its Fourier decomposition, we find
\begin{equation}
\begin{split}
g_{p\theta, n} = \frac{e^{i n M_0(0)}}{2(1-e_0^2)} \bigg(&\frac{e_0^2}{2} [J_{n-2}(n e_0) + J_{n+2}(n e_0)] \\ 
&- 2e_0 [J_{n-1}(ne_0) + J_{n+1}(n e_0)] \\
&+ (2 +e_0^2) J_n(ne_0) \bigg).
\end{split}
\end{equation}
Next, we consider $g_{er}$, for which we have
\begin{equation}
g_{er, n} = \frac{i \sqrt{1-e^2}e^{i n M_0(0)}}{2} \left[J_{n+1}(n e_0) - J_{n+1}(n e_0)\right].
\end{equation}
Last, we consider $g_{e\theta}$. We have 
\begin{equation}
\begin{split}
g_{e\theta, n} = \frac{e^{i n M_0(0)}}{2} \bigg(&\frac{e_0}{2} [J_{n-2}(n e_0) + J_{n+2}(n e_0)] \\ 
& +2 [J_{n-1}(ne_0) + J_{n+1}(n e_0)] .\\
&-3e_0 J_n(ne_0) \bigg).
\end{split}
\end{equation}
Hence, we find that each of these functions generically has nonzero support at each frequency harmonic with the orbital frequency $1/P_0$. 

However, note that as $e_0 \rightarrow 0$, only $J_0(n e_0)$ will survive. Hence we find, consistent with \cite{Blas:2021mpc, Blas:2021mqw}, that circular orbits will only go on resonance with the fundamental frequency and the first two harmonics. Moreover, as $e_0 < 1$, as we consider only bound orbits, the asymptotic behavior of $J_n (n e_0)$ as $n \rightarrow \infty$ will be exponentially suppressed, and hence perturbations for higher harmonics beyond the turning point of $J_n( n e_0)$ will not make important contributions. However, when this occurs will depend strongly on $e_0$.

%%%%%%%%%%%%%%%%%%%%%%%%%%%%%%%%%%%%%%%%%%%%%%%%%%%%%%%%%%%%%%%
\section{Fisher information for multivariate Gaussians}
\label{app:FisherInformation}
%%%%%%%%%%%%%%%%%%%%%%%%%%%%%%%%%%%%%%%%%%%%%%%%%%%%%%%%%%%%%%%

We consider data that are described by a multivariate Gaussian for a set of parameters $\bm{\theta}$,
\begin{equation}
\begin{split}
   &\mathcal{L}(\bm{d} | \bm \theta) = \frac{\exp \left[ -\frac{1}{2}\left(\bm{d}-\bm{\mu}\right)^T \bm{\Sigma}^{-1}\left(\bm{d}-\bm{\mu}\right)\right]}{\left| 2\pi \bm \Sigma \right|^{-1/2}},
\end{split}
\label{eq:FullLikelihood}
\end{equation}
where $\bm{\mu}$ and $\bm{\Sigma}$ are the hypothesized mean and covariance parametrized by $\bm{\theta}$, and $\bm{d}$ are the data which are drawn from some distribution with true mean $\bm{\mu}^t$ and true variance $\bm{\Sigma}^t$ resulting from true parameters $\bm{\theta}^t$. To project sensitivities in the asymptotic limit to arbitrary parameters $\bm{\theta}$, we will evaluate the Fisher information matrix, given by
\begin{equation}
\bm{I}_{ij} = \left\langle \left(\frac{\partial}{\partial \bm{\theta}_i} \log\mathcal{L} \right)\left(\frac{\partial}{\partial \bm{\theta}_j} \log\mathcal{L} \right) \right\rangle.
\end{equation}
While this calculation is cumbersome, the ultimate result is rather simple, with
\begin{equation}
\begin{split}
\bm{I}_{ij}(\bm\theta) &= \frac{\partial \bm \mu^T}{\partial \bm{\theta}_i}\bm{\Sigma}^{-1} \frac{\partial \bm \mu}{\partial \bm{\theta}_j} \\
&+ \frac{1}{2} \mathrm{Tr}\left[\bm\Sigma^{-1} \frac{\partial \bm{\Sigma}}{\partial \bm{\theta}_i} \bm\Sigma^{-1} \frac{\partial \bm{\Sigma}}{\partial \bm{\theta}_j} \right].
\end{split}
\label{eq:GaussianFisher}
\end{equation}
The asymptotic sensitivities to parameters $\bm{\theta}_i$ at the Cramer-Rao limit are then given by 
\begin{equation}
    \sigma_{\bm{\theta}_i}^2 =\left[  \bm{I}^{-1}( \bm{\theta}^t) \right]_{ii},
\end{equation}
again under assumption of the true underlying parameters $\bm{\theta}^t$.

%%%%%%%%%%%%%%%%%%%%%%%%%%%%%%%%%%%%%%%%%%%%%%%%%%%%%%%%%%%%%%%%%%%%%%%%%%%%%%%%%%%%%%%%%%%%%%%%%%
\section{Recovering secularly averaged sensitivities}
\label{app:RecoveringSecular}
%%%%%%%%%%%%%%%%%%%%%%%%%%%%%%%%%%%%%%%%%%%%%%%%%%%%%%%%%%%%%%%%%%%%%%%%%%%%%%%%%%%%%%%%%%%%%%%%%%

The calculations we have performed in this work go considerably beyond those considered in \cite{Blas:2021mpc, Blas:2021mqw,Blas:2016ddr,Blas:2019hxz} in terms of detail and predictive power. Nonetheless, it is instructive to consider how one might recover the results of those earlier works in a simplified limit of our calculations. 

\begin{figure}[!htb]  
    \begin{center}
    \includegraphics[width=0.49\textwidth]{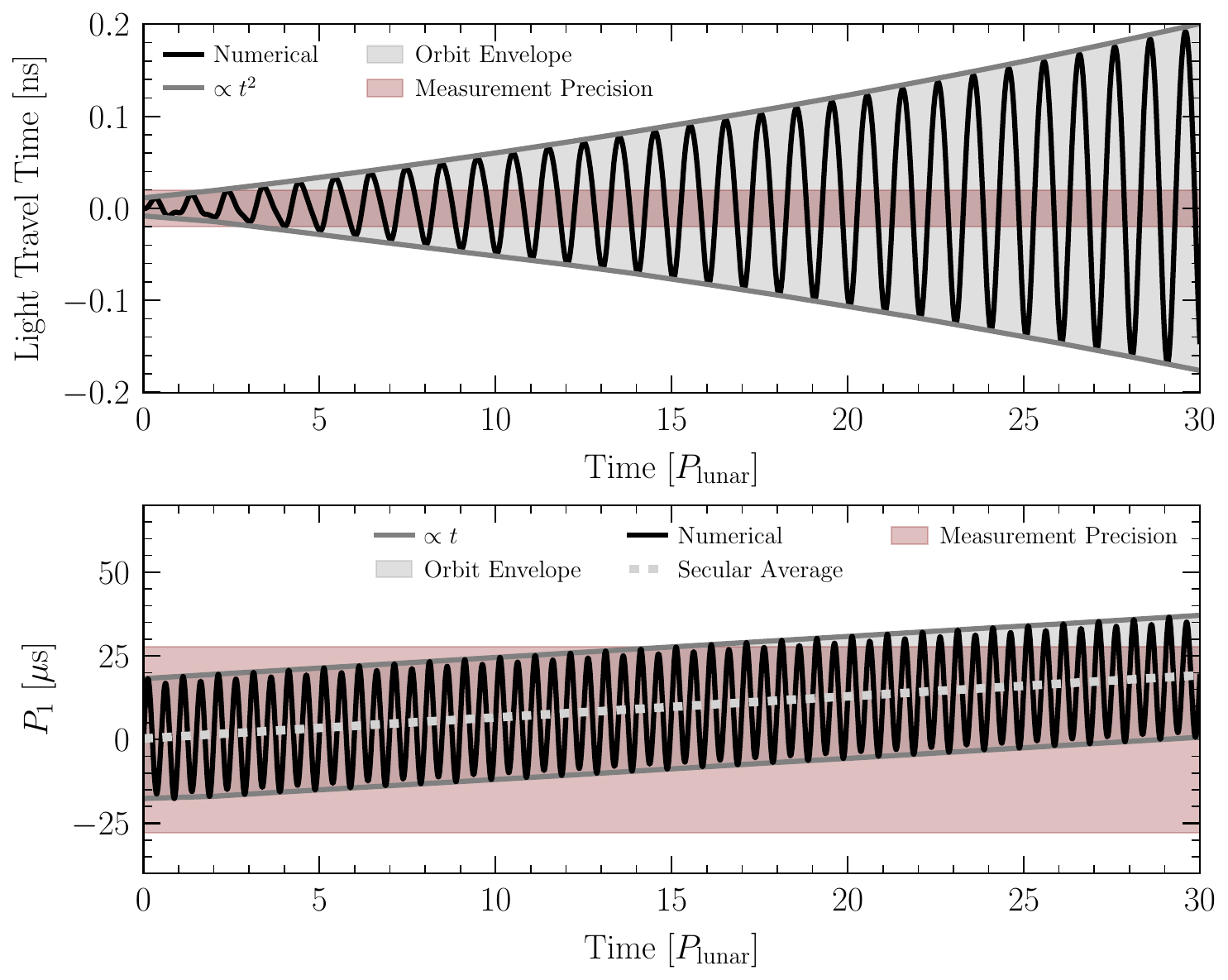}
    \caption{(\textit{Above}) In black, we depict the time evolution of the light-travel-time perturbation for the Earth-Moon system when acted upon by an acceleration in the azimuthal direction of the orbit with a magnitude of $1\,\mu\mathrm{m}/\mathrm{day}^2$. Grey lines depict the envelope of the growth of the perturbation in time, which scales as $t^2$, while the grey band illustrates the region within which the perturbation oscillates over a single orbit. The red band illustrates the precision of a single light-travel-time measurement, assumed to correspond to a distance perturbation precision of $3\,\mathrm{mm}$. (\textit{Below}) As in the top panel, but inspecting the time-evolution of the perturbation to the binary period. A less rapid linear growth of the perturbation is realized, with lesser relative sensitivity via the measurement precision of $3\,\mathrm{mm}$ related to a period sensitivity via the relation in Eq.~\eqref{eq:SensitivityAnsatz} made in \cite{Blas:2021mpc, Blas:2021mqw}. Moreover, the analysis of \cite{Blas:2021mpc, Blas:2021mqw} made use only of the secular evolution of the period perturbation, which we illustrate on our plot by downsampling averaging the numerical result in black over each period. Further sensitivity will be lost in this approach due to the loss of resolution of the oscillatory features.}
    \label{fig:SecularComparison}
    \end{center}
\end{figure}

In those works, sensitivity was estimated based on perturbations to the orbital period $P$. In terms of the semilatus rectum and the true eccentricity, the period is given by 
\begin{equation}
    P = \sqrt{\frac{4 \pi^2}{GM} \left(\frac{p}{1-e^2} \right)^{3}}.
\end{equation}
Hence, the first-order perturbation to the period is given by
\begin{equation}
P_1 = P_0 \left(\frac{3}{2p_0}p_1 + \frac{3 e_0}{1-e_0^2} e_1 \right)
\end{equation}
From our results of Sec.~\ref{sec:AnalyticResults}, it hence follows that, unlike the light-travel-time or pulse time-of-arrival, the perturbation to the period $P_1$ will grow at most linearly in $t$ when on-resonance as it does not depend on the true anomaly perturbation $f_1$. 

We note that the orbital period is not a directly observable quantity. By comparison, the sensitivity calculation we have performed in this work has been careful to construct directly observable quantities from the predicted perturbed orbits. Nonetheless, we proceed to compare them by adopting the ansatz of period perturbation sensitivities made in \cite{Blas:2021mpc, Blas:2021mqw}, in which an Earth-Moon distance measurement with precision $\sigma$ could be related to a measurement of the period by
\begin{equation}
\sigma_P = \frac{3 (1 - e_0^2) P_0}{2 p_0} \sigma\,.
\label{eq:SensitivityAnsatz}
\end{equation}
Since, by assumption, both the light-travel-time and the period perturbation sensitivity are constant in time, we see that sensitivity calculations correctly making use of the light-travel-time will achieve more rapid growth than those that incorrectly use the period perturbation. For instance, for a deterministic GW, sensitivity via the light-travel-time scales as $t^{5/2}$ while it scales as $t^{3/2}$ from the period perturbation.

To illustrate this difference, we refer back to the isolated Earth-Moon binary, which we perturb by an acceleration of the form
\begin{equation*}
    \bm{a}_\mathrm{sig}(t) = \cos(2 \pi \nu t) \hat{\bm{\theta}}(\tuple_0(t)) \, \, \mu\mathrm{m}\, \mathrm{day}^{-2}\,.
\end{equation*}
with $\nu = 2 / P_0$. We then compute the perturbed orbital elements and map them into the first-order perturbations to the light-travel time, relevant for the analysis performed in this work, and to the orbital period, relevant for the analysis performed in \cite{Blas:2021mpc, Blas:2021mqw}. The time evolution of those perturbations for the first 30 orbits starting from an unperturbed initial condition is shown in Fig.~\ref{fig:SecularComparison}. We clearly see the quadratic growth of the light-travel-time as compared to the linear growth of the period perturbation.

To directly compare to the results of \cite{Blas:2021mpc, Blas:2021mqw}, we make identical assumptions of measurement precision in the light-travel time corresponding to distance perturbation measurement of $\sigma = 3\,\mathrm{mm}$, which we relate to a period perturbation precision via Eq.~\eqref{eq:SensitivityAnsatz}. We then illustrate single-measurement precision to each perturbation with a light red band in Fig.~\ref{fig:SecularComparison}. The properly calculated sensitivity to the light-travel-time perturbation is considerably greater than the incorrectly calculated sensitivity to the period perturbation, with the relative difference growing in time due to their different scaling with time in their evolution.

%%%%%%%%%%%%%%%%%%%%%%%%%%%%%%%%%%%%%%%%%%%%%%%%%%%%%%%%%%%%%%%%%%%%%%%%%%%%%%%%%%%%%%%%%%%%%%%%%%
\section{Scalar fluctuations in the coherent limit}
\label{app:MonochromaticCoherence}
%%%%%%%%%%%%%%%%%%%%%%%%%%%%%%%%%%%%%%%%%%%%%%%%%%%%%%%%%%%%%%%%%%%%%%%%%%%%%%%%%%%%%%%%%%%%%%%%%%

Under the assumption of a Gaussian field description of the ULDM, Eq.~\eqref{eq:PotentialCorrelator} fully specifies the statistics of the fluctuating potential $\psi$. Still, over time intervals shorter than a coherence time, the statistics of the fluctuating potential $\ddot \psi$ admit a simpler description. Regarding the statistics of $\ddot \psi$, being a non-linear variable in a Gaussian field, they will be non-Gaussian. For the purposes of developing projected sensitivities, we neglect the intrinsic non-gaussianity and describe $\psi$ by its moment-matched Gaussian statistics. We note that this moment-matched approximation is made in all analogous PTA searches for ULDM \cite{Porayko:2018sfa, EPTA:2023sfo, EuropeanPulsarTimingArray:2023egv, NANOGrav:2023hvm}.

\textcolor{black}{
A recent study~\cite{Li:2025xlr} examined the role of ULDM non-Gaussianity in PTA datasets, finding its largest impact on statistics and inferred sensitivities that combine correlated timing residuals across multiple observational targets, while the single-target residual power was well approximated by a Gaussian description. The latter regime is the most relevant one for the forecasts developed here. Except for deliberately engineered artificial-satellite constellations with matched orbital periods, as discussed for Gaussian SGWBs in App.~\ref{app:NetworkCorrelations}, our target set does not rely on coherently combining many identical orbital responses to an ULDM signal. Instead, the projected reach is typically dominated by the target with the strongest resonant response. The results of Ref.~\cite{Li:2025xlr} therefore support the expected accuracy of the Gaussian approximation for the single-target sensitivities emphasized in this work. We leave a more complete treatment of sensitivity corrections from the exact non-Gaussian statistics of $\psi$ to future work.
}

Hence, taking it as a Gaussian random field, we can write $\ddot \psi$ as 
\begin{equation}
\begin{split}
\ddot \psi(t) = \int_0^\infty  \frac{\mathrm{d}\omega}{2\pi} \bigg[&(C_A(\omega) + C_B(\omega)) \cos(\omega t) \\
&+ (S_A(\omega) + S_B(\omega))\sin(\omega t) \bigg]
\end{split}
\end{equation}
where $C_A(\omega)$, $C_B(\omega)$, $S_A(\omega)$, and $S_B(\omega)$ are Gaussian random variables that are uncorrelated with each other and satisfy
\begin{equation}
\begin{gathered}
\langle C_A(\omega) C_A(\omega')\rangle = \langle S_A(\omega)S_A(\omega')\rangle =  A(\omega)\delta(\omega - \omega') \\ 
\langle C_B(\omega) C_B(\omega')\rangle = \langle S_B(\omega)S_B(\omega')\rangle =  B(\omega) \delta(\omega - \omega')
\end{gathered}
\end{equation}
where $A(\omega)$ and $B(\omega)$ are as defined in Eq.~\eqref{eq:ULDMPowerComponents}.

Recall that $A(\omega)$ provides support at frequencies between $0$ and $m_\mathrm{DM} \sigma^2 = 1/\tau_\mathrm{coh}$, while $B(\omega)$ provides support at frequencies between $2m_\mathrm{DM}$ and $2 m_\mathrm{DM}+2m_\mathrm{DM} \sigma_0^2 =2 m_\mathrm{DM}+1/\tau_\mathrm{coh}$. Then, in the limit of $t \ll \tau_\mathrm{coh}$ we have 
\begin{widetext}
\begin{equation}
\begin{split}
\ddot \psi(t) &= \int_0^{1/\tau_\mathrm{coh}} \frac{\mathrm{d} \omega}{2 \pi} \left[ C_A(\omega) \cos(\omega t) + S_A \sin(\omega t) + C_B(\omega) \cos(2 m_\mathrm{DM}t + \omega t) + S_B(\omega) \sin(2 m_\mathrm{DM}t + \omega t)  \right] \\
&\approx \int_0^{1/\tau_\mathrm{coh}} \frac{\mathrm{d} \omega}{2 \pi} \left[ C_A(\omega) + C_B(\omega) \cos(2 m_\mathrm{DM}t) + S_B(\omega) \sin(2 m_\mathrm{DM}t)  \right].
\end{split}
\end{equation}
\end{widetext}
Making the definitions:
\begin{equation}
\begin{gathered}
\tilde{C}_{A,B} = \int \frac{\mathrm{d} \omega}{2 \pi} C_{A,B}(\omega), ~~~ 
\tilde{S}_B = \int \frac{\mathrm{d} \omega}{2 \pi} S_B(\omega). \\
\end{gathered}
\end{equation}
then we have
\begin{equation}
    \ddot \psi(t) \approx   \tilde C_{A} + \tilde C_B \cos(2 m_\mathrm{DM}t) + \tilde S_{B} \sin(2 m_\mathrm{DM}t).
\end{equation}
Hence we see that $\ddot \psi$ is well-approximated by a plane-wave mode at frequency $2 m_\mathrm{DM}$ with a constant contribution coming from $\tilde C_{A}$. The non-fluctuating piece, having not undergone a single-low frequency oscillation in the short time considered here, is observationally unimportant for a single binary system\footnote{While the low-frequency component of $\ddot \psi$ does not fluctuate on these short timescales, it will vary spatial on distances greater than the coherence length of $\ddot \psi$ as inherited from the ULDM field, which may have observational relevance when comparing multiple systems.}. Hence, we neglect it in what follows and consider only 
\begin{equation}
\ddot \psi(t) = \tilde C_B \cos(2 m_\mathrm{DM}t) + \tilde S_{B} \sin(2 m_\mathrm{DM}t).
\end{equation}

Now let us inspect the statistics of the $\tilde C_B$ and $\tilde S_B$ further. Recall that $C_B(\omega)$ is Gaussian distributed, so $\tilde C_{B}$, as the integral over Gaussian distributed variables, must be as well. Identical logic applies to $\tilde S_{B}$ (and to $\tilde C_A$). In particular, we have
\begin{equation}
\begin{gathered}
    \langle \tilde C_B\rangle =  \langle \tilde S_B\rangle =  0, \\
    \langle \tilde C_B^2\rangle = \langle \tilde S_B^2\rangle = \frac{1}{4 \pi^2} \int \mathrm{d}\omega B(\omega),
\end{gathered}
\end{equation}
with $\tilde C_{B}$ and $\tilde S_B$ uncorrelated with each other. Since $\tilde C_B$ and $\tilde S_B$ follow identical Gaussian statistics, we may rewrite $\ddot \psi(t)$ as 
\begin{equation}
\label{eq:coh_psi}
\ddot \psi(t) = \alpha \left[\int\frac{\mathrm{d}\omega}{4 \pi^2} B(\omega) \right]^{1/2}\cos(2 m_\mathrm{DM}t + \varphi),
\end{equation}
where $\alpha$ is a Rayleigh-distributed variable with scale parameter $1$ and $\varphi$ is uniformly distributed on $[0, 2\pi)$, see, \textit{e.g.}, \cite{Foster:2017hbq} or App.~\ref{app:MonochromaticCoherence}.

Notably, this treatment generalizes and puts on rigorous footing that of \cite{Khmelnitsky:2013lxt}. Now, consider a $\ddot \psi(t)$ for given values of $\alpha$ and $\varphi$. Then we may treat it ``deterministically" in the sense that we may compute the induced perturbations given the values of $\alpha$, $\varphi$, and any other possibly relevant parameters $\bm{\theta}$. We will denote these induced perturbations by $\bm{\mu}(t | \alpha, \varphi, \bm{\theta}).$ 

Assuming observations with Gaussian distributed error $\sigma$ at a set of times $\{t_i\}$, we can make a treatment of observed data $\bm{d}$ using the likelihood 
\begin{equation}
    \mathcal{L}(\bm{d} | \alpha, \varphi, \bm{\theta}) = f_\alpha(\alpha)f_\varphi(\varphi)\prod_i \bm{\Phi}(\bm{d}_i | \bm{\mu}_i(\alpha, \varphi, \bm{\theta}), \sigma^2)
    \label{eq:MonochromaticLikelihood}
\end{equation}
where $f_\alpha$ is the probability density function for the Rayleigh-distributed $\alpha$, $f_\varphi$ is the probability density function for the uniformly distributed $\varphi$, and $\bm{\Phi}(d | \mu, \sigma^2)$ is the Gaussian probability density function for $d$ given mean $\mu$ and variance $\sigma^2$. We now emphasize two important points. First, the likelihood construction of Eq.~\eqref{eq:MonochromaticLikelihood} is valid only for observations that are much shorter than the coherence time. Given that first condition, although the likelihood in Eq.~\eqref{eq:MonochromaticLikelihood} differs from that of Eq.~\eqref{eq:FullLikelihood} in App.~\ref{app:FisherInformation}, they are fundamentally constructed from the same underlying Gaussian statistics of the ULDM and scalar potential, and so are equivalent in terms of their information content and statistical power.

%%%%%%%%%%%%%%%%%%%%%%%%%%%%%%%%%%%%%%%%%%%%%%%%%%%%%%%%%%%%%%%%%%%%%%%%%%%%%%%%%%%%%%%%%%%%%%%%%%
\section{Satellite networks and orbital correlations}
\label{app:NetworkCorrelations}
%%%%%%%%%%%%%%%%%%%%%%%%%%%%%%%%%%%%%%%%%%%%%%%%%%%%%%%%%%%%%%%%%%%%%%%%%%%%%%%%%%%%%%%%%%%%%%%%%%

An interesting future possibility to increase the chances of detection and confirmation of the origin of any
possible anomaly is the use of several probes sensitive to the same background. As a first exercise in this direction, we now consider the case where we have multiple satellites and evaluate the correlated evolution of those perturbed orbits. This will require only a modest extension of the previous computations. For each satellite, which we index with a capitalized superscript, we can compute a mapping of orbital perturbations to distance perturbations $\bm{T}^A$, a fundamental matrix for the first-order perturbations $\Phit^A$, and the solutions to the covariance integrals $\First^A$. We can then evaluate the correlation between the observables for arbitrary satellites driven by, \textit{e.g.}, a SGWB, as
\begin{equation}
\begin{split}
&\bm{\Sigma}_{AB}(t, t') = \bm{T}_{\alphat }^A(t)  \bm{T}_{\betat  }^B(t') \langle \tuple_1^{\alphat , A} (t) \tuple_1^{\betat  ,B}(t') \rangle \\
&~~= 3 \pi H_0^2 C_{ij,lm} \bm{T}_{\alphat }^A(t)  \bm{T}_{\betat  }^B(t') \Phit^A_{\alphat (p)}(t)\Phit^B_{\betat  (q)}(t')\\
&~~~~~ \times \int \mathrm{d}\nu \nu \Omega_\mathrm{GW}(\nu) 
\First^A_{(p)ij, d}(t |\nu) \First^B_{(q)lm, d}(t' |\nu)\,.
\end{split}
\end{equation}
Using this satellite cross-covariance, we can construct the complete covariance for perturbations induced in a network of satellites experiencing the same realization of a stochastic perturbation and perform a similar (though higher dimensional) analysis using our likelihood formalism. A key advantage of this scenario is that can allow for the leveraging of the particular correlation structure associated with GWs or ULDM across satellites to disentangle signals from confounding backgrounds, though we leave a more thorough examination of this possibility to future work. 

\bibliography{refs}
\end{document}